\documentstyle[seceq,epsf,subfigure,wrapfig]{ptptex}

\makeatletter
\@ifundefined{lesssim}{\def\lesssim{\mathrel{\mathpalette\vereq<}}}{}
\@ifundefined{gtrsim}{\def\gtrsim{\mathrel{\mathpalette\vereq>}}}{}
\def\vereq#1#2{\lower3pt\vbox{\baselineskip1.5pt \lineskip1.5pt
\ialign{$\m@th#1\hfill##\hfil$\crcr#2\crcr\sim\crcr}}}
\makeatother

\title{
Physical Processes in Naked Singularity Formation
}

\author{
Tomohiro {\sc Harada},$^{1,}$
\footnote{E-mail: harada@gravity.phys.waseda.ac.jp}
Hideo {\sc Iguchi}$^{2,}$\footnote{E-mail: iguchi@th.phys.titech.ac.jp}
and Ken-Ichi {\sc Nakao}$^{3}$\footnote{E-mail: knakao@sci.osaka-cu.ac.jp}
}

\inst{
$^{1}$Department of Physics, Waseda University, Tokyo 169-8555, Japan
\\
$^{2}$Department of Physics, Tokyo Institute of Technology, Tokyo 152-8550, Japan \\
$^{3}$Department of Physics, Osaka City University, Osaka 558-8585, Japan
}

\recdate{
December 28, 2001}

\abst{
Gravitational collapse is one of the most fruitful subjects 
in gravitational physics. 
It is well known that singularity formation is 
inevitable in complete gravitational collapse. 
It was conjectured that such a singularity should be hidden
by horizons if it is formed from generic initial data with 
physically reasonable matter fields. Many possible
counterexamples to this conjecture have 
been proposed over the past three
decades, although none of them has proved to be 
sufficiently generic.
In these examples,
there appears a singularity that is not hidden by horizons. 
This singularity is called a `naked singularity.'
The appearance of a naked singularity represents the 
formation of an observable 
high-curvature, strong-gravity region.
In this paper we review examples of
naked singularity formation and recent progress 
in research of observable physical processes -
gravitational radiation
and quantum particle creation -
from a forming naked singularity.}

\begin{document}

\maketitle
\tableofcontents

\section{Introduction: Review of naked singularity formation}
\label{sec:intro}
\subsection{Introduction}
According to the general theory of relativity, which is a classical gravity 
theory, after undergoing supernova explosion in the last stage of its 
evolution, a star with a mass dozens of times larger than the solar mass will 
contract without limit, due to its strong gravity, and form 
a ``domain'' called a ``spacetime singularity.'' 
Spacetime singularity formation is a very general phenomenon, 
not only in the gravitational collapse of stars of very large but also in 
physical processes in which the general theory of relativity plays an 
important role. In fact, it was proved 
by Hawking and Penrose~\cite{Penrose:1965wq,Hawking:1967,Hawking:1970sw}
that the appearance of spacetime singularity is 
generic, i.e., spacetime singularities appear for any spacetime symmetry.
However, the singularity theorems of Hawking and Penrose 
only prove the causally geodesic incompleteness
of spacetime and say nothing about the detailed features of 
the singularities themselves. For example, we cannot obtain information 
about how the spacetime curvature and the energy density diverge 
in a spacetime singularity from these theorems.  

Spacetime singularities can be classified into two kinds, according to 
whether or not they can be observed. A spacetime singularity
that can be observed is called a ``naked singularity'', while a typical 
example of a spacetime singularity that cannot be observed is a 
black hole. A singularity is a boundary of spacetime. Hence, 
in order to obtain solutions of hyperbolic field equations for 
matter, gauge fields and spacetimes themselves 
in the causal future of a naked singularity, we need to impose 
boundary conditions on it. However, we do not yet know any law of physics 
that determines reasonable boundary conditions on singularities.
Therefore, the existence of a naked singularity implies behavior 
that cannot be predicted with our present knowledge. 

Is such a naked singularity formed in our universe?
With regard to this question, Penrose proposed the so-called cosmic censorship 
conjecture.~\cite{Penrose:1969pc,Penrose:1979}
This conjecture represents one of the most important unsolved problems in
general relativity. Its truth is often assumed in the analysis
of physical phenomena in strong gravitational fields.
There are two versions of 
this conjecture. The weak conjecture states that all singularities in
gravitational collapse are hidden within black holes. This conjecture
implies the future predictability of the spacetime outside the event
horizon. The strong conjecture asserts that no singularity 
visible to any observer can exist.  
It states that all physically reasonable spacetimes are globally hyperbolic.
Unfortunately, no one has succeeded in giving a mathematically rigorous 
and provable formulation of either versions of 
the cosmic censorship conjecture.

If naked singularities are formed frequently 
in our universe, then they have a very important meaning in the experimental
study of physics in high-energy, high-density regimes.
Solutions of theoretical models exhibiting naked singularity 
formation may be nothing more than unrealistic behavior of 
toy models and cannot be considered as providing proof of 
the existence of naked singularities in our universe. 
However, some such solutions are likely to become 
important footholds from which we can 
advance research of spacetime singularities. 
In fact, the possibility of naked singularity formation 
in a large linear collider has recently been suggested 
in the context of the scenario of large extra dimensions.~\cite{Casadio:2001}
In this paper we review recent progress in theoretical research 
on naked singularity formation and
physical processes contained therein. In the remaining 
of this section, we outline
examples of naked singularities in the case of 
spherically symmetric gravitational 
collapse. We also outline the hoop conjecture and research on
axisymmetric and cylindrical collapse. In \S \ref{sec:perturb} we 
review recent works on nonspherical perturbations of spherical 
collapse and the gravitational radiation from a forming naked singularity.
In \S \ref{sec:quantum} we review quantum particle creation from 
a forming naked singularity.
We use units in which $G=c=\hbar=1$.

\subsection{Spherical dust collapse}
\label{sec:dust}
In order to examine the validity of the cosmic censorship
conjecture, it is clear that we have to deal with 
dynamical spacetimes. 
In this context, we often assume some kind of symmetries for
the spacetime, because the introduction of symmetry 
makes the analysis much easier. 
Among such symmetric spacetime, spherically symmetric spacetimes have been 
the most widely studied partly because for them 
the analysis becomes much simpler
due to the absence of gravitational waves, 
and partly because we can expect that some class of 
realistic gravitational collapse 
can be treated as spherically symmetric with small deviations from it.
Here we review spherically symmetric gravitational collapse
and the appearance of naked singularities 
in spherically symmetric spacetimes.

\subsubsection{Spherically symmetric spacetime}
\label{sec:generalspherical}
Before restricting matter fields, we present the 
Einstein equation for a general spherically symmetric 
spacetime.
In spherically symmetric spacetime, without loss of generality,
the line element can be written in diagonal form as
\begin{equation}
  ds^2=-e^{2\nu(t,r)}dt^2+e^{2\lambda(t,r)}dr^2+R^2(t,r)(d\theta^2
  +\sin^2\theta d\phi^2).
  \label{eq:lineelement}
\end{equation}
Here we adopt a comoving coordinate system, which is possible for
matter fields of type I. (See Ref.~\citen{he1973} for classification 
of matter fields.)
In this coordinate system,
the stress-energy tensor $T^{\mu}_{~\nu}$ that is 
the source of a spherically symmetric gravitational field
must be of the form
\begin{eqnarray}
  T^{\mu}_{~\nu}&=& \left(\begin{array}{cccc}
    -\rho & 0 & 0 & 0 \\
    0 & \Sigma & 0 & 0 \\
    0 & 0 & \Pi & 0 \\
    0 & 0 & 0 & \Pi 
    \end{array}\right),
  \label{eq:generalstressenergy}
\end{eqnarray}
where $\rho(t,r)$, $\Sigma(t,r)$ and $\Pi(t,r)$ are
the energy density, radial stress and
tangential stress, respectively.
If we consider a perfect fluid, which is described by
\begin{equation}
  T^{\mu\nu}=(\rho+P)u^{\mu}u^{\nu}+Pg^{\mu\nu},
   \label{eq:perfectfluid}
\end{equation}
then the stress is isotropic, i.e.,
\begin{equation}
  \Sigma=\Pi=P.
\end{equation}

>From the Einstein equation and the equation of 
motion for the matter fields, we obtain
\begin{eqnarray}
  m^{\prime}&=&4\pi\rho R^2 R^{\prime}, 
  \label{eq:mprime} \\
  \dot{m}&=&-4\pi\Sigma R^2 \dot{R},
  \label{eq:mdot}\\
  \dot{R}^{\prime}&=&\dot{R}\nu^{\prime}+R^{\prime}\dot{\lambda},
  \label{eq:rprime}\\
  \Sigma^{\prime}&=&-(\rho+\Sigma)\nu^{\prime}-2(\Sigma-\Pi)
  \frac{R^{\prime}}{R}, 
  \label{eq:sigmaprime}\\
  m&=&\frac{R}{2}\left(1-R^{\prime 2}e^{-2\lambda}+\dot{R}^2e^{-2\nu}
    \right)
  \label{eq:m},
\end{eqnarray}
where $m=m(t,r)$ is the Misner-Sharp mass,~\cite{ms1964}
and the prime and dot denote partial derivatives 
with respect to $t$ and $r$, respectively.

Here we stipulate the existence of an apparent horizon,
which is defined as the outer boundary
of a connected component of the trapped region.
The important feature of the apparent horizon is that,
if the spacetime is strongly asymptotically predictable 
and the null convergence
condition holds,
the presence of the apparent horizon 
implies the existence of an event horizon
outside or coinciding with it.
If the connected component of the trapped region 
has the structure of a manifold with boundaries,
then the apparent horizon is an outer marginally 
trapped surface with vanishing expansion.
(See Ref.~\citen{he1973} for the definitions and proofs.)
Along a future-directed outgoing null geodesic, the relation 
\begin{equation}
  \frac{dR}{dt}=\dot{R}+R^{\prime}\frac{dr}{dt}
  =e^{\nu}\left(\pm\sqrt{-1+\frac{2m}{R}+R^{\prime 2} e^{-2\lambda}}
  +R^{\prime}e^{-\lambda}\right)
\label{eq:expansion}
\end{equation}
is satisfied,
where the upper and lower signs correspond to expanding and 
collapsing phases, respectively, and we assume $R^{\prime}>0$.
Therefore, in the expanding phase,
there is no apparent horizon.
In the collapsing phase, on a hypersurface of constant $t$, 
the two-sphere $R=2m$ is an apparent horizon.
The region $R<2m$ is trapped ($dR/dt<0$), while
the region $R>2m$ is not trapped ($dR/dt>0$). 

Here we should describe singularities that may appear 
in spherically symmetric collapse.
A shell-crossing singularity is one characterized by
$R^{\prime}=0$ and $R>0$, while a shell-focusing singularity
is one characterized by $R=0$.
Also a central singularity is one characterized by $r=0$,
while a non-central singularity is one
characterized by $r>0$, where $r=0$ is chosen as the symmetric center.
It is noted that, since 
\begin{eqnarray}
  R^{\mu}_{~~\mu}&=&8\pi(\rho-\Sigma-2\Pi) \\
  R^{\mu\nu}R_{\mu\nu}&=&64\pi^2(\rho^2+\Sigma^2+2\Pi^2) 
\end{eqnarray}
are satisfied,
the divergence of the energy density or stress directly implies
a scalar curvature singularity.  
As for strength of singularities, two conditions are 
often used; one is the strong curvature condition proposed by 
Tipler,~\cite{tipler1977} and the other is the limiting focusing condition
proposed by Kr\'olak.~\cite{krolak1987} 
The former condition is stronger than the latter.
(See also Ref.~\citen{clarke1993}.)
These conditions are defined in terms of the speed of growth of
the spacetime curvature along a geodesic that terminates
at the singularity.
A shell-crossing naked singularity
is known to be weak in terms of both conditions.
It is expected that a spacetime with 
weak singularity can be extended further
in a distributional sense, although it is not known
how this is possible in general situations.  

\subsubsection{Dust collapse}
\label{sec:LTB}
First we consider a dust fluid, which is defined as a pressureless fluid.
This model has been most widely studied,  
partly because of the existence of an exact solution,
which is called the Lema\^{\i}tre-Tolman-Bondi (LTB) 
solution,~\cite{lemaitre1933,tolman1934,bondi1947}
and partly because it provides a nontrivial 
interesting example of naked singularity formation.
For this reason, we describe this model in some detail here.  

We restrict the matter content of the model to a dust fluid. Therefore 
\begin{eqnarray}
  \Sigma&=&0, \\
  \Pi&=&0.
\end{eqnarray}
Then, Eqs.~(\ref{eq:mprime})--(\ref{eq:m}) can be integrated as
\begin{eqnarray}
  m&=&\frac{F}{2},
  \label{eq:mconservetb}\\
  \rho&=&\frac{F^{\prime}}{8\pi R^2 R^{\prime}}, 
  \label{eq:energydensitycomovingtb}\\
  e^{2\lambda}&=&\frac{R^{\prime 2}}{1+f}, 
  \label{eq:grrtb} \\
  \nu&=&0,
  \label{eq:lapsetb}\\
  \dot{R}^2 &=&f+\frac{F}{R},
  \label{eq:energyofparticletb}
\end{eqnarray}
where the arbitrary functions $F=F(r)$ and $1+f=1+f(r)>0$ are 
twice the conserved
Misner-Sharp mass and the specific energy, respectively.
In Eq.~(\ref{eq:lapsetb}), 
we have used the rescaling freedom of the time coordinate.
This means that a synchronous comoving coordinate system
is possible.
Equation~(\ref{eq:energyofparticletb}) can be integrated to yield
\begin{equation}
  t=\pm\left[\frac{R^{3/2}}{\sqrt{F}}G\left(-\frac{fR}{F}
  \right)\right]^{R}_{R^{0}}, 
\label{eq:tandr}
\end{equation}
where $R^{0}(r)$ and $G(y)$  are defined as 
\begin{eqnarray}
  R^{0}(r)&\equiv& R(0,r), \\ 
  G(y)&\equiv&\left\{
    \begin{array}{rl}
      \displaystyle{\frac{\mbox{Arcsin}\sqrt{y}}{y^{3/2}}}
        -\displaystyle{\frac{\sqrt{1-y}}{y}}
      ,\quad&\quad \mbox{for $0<y\le 1$} \\
      \displaystyle{\frac{2}{3}},\quad\quad\quad\quad\quad\quad 
      &\quad \mbox{for $y=0$} \\
      \displaystyle{\frac{-\mbox{Arcsinh}\sqrt{-y}}{(-y)^{3/2}}}
        -\displaystyle{\frac{\sqrt{1-y}}{y}},
      &\quad \mbox{for $y<0$},
    \end{array}\right.,\\
  \label{eq:functiong}
\end{eqnarray}
the quantity $[Q(R)]^R_{R^0}$ is defined as 
\begin{equation}
  \left[Q(R)\right]^R_{R^0}\equiv Q(R)-Q(R^0),
\end{equation}
and the upper and lower signs in Eq.~(\ref{eq:tandr}) 
correspond to expanding and collapsing phases, respectively.
Hereafter, our main concern is with the collapsing phase.

Assuming that $R$ is initially a monotonically increasing function 
of $r$, and rescaling the radial coordinate $r$, 
we identify $r$ with the circumferential radius $R$ 
on the initial space-like hypersurface $t=0$.
Then, regularity of the center requires
\begin{eqnarray}
  f(0)&=&0, \\
  R(t,0)&=&0, \\
  \frac{F(r)}{r^3} &<& \infty~~\mbox{at}~~r\to 0.
\end{eqnarray}
The solution can be matched with the Schwarzschild spacetime
at an arbitrary radius $r=r_b$ if we identify the Schwarzschild
mass parameter with $F(r_b)/2$.

It is helpful for later use to write down the solution
for marginally bound collapse, i.e., the case of $f(r)=0$.
This solution is 
\begin{equation}
  \label{eq:radius}
  R = \left(\frac{9F}{4}\right)^{1/3}(t_{s}-t)^{2/3},
\end{equation}
where $t_{s}=t_{s}(r)$ is given by
\begin{equation}
  \label{t0}
  t_{s}(r) = \frac{2}{3\sqrt{F}}r^{3/2}.
\end{equation}
Then, $R$ can also be written as
\begin{equation}
R=r \left(1-\frac{3}{2}\sqrt{\frac{F}{r^3}} t\right)^{2/3}.
\end{equation}
The matching condition with the Schwarzschild spacetime
yields the relation between the Schwarzschild time coordinate $T$ and 
the synchronous comoving time coordinate $t$ at $r=r_{b}$ as
\begin{equation}
T=t-2\sqrt{2MR}+2M\ln\frac{\sqrt{R}+\sqrt{2M}}
{\sqrt{R}-\sqrt{2M}},
\label{matching}
\end{equation}
where an integral constant has been absorbed through the 
redefinition of $T$. 

Again we go back to general LTB solutions.
These solutions allow shell-crossing singularities 
which may be naked.~\cite{ysm1973}
Hereafter, we concentrate on shell-focusing singularities.
Equation~(\ref{eq:energyofparticletb}) implies that every
mass shell labelled by $r$ that is initially collapsing
inevitably results in shell-focusing singularity.
It is easily found that the time at which the 
shell-focusing singularity appears, $t_{s}(r)$,  
and the time of the apparent horizon, $t_{AH}(r)$, 
are given by
\begin{eqnarray}
  t_{s}(r)&=&\frac{r^{3/2}}{\sqrt{F}}G\left(-\frac{fr}{F}\right), \\
  t_{AH}(r)&=&t_{s}(r)-FG(-f).
\end{eqnarray}
Therefore, a shell-focusing singularity that appears at $r>0$ is 
in the future of the apparent horizon. 

A non-central shell-focusing singularity is not naked.
Indeed, suppose that a light 
ray emanates from a shell-focusing singularity
at some $r_{1}>0$, which is given by $t=t(r)$.
Then, by continuity there must exist an $\epsilon>0$
such that for $r_{1}<r<r_{1}+\epsilon$ a light ray 
with positive expansion is later than the
apparent horizon and earlier than the shell-focusing singularities, 
since the apparent horizon is earlier than the 
shell-focusing singularities everywhere but at the center. 
This implies that $0<R(t(r),r)<F(r)$ and $dR/dt(t(r),r)>0$.
By Eq.~(\ref{eq:expansion}) these relations lead to a contradiction.
We thus conclude that shell-focusing singularities 
(except possibly for central shell-focusing singularities) are 
not visible to an observer. Therefore it is sufficient 
to consider central shell-focusing singularities 
in order to examine whether or not strong naked
singularities exist.
By the above argument, a light ray that emanates
from a singularity must lie 
in the past of the apparent horizon.

The LTB solution from generic regular initial
data results in a shell-focusing naked singularity at 
the center, $r=0$.
In order to show the existence of a naked singularity,
we investigate the geodesic equation for an
outgoing radial null geodesic that emanates from the singularity.
For this purpose, we derive the root
equation that probes the naked singularity, 
following Joshi and Dwivedi.~\cite{jd1993}
An outgoing radial null geodesic is given as
\begin{equation}
  \frac{dr}{dt}=e^{\nu-\lambda}.
  \label{eq:nulltb}
\end{equation}
Here we define
\begin{equation}
  x\equiv \frac{R}{r^{\alpha}},
  \label{eq:defx}
\end{equation}
where $\alpha>1$ is determined by requiring $x$ to have a positive
finite limit $x_0$ as $r\to 0$.
Note that the regular center corresponds to $\alpha=1$.
Then, from l'Hospital's rule, we obtain
\begin{eqnarray}
  x_0&=& \lim_{r\to 0}\frac{R}{r^{\alpha}} \nonumber \\
  &=&\left.\lim_{r\to 0}\frac{1}{\alpha
  r^{\alpha-1}}\frac{dR}{dr}\right|_{R=x_0 r^{\alpha}} \nonumber
  \\
  &=&\left.\lim_{r\to 0}\frac{1}{\alpha r^{\alpha-1}}\left(R^{\prime}
    +e^{\lambda-\nu}\dot{R}\right)\right|_{R=x_0 r^{\alpha}}.
\end{eqnarray}
Substituting the LTB solution,
we obtain
\begin{equation}
  x_0=\lim_{r\to 0}\left.\frac{R^{\prime}}{\alpha r^{\alpha-1}}
  \left(1-\frac{\sqrt{f+\frac{F}{R}}}{\sqrt{1+f}}\right)
  \right|_{R=x_0 r^{\alpha}}.
\end{equation}
In order to obtain the root equation for the LTB solution,
we must have an explicit expression for $R^{\prime}$.
By differentiating both sides of Eq.~(\ref{eq:tandr})
with respect to $r$,
we obtain such an expression of $R^{\prime}$, after a straightforward 
but rather lengthy calculation, as
\begin{equation}
  R^{\prime}=N(x,r) r^{\alpha-1},
  \label{eq:rprimebehavior}
\end{equation}
where $N(x,r)$ is given by
\begin{equation}
  N(x,r)\equiv (\eta-\beta)x+\left[\Theta-\left(\eta-\frac{3}{2}\beta
      \right)
      x^{3/2}G(-Px)\right]\sqrt{P+\frac{1}{x}},
\end{equation} 
with 
\begin{eqnarray}
  \eta(r)&\equiv&\frac{rF^{\prime}}{F}, \\
  \beta(r)&\equiv&\left\{\begin{array}{cc}
    \displaystyle \frac{rf^{\prime}}{f},& {\mathrm for}~~f\ne 0, \\ 
    0,                    & {\mathrm for}~~f=0, \end{array}\right. \\
  p(r)&\equiv&\frac{rf}{F}, \\
  P(r)&\equiv&pr^{\alpha-1}, \\
  \Lambda(r)&\equiv&\frac{F}{r^{\alpha}}, \\
  \Theta(r)&\equiv&\frac{1+\beta-\eta}{(1+p)^{1/2}r^{3(\alpha-1)/2}}
  +\frac{\left(\eta-\frac{3}{2}\beta\right)G(-p)}{r^{3(\alpha-1)/2}}.
\end{eqnarray}
Therefore, the desired equation is 
\begin{equation}
  x_0=\frac{N(x_0,0)}{\alpha}\lim_{r\to 0}
  \left(1-\frac{\sqrt{f+\frac{\Lambda}{x_0}}}{\sqrt{1+f}}\right).
  \label{eq:rooteq}
\end{equation}
Note that $\alpha$ should be determined by the requirement
that $\Theta(r)$ have a finite limit as $r\to 0$. 

For simplicity, we assume that $F(r)$ and $f(r)$ are of the forms
\begin{eqnarray}
  F(r)&=&F_3r^3+F_5r^5+F_7 r^7+\cdots,
  \label{eq:Fodd}\\
  f(r)&=&f_2r^2+f_4r^4+f_6 r^6+\cdots.
  \label{eq:feven}
\end{eqnarray}
This implies that the density and specific
energy fields are initially 
not only finite but also analytic at the symmetric center.
That is, the initial density and specific energy profiles
are analytic functions with respect to the locally Cartesian 
coordinates.
Hereafter we assume $F_3>0$, which ensures the positiveness of 
the central energy density at $t=0$.  
For marginally bound collapse, which is defined by $f=0$, 
a positive finite root of Eq.~(\ref{eq:rooteq}) 
is obtained for $F_5<0$ as
\begin{equation}
  x_0=\left(-\frac{F_5}{2F_3}\right)^{2/3},
   \label{eq:x0}
\end{equation}
with $\alpha=7/3$.
$F_5<0$ means $\rho^{\prime\prime}(0,0)<0$.
Therefore, there exists a naked singularity
in marginally bound collapse
with $\rho^{\prime\prime}(0,0)<0$ initially.
The nakedness of the singularity in this spacetime 
was found numerically by Eardley and Smarr~\cite{es1979}
and proved by Christodoulou.~\cite{christodoulou1984}
For marginally bound collapse with $F_5=0$, 
it is easily found that the root equation (\ref{eq:rooteq})
has no positive finite root for any $\alpha>1$.
Therefore, in this case,
the singularity is not naked. 
For homogeneous marginally bound collapse, 
the singularity is not naked, because $F_5=0$. 
For the non-marginally bound case $f\ne 0$,
a similar but more complicated 
condition for the appearance of a naked singularity
is obtained, and it was shown that the appearance of 
a naked singularity is generic in the space of 
this class of functions.~\cite{christodoulou1984,sj1996,jj1997}

We can also consider a more general class of $F(r)$ and $f(r)$ 
of the forms
\begin{eqnarray} 
  F(r)&=&F_3r^3+F_4 r^4+F_5r^5+F_6 r^6+F_7 r^7+\cdots,
  \label{eq:Finteger}\\
  f(r)&=&f_2r^2+f_3 r^3+f_4r^4+f_5 r^5+f_6 r^6+\cdots.
  \label{eq:finteger}
\end{eqnarray}
This class of the functions corresponds 
to initial density and specific energy 
distributions that are finite but not analytic in general 
with respect to locally Cartesian coordinates.
It was shown that a naked singularity also 
appears from a generic data set
in this extended space of 
functions.~\cite{sj1996,jj1997}
For marginally bound collapse with $F_3>0$, $F_4=F_5=0$ and 
$F_6 < -(26+15\sqrt{3})F_{3}^{5/2}/2 $,
Eq.~(\ref{eq:rooteq}) has a finite positive root $x_{0}$
with $\alpha=3$, and hence the singularity is naked, where
$x_{0}$ is given by the root of some quadratic equation.

We can also consider self-similar dust collapse.
Self-similarity requires all dimensionless quantities 
to be functions of $\bar{r}/t$ for some comoving coordinates $(t,\bar{r})$,
where $\bar{r}$ is different from $r$ in general. 
This requirement implies the functional forms 
$F(\bar{r})=\lambda \bar{r}$ and $f(\bar{r})=0$, where $\lambda$ is constant.
It is found that the singularity is naked for $0<\lambda\le 6(26-15\sqrt{3})$
and not naked for larger values of $\lambda$.
In fact, through some manipulation, it is found that 
the self-similar case can be classified into the 
marginally bound case with $F_3>0$, $F_4=F_5=0$ and $F_{6}<0$,
which implies that an analytic initial density profile is not allowed
for the self-similar case.

With regard to the global visibility of a naked singularity,
we can give a simple answer.
If the Taylor expansions of $F(r)$ and $f(r)$ around the center 
are such that a naked singularity appears,
we can immediately construct not only spacetimes
with globally naked singularities but also
those with locally naked singularities 
by choosing the functions $F(r)$ and $f(r)$.
In other words, local visibility is determined only by the 
central expansions of the two arbitrary functions,
while global visibility depends on
their functional forms in the whole range $0<r<\infty$.
In self-similar dust collapse, if a singularity is naked, then it
is globally naked.

In contrast to shell-crossing singularities,
shell-focusing naked singularities satisfy
the limiting focusing condition for $\alpha=7/3$,~\cite{newman1986}
and even both the limiting focusing and strong curvature
conditions for $\alpha=3$ along the first
radial null geodesic from the singularity.~\cite{sj1996,jj1997}
Irrespective of the value of $\alpha$,
the shell-focusing singularity satisfies both conditions 
for time-like geodesics.~\cite{djd1999}
For the marginally bound case, the redshift of the first light 
is finite for $\alpha=7/3$~\cite{christodoulou1984}
but infinite for $\alpha=3$.~\cite{dwivedi1998}

The LTB spacetime with a shell-focusing singularity 
is inextendible.~\cite{es1979} Detailed analysis
shows that a naked singularity which may
appear in spherical dust collapse is ingoing 
null.~\cite{christodoulou1984,newman1986}
This means that there exists a one-parameter family of 
outgoing radial null geodesics that emanate from the singularity, 
while there exists only one ingoing radial null geodesic 
that terminates at the singularity. 
Very recently, it was shown that nonradial null geodesics,
which have nonzero angular momentum, can emanate from
a singularity if and only if a radial null geodesic
emanates from the singularity.~\cite{mn2001} The appearance 
of the naked singularity to a distant observer
through these nonradial null geodesics has been discussed.~\cite{djd2001}
 
As we have seen above, 
the collapse of an inhomogeneous dust ball, which is given by
the LTB solution, results in a shell-focusing 
naked singularity from generic regular initial data.
The collapse of a homogeneous dust ball, which is called
the Oppenheimer-Snyder solution,~\cite{os1939} 
results in a covered singularity.
Though the Oppenheimer-Snyder solution was
thought to be a typical example of complete gravitational
collapse, the absence of a naked singularity in this solution 
turns out to be atypical in general spherically symmetric
dust collapse.

\subsection{Spherical collapse of realistic matter fields}
It is clear that a dust fluid is not a good matter model, because 
the effects of pressure would not be negligible 
in actual singularity formation.
Here we briefly review several examples 
of gravitational collapse in the presence of 
the effects of some kind of pressure, 
in the context of singularity formation. 
We do not describe the details of each example, 
because this is not the aim of this paper.

\subsubsection{Perfect fluid collapse}
The perfect fluid matter model is one of the most 
natural ways of introducing matter pressure.
If the pressure is bounded from above, 
there can appear a shell-crossing naked 
singularity,~\cite{mys1974} which is gravitationally
weak. We will see below that there can appear a 
strong curvature naked singularity
even with unbounded pressure. 
 
Ori and Piran~\cite{op1987,op1988,op1990} 
investigated the self-similar, spherically symmetric 
and adiabatic gravitational collapse of a perfect fluid with a barotropic
equation of state.
Employing a self-similarity assumption, the equation of state can be 
restricted to the form $P=k\rho$, and 
the Einstein field equations are reduced to
a system of ordinary differential equations.
For $0<k\le 0.4$, Ori and Piran found numerically 
that there is a discrete set of 
self-similar solutions that allow analytic 
initial data beyond a sonic point.
These solutions can be labelled by the 
number of zeroes in the velocity
field of world lines of constant circumferential radius
relative to the fluid element. 
There exists a pure collapse solution among these
self-similar solutions, which they call the general relativistic
Larson-Penston solution.
They showed that a central naked singularity 
forms in this Larson-Penston solution for $0<k \lesssim 0.0105$.
They also showed that there are naked-singular
solutions with oscillations in the velocity field
for $0<k\le 0.4$.
The results of this work were confirmed and extended
to $0<k\le 9/16$.~\cite{fh1993}
This naked singularity is ingoing null.
It was shown that this naked singularity 
satisfies both the limiting focusing condition and
the strong curvature condition 
for the first null ray.~\cite{lake1988,wl1989}

However, it is obvious that 
all initial data sets from which self-similar
spacetimes develop occupy zero measure in the
space of all spherically symmetric regular initial data sets.
Taking this fact into consideration, there have been discussions that
the emergence of the naked singularity may be
an artifact of the assumption of self-similarity.
In order to judge the necessity of self-similarity assumption,
Harada~\cite{harada1998} numerically simulated the spherically
symmetric and adiabatic gravitational 
collapse of a perfect fluid with the same 
equation of state $P=k\rho$ without this assumption.
Since null coordinates were used in these simulations, 
he could detect naked singularities, not relying upon 
the absence of an apparent horizon.
The result was that naked singularities develop from
rather generic initial data sets for $0<k\lesssim 0.01$,
which is consistent with the result of Ori and Piran 
obtained using the self-similarity assumption.
In fact, through further numerical simulations by 
Harada and Maeda,~\cite{hm2001} it was found 
that generic spherical collapse approaches 
the Larson-Penston self-similar solution 
in the region around the center, at least for $0<k\le 0.03$. 
This finding is supported by a linear stability analysis 
of self-similar solutions.
In other words, the Larson-Penston self-similar solution
is an attractor solution in the spherical gravitational collapse
of a perfect fluid with the equation of state 
$P=k\rho$, at least for $0<k\le 0.03$.
This work provided the first evidence of the generic nature  
of the appearance of naked singularities  
in spherically symmetric spacetimes with perfect fluids.
This work also provided the first nontrivial 
evidence of the attractive 
nature of a self-similarity solution
in gravitational collapse. 
Although the final fate of the generic spherical 
collapse of a perfect fluid with larger 
values of $k$ is not known, 
Harada~\cite{harada2001} analytically
showed that the Larson-Penston self-similar 
solution is no longer stable, because of the so-called 
kink instability, for $k\gtrsim 0.036$.

Undoubtedly, self-similarity plays an important role
in certain circumstances of gravitational collapse
in this attractive case and also in the critical case described below.
In particular, the former is direct evidence 
supporting the self-similarity hypothesis
that spherically symmetric
spacetime might naturally evolve from complicated 
initial conditions into a self-similar form,
which was originally proposed in the context of 
cosmological evolution.~\cite{carr2000}
(For a recent review of a spherically symmetric 
self-similar systems with perfect fluids,
see Ref.~\citen{cc1999}.)

\subsubsection{Critical collapse}
Critical behavior in gravitational
collapse was discovered by Choptuik.~\cite{choptuik1993}
He investigated the threshold between collapse to a black hole
and dispersion to infinity in a spherically symmetric 
self-gravitating system of a massless scalar field, i.e., 
an Einstein-Klein-Gordon system.
He found so-called critical behavior,
such as a scaling law for the formed black hole mass,
which is analogous to that in statistical physics. 
He also found that there is a discrete self-similar solution that sits 
at the threshold of black hole formation, which is called 
a ``critical solution.'' Similar phenomena have been observed 
in the collapse of various kinds of matter fields, 
for example, axisymmetric gravitational waves,~\cite{ae1993} 
radiation fluid~\cite{ec1994} and 
more general perfect fluids.~\cite{nc2000a,nc2000b} 
A renormalization group approach applied by Koike et al.~\cite{kha1995}
gives a simple physical explanation of the critical phenomena and 
showed that the critical solution is characterized by a single
unstable mode.
Recently, critical behavior was found to possibly exist 
even in the Newtonian collapse of an isothermal gas.~\cite{mh2001}
(For a recent review of critical phenomena in gravitational 
collapse, see Ref.~\citen{gundlach1999}.)

A consequence of the mass scaling law for a black hole
is the appearance of a ``zero-mass black hole.''
A zero-mass black hole can be regarded as a naked singularity, 
because the curvature strength on the black hole horizon 
is proportional to the inverse square of the black hole mass.
It is obvious that zero-mass black hole formation
is unstable, because it is a result of exact fine-tuning.
On the other hand, if we take the limitation of 
general relativity into consideration, the future predictability 
of classical theory breaks down for a finite-measure
set of parameter values in this model. 
The critical collapse of a scalar field 
provides a very important example of 
naked singularity because it is the first example of 
a naked singularity that develops from regular initial data
in the collapse of elementary fields.

Several theorems regarding the appearance 
and instability (non-generic nature) of naked singularities 
in a spherically symmetric Einstein-Klein-Gordon system
have been proved by 
Christodoulou.~\cite{christodoulou1987,christodoulou1991,christodoulou1993,christodoulou1994,christodoulou1999}

\subsubsection{Collapse of collisionless particles}
One interesting example of spherical collapse is a
spherical self-gravitating system of counterrotating particles,
i.e., an Einstein cluster.
The static system was considered by Einstein,~\cite{einstein1915}
and the corresponding dynamical system was considered by
Datta,~\cite{datta1970} Bondi,~\cite{bondi1971} and Evans.~\cite{evans1976}
This system can be constructed by putting
infinitely many collisionless particles orbiting
around the symmetric center
with a single radial velocity at any given radius,
so that the system is spherically symmetric.
Although each particle has conserved angular momentum,
the total angular momentum vanishes, due to the spherical symmetry.
This system is an example of a matter field with
vanishing radial stress.~\cite{magli1997,magli1998,hni1999}
The metric functions for a dynamical cluster of counterrotating
particles are written in terms of an elliptic
integral.~\cite{magli1998,hin1998}
This system has three arbitrary functions, which determine
the initial mass distribution $m(r)$, the energy distribution $f(r)$, and
the angular momentum distribution $L(r)$.
These three are all conserved in this system.
It was shown that there appears a naked singularity for some class of
these arbitrary functions that corresponds to regular
initial data,~\cite{hin1998,jm2000}
although this appearance is not generic
in the space of all regular initial data sets.
In particular, for marginally bound collapse with
some specific angular momentum distribution $m(r)=4L(r)$,
the metric functions can be expressed in terms of elementary functions
alone,
and this collapse results in naked singularity
formation, irrespective of the initial density profile.
Detailed analysis shows that this naked singularity is
time-like, unlike naked singularities
in spherical dust collapse.~\cite{khi2000}
In fact, this spacetime dynamically asymptotically
approaches the static model
with a central time-like naked singularity.
This naked singularity satisfies the limiting focusing condition
for the first null ray, and
even the strong curvature condition
for a time-like geodesic.~\cite{hni1999,khi2000}

The above described model is a special realization of a
self-gravitating system
of collisionless particles, i.e., an Einstein-Vlasov system.
This system can be described by a distribution function,
which obeys the Vlasov equation.
(See Ref.~\citen{rendall2000} for a recent review of this system.)
The global existence theorem of regular solutions with $C^{1}$
initial data for the distribution function in the Newtonian
counterpart of this system, the Poisson-Vlasov system,
has been proved.~\cite{pfaffelmoser1992}
This implies that the singularity which may form for $C^{1}$
initial data in the Einstein-Vlasov system
is not ``matter-generated.''~\cite{rendall1992}
For a spherically symmetric Einstein-Vlasov system
with $C^{1}$ initial data,
the global existence theorem of regular solutions
with small initial data~\cite{rr1992}
and the regularity theorem away from the
symmetric center~\cite{rrs1995} have been proved.

\subsubsection{Other examples}
We should mention several additional examples of naked singularities.
Although we have restricted our attention to type I matter,
there exists a spherical example of a naked singularity with type II matter.
If we consider imploding ``null dust'' into the center,
the spacetime is given by the Vaidya
metric.~\cite{vaidya1943,vaidya1951a,vaidya1951b}
For this spacetime, it has been shown that naked singularity
formation is possible from regular initial data.~\cite{jd1992}

There exists a spherical ``quasi-spherical'' solution in the case of
a dust fluid, which is called the Szekeres solution.~\cite{szekeres1975}
This solution can be regarded as a deformation of
spherical dust collapse. It has been shown that shell-focusing
singularities are possible
in this solution, and the
conditions necessary for the appearance of
naked singularities are very similar to those in the case of spherical
dust collapse.~\cite{jk1996}
The global visibility of this singularity was also
examined.~\cite{djj1998}

There have been many analyses with a set of assumptions and
the conditions for the appearance of a naked singularity
are written down in terms of
the energy density, radial stress and tangential stress.
In these analyses, in general the conclusion is
that naked singularities are possible for generic matter
fields that satisfy some energy conditions.
In this approach, this conclusion is very natural, because
there remains great freedom in the choice of matter fields, even if
some energy condition is imposed.

Finally, we should mention locally naked singularities
in black hole spacetimes.
It is well known that a Reissner-Nordstr\"om
black hole has a time-like naked singularity in the interior
of the event horizon.
There also exist locally naked singularities
in the general Kerr-Newmann-de Sitter family of
black hole spacetimes. Although these singularities do not
violate the weaker version
of the cosmic censorship conjecture,
they are inconsistent with the stronger version.
For these locally naked singularities, 
there has been a great amount of evidence suggesting that
the Cauchy horizon is unstable and that it might be replaced by
covered null singularities in the presence of
perturbations.~\cite{ch1982,pi1990,bp1992,bmm1998,hp1998,ori1998,ori1999}

\subsection{Hoop conjecture: axisymmetric and cylindrical collapse}
\label{sec:hoop}
An important conjecture concerning black hole formation following
gravitational collapse was made by Thorne.\cite{Thorne:1972ji}
This so-called hoop conjecture states that black holes with horizons form
when and only when a mass $M$ gets compacted into a region whose
circumference in every direction in $C \lesssim 4 \pi M$.
He analyzed the causal structure of cylindrically symmetric spacetimes
and found remarkably different nature from spherically symmetric spacetimes.
The event horizon cannot exist in cylindrically symmetric spacetime.
Therefore, if an infinitely long cylindrical object gravitationally
collapses to a singularity, it becomes a naked singularity, not a black hole
covered by an event horizon. 
Then, it is natural to ask what happens 
if a very long but finite object collapses.
The hoop conjecture was derived from such
a thought experiment. It should be noted that the hoop conjecture itself
does not assert that a naked singularity will not appear. It is a conjecture
regarding the necessary and sufficient condition for black hole formation.

As we have reviewed above, many examples of naked singularity
formation are obtained from the analysis of spherically
symmetric spacetime. These examples are consistent with the
hoop conjecture.
The circumference $C$ corresponding to a radius $R$ centered
at the symmetric center is $2\pi R$.
For a central naked singularity of a spherically symmetric spacetime,
the ratio of $C$ to the gravitational mass $M$ within $R$
can be estimated as
\begin{equation}
 \lim_{r \rightarrow 0}\frac{2\pi R}{4\pi M}
= \lim_{r \rightarrow 0}\frac{R}{2 M}
= \lim_{r \rightarrow 0}\frac{r^\alpha}{2 M_c r^3} > 1,
\end{equation}
where $M_c$ is a constant and $1<\alpha \leq 3$.
In this case the condition $C > 4\pi M$ coincides with the
condition that the center is not trapped.
Therefore the appearance of a central
naked singularity in spherical gravitational collapse does not constitute
a counterexample to the hoop conjecture.

As for the nonspherically symmetric case, several studies support
the hoop conjecture.
A pioneering investigation is the numerical simulation
of the general relativistic collapse of axially
symmetric stars.~\cite{Nakamura:1982,nmms1982} From
initial conditions that correspond to a
very elongated prolate fluid with sufficiently
low thermal energy, the fluid collapses with this elongated form maintained.
In the numerical simulation, evidence of black hole formation
was not found.  This result suggests that such an elongated object
might collapse to a naked singularity, and in this case it would provide
evidence supporting the hoop conjecture.

Shapiro and Teukolsky numerically studied the evolution of a
collisionless gas spheroid with fully general relativistic
simulations.~\cite{Shapiro:1991,Shapiro:1992} In their calculations,
spacetimes describing collapsing gas spheroid were foliated by using the
maximal time slicing.
They found some evidence that a prolate spheroid
with a sufficiently elongated initial configuration and
even with a small angular momentum, might form naked singularities.
More precisely, they found that when a spheroid is highly prolate, a
spindle singularity forms at the pole, where the numerical evolution cannot
be continued. They also found that the singular region extends
outside the matter region by showing that the Riemann invariant grows there.
Then, they searched for trapped surfaces, but found that they do not exist.
They considered these results as indicating
that the spindle might be a naked singularity. However, the absence of
a trapped surface on their maximal time slicing does not necessarily
mean that the singularity is indeed naked. To be established whether it is
naked or not, we would need to investigate
the region of the spacetime future of the singularity.

Wald and Iyer~\cite{Wald:1991} proved that even in Schwarzschild
spacetime, it is
possible to choose a time slice that comes arbitrarily close to the
singularity, yet for which no trapped surfaces exists in its past.
A simple analytical counterpart of the model of the prolate collapse
studied numerically by Shapiro and Teukolsky is provided by the
Gibbons-Penrose construction.~\cite{Barrabes:1991}
This construction considers a thin shell
of null dust collapsing inward from a past null infinity.
Pelath, Tod, and Wald~\cite{Pelath:1998am} gave an explicit
example in which trapped surfaces are present on the shell, but none
exist prior to the last flat slice, thereby explicitly showing that the
absence of trapped surfaces on a particular, natural slicing does not
imply the absence of trapped surfaces in spacetime.

It should be noted that there are problems in the hoop conjecture to prove
it as a mathematically unambiguous theorem, such as how we should define
the mass of the object and the length of the hoop.
Although we have not found a counterexample to the hoop conjecture,
it is necessary to obtain a complete solution to these problems.

\section{Naked singularity formation beyond spherical symmetry}
\label{sec:perturb}
Many of the examples of naked singularity formation have been obtained
by the analyses under the assumption of spherical symmetry.
Investigation of the
nonspherical systems have great significance in regard to
gravitational wave radiation emitted from a forming naked singularity, the
stability
of the `nakedness' of a naked singularity, and the hoop conjecture.
In this section we describe the efforts made to go beyond
the case of spherical symmetry.

\subsection{Linear perturbation analyses of the
Lema\^{\i}tre-Tolman-Bondi spacetime}
\label{sec:gr-pertub}
As mentioned in the previous section, central shell-focusing naked
singularities should appear from `generic' initial data in the LTB spacetime.
However, there are some unrealistic assumptions in this model, 
e.g., pressureless dust matter, spherical symmetry, and so on.
Here we consider whether spherical symmetry is essential to 
the appearance of a shell-focusing naked singularity in the LTB
spacetime.  
At the same time, we investigate whether a naked singularity can be a
strong source of gravitational wave bursts. 
For this purpose, we introduce asphericity into the LTB spacetime
using a linear perturbation method. 
The angular dependence of perturbations is decomposed into series of
tensorial spherical harmonics. Spherical harmonics are
called even parity if they have parity $(-1)^l$ under spatial 
inversion and odd parity if
they have parity $(-1)^{l+1}$. Even and odd perturbations are decoupled 
from each other in the linear perturbation analysis.

We consider both the odd- and even-parity modes of these 
perturbations in the 
marginally bound LTB spacetime and examine the stability 
of the nakedness of 
that naked singularity with respect to those linear perturbations.
We also attempt to investigate whether the naked singularity
is a strong source of gravitational radiation.
\cite{Iguchi:1998qn,Iguchi:1999ud}

\subsubsection{Odd-parity perturbations}
\label{sec:odd}
Here we consider a marginally bound collapse background for
simplicity. (See \S \ref{sec:generalspherical} and \ref{sec:LTB} 
for the background solution.)
We use gauge invariant perturbation formalism for a general
spherically symmetric spacetime established by Gerlach and Sengupta. 
\cite{Gerlach:1979rw,Gerlach:1980tx}
Their formalism is described in Appendix \ref{sec:gauge}.
Hereafter we follow the notation and conventions given there.
The linearized Einstein equations (\ref{GS-9a}) and (\ref{GS-9b}) can
be rewritten as
\begin{eqnarray}
  \label{wave-eq}
  \ddot{\psi_s}-\frac{1}{R'^2}{\psi_ s''}
  &=& \frac{1}{R'^{2}}\left(6\frac{R'}{R}
  -\frac{R''}{R'}\right){\psi_s'}-\left(6\frac{\dot{R}}{R}
  +\frac{\dot{R'}}{R'}\right)\dot{\psi_s} 
  -4\left[\frac{\dot{R'}}{R'}\frac{\dot{R}}{R}
  +\frac{1}{2}\left(\frac{\dot{R}}{R}\right)^2\right]\psi_s  \nonumber \\
& &-\frac{16\pi}{R'R^2}\partial_{r}
\left(\frac{r^2 \rho (r) U(r)}{R'R^2}\right),
\end{eqnarray}
where $\rho (r) = \bar{\rho}(0,r)$ is the background density profile at $t=0$,
and $U(r)$ characterizes the perturbation of the four-velocity as
\begin{equation}
 \delta u_{\mu} = (0,0,U(r)S_A),
\end{equation}
where $S_{A}$ is the odd-parity vector harmonic function.
Here the dot and prime denote partial derivatives with respect to 
$t$ and $r$, respectively.
We introduce the gauge invariant variable 
\begin{equation}
 \label{psis}
 \psi_s \equiv \frac{1}{A}\left[ {\left
 ( \frac{k_1}{R^2}\right)^{\cdot}} -\left( \frac{k_0}{R^2} \right)'\right],
\end{equation} 
where $A\equiv e^{\lambda}=R'$ in the line element (\ref{eq:lineelement}). 
We solve this partially differential equation numerically. 

Let us consider the regularity conditions for the background 
metric functions and gauge-invariant perturbations at $r=0$. 
Hereafter, we restrict ourselves to the axisymmetric case, 
i.e., $m=0$. Note that this restriction does not lessen the generality 
of our analysis. Further, we consider only the case in which 
the spacetime is regular before the appearance of the singularity. 
This means that, before naked singularity 
formation, the metric functions $R(t,r)$ and $A(t,r)$ behave 
near the center according to 
\begin{eqnarray}
  \label{r-0-R}
  R &\longrightarrow& R_c(t)r+O(r^3), \\
  \label{r-0-A}
  A &\longrightarrow& R_{c}(t)+O(r^{2}).
\end{eqnarray}
To investigate the regularity conditions of the gauge-invariant 
variables $k_a$ and $L_0$, 
we follow Bardeen and Piran.~\cite{Bardeen:1983} 
The results are  
\begin{eqnarray}
  \label{r-0-L}
  L_0 &\longrightarrow& L_c(t)r^{l+1}+O(r^{l+3}), \\
  \label{r-0-k0}
  k_0 &\longrightarrow& k_{0c}(t)r^{l+1}+O(r^{l+3}), \\
  \label{r-0-k1}
  k_1 &\longrightarrow& k_{1c}(t)r^{l+2}+O(r^{l+4}). 
\end{eqnarray}
 From Eqs.~(\ref{psis}), (\ref{r-0-R}), (\ref{r-0-A}), (\ref{r-0-k0}) and 
(\ref{r-0-k1}), we find that $\psi_s$ behaves near the center as 
\begin{eqnarray}
  \label{r-0-psis}
  \psi_{s} &\longrightarrow& 
 \psi_{sc}(t)r^{l-2}+O(r^l)
 ~~~~~~~~~~~~\mbox{for $l\geq 2$}, \\
  \label{r-0-psis1}
  \psi_{s} &\longrightarrow& \psi_{sc(t)}r+O(r^3)
  ~~~~~~~~~~~~~~~~\mbox{for $l=1$}.
\end{eqnarray}
In the case $l\geq2$, the coefficient $\psi_{sc}(t)$ 
is related to $R_{c}(t)$ and $k_{0c}(t)$ as  
\begin{equation} 
  \label{relation-psis}
\psi_{sc}(t)=-(l-1)\frac{k_{0c}(t)}{R_{c}^{3} (t)}.
\end{equation}
 From the above equations, we note that only the quadrupole mode, 
$l=2$, of $\psi_{s}$ does not vanish at the center. 

We now comment on the behavior of the matter 
perturbation variable $L_0$ around 
the naked singularity on the slice $t=t_0$.
The regularity conditions for $L_0$ and $\bar{\rho}$
determine the behavior of $U(r)$ near the center as
\begin{equation}
  \label{dJdr-c}
  U(r) \propto r^{l+1}.
\end{equation}
This property does not change even if a central singularity
appears. However, the $r$ dependence of $R$ and 
$A$ near the center changes at 
that time. Assuming a mass function $F(r)$ of the form
\begin{equation}
  F(r)=F_3+F_{n+3}r^{n+3}+ \cdots,
\end{equation}
we obtain the relation 
\begin{equation}
  t_s(r) = t_0 + t_n r^n +\cdots
\end{equation}
 from Eqs.\ (\ref{eq:energydensitycomovingtb}) and (\ref{t0}),
where $n$ is a positive even integer and $t_0$ and $t_n$ are constants. 
After substituting this 
relation into Eq.\ (\ref{eq:radius}),
we obtain the behavior of $R$ and $A$ around the central singularity as  
\begin{equation}
  \label{singr}
  R(t_0,r) \propto r^{1+ {2 \over 3} n} 
\end{equation}
and
\begin{equation}
  \label{singa}
  A(t_0,r) \propto r^{{2\over3} n}
\end{equation}
on the slice $t=t_0$.
As a result, we obtain the $r$ dependence 
of $L_0$ around the center when the naked singularity appears as
\begin{equation}
  L_0(t_0,r) \propto r^{l-2n+1}.
\end{equation}
For example, if $l=2$ and $n=2$, then $L_0$ is inversely proportional to $r$
and diverges toward the central naked singularity. Therefore, the source term
of the wave equation is expected to have a large magnitude around the naked 
singularity. Thus the metric perturbation variable $\psi_s$ as well as 
the matter variable $L_0$ may diverge toward the naked singularity.

In place of the $(t,r)$ coordinate system, we 
introduce a single-null coordinate system, $(u,\tilde{r})$,
where $u$ is an outgoing null coordinate chosen so that it
agrees with $t$ at the symmetric center, and we choose $\tilde{r}=r$.
We perform the numerical integration along two characteristic
directions. Therefore we use a double null grid in the numerical
calculation. In this new coordinate system, $(u,\tilde{r})$, 
Eq.\ (\ref{wave-eq}) 
is expressed in the form 
\begin{eqnarray}
  \label{dphis/dlambda}
  \frac{d\phi_s}{du} &=&
  -\frac{\alpha}{R}\left[3R'+{1\over2}R\dot{R}\dot{R'}
  -\frac{5}{4}\dot{R}^{2}R'\right]\psi_s 
  -\frac{\alpha}{2}\left[\frac{R''}{R'^2}
  -\frac{2}{R}\left(1-\dot{R}\right)
   \right]\phi_s \nonumber \\
    & &
   -\frac{8\pi \alpha}{R}\left(
   \frac{r^2 \rho (r) U(r)}{R'R^2}\right)',\\
  \label{delpsi/delrprime}
  \partial_{\tilde{r}}\psi_{s}&=& \frac{1}{R}\phi_s
  -3\frac{R'}{R}
\left(1+\dot{R}\right)\psi_s,
\end{eqnarray}
where the ordinary derivative on the left-hand side of
Eq.\ (\ref{dphis/dlambda}) and the partial derivative on the left-hand
side of Eq.\ (\ref{delpsi/delrprime}) are given by
\begin{eqnarray}
  \frac{d}{du} &=& \partial_u +
  \frac{d\tilde{r}}{du}\partial_{\tilde{r}}
   = \partial_u-\frac{\alpha}{2R'}\partial_{\tilde{r}} \nonumber  \\ 
  &=&\frac{\alpha}{2}\partial_t-\frac{\alpha}{2R'}\partial_{r},\\
  \label{r'a}
  \partial_{\tilde{r}} &=& -\frac{(\partial_r u)_t}{(\partial_t
    u)_r}\partial_{t}+\partial_{r}=
    R'\partial_{t}+\partial_{r}.
\end{eqnarray}
Also, $\phi_s$ is defined by 
Eq.\ (\ref{delpsi/delrprime}) and $\alpha$ is given by
\begin{equation} 
  \alpha\equiv {1\over (\partial_{t} u)_r}.
\end{equation}

We assume $\psi_s$ vanishes on the initial null hypersurface. Therefore,
there exist initial ingoing waves that offset the waves produced by the
source term on the initial null hypersurface. In
Ref. \citen{Iguchi:1998qn}, it is confirmed
that this type of initial ingoing waves propagate through the dust 
cloud without net amplification, even when they pass through 
the cloud just before the appearance of the naked singularity. 
Therefore these initial ingoing waves are not significant to the results 
of this perturbation analysis. 

We adopt the initial rest mass density profile 
\begin{equation}
  \label{density}
  \rho(r)=\rho_0 \frac{1+ \exp\left(-\frac{1}{2}\frac{r_1}{r_2}\right)}
  {1+ \exp\left(\frac{r^n -r_1^n}{2 r_1^{n-1}r_2}\right)},
\end{equation}
where $\rho_0$, $r_1$ and $r_2$ are positive constants and $n$ is a 
positive even
integer. With this form, the dust fluid spreads all over the space. However,
if $r \gg r_1,r_2$, then $\rho(r)$ decreases exponentially, so that the dust
cloud is divided into a core part and an envelope, which can be
considered as essentially the vacuum region. We define a core radius as
\begin{equation}
  \label{cradious}
  r_{\mbox{\scriptsize core}}=r_1+\frac{r_2}{2}.
\end{equation}
If we set $n=2$, there appears a central naked
singularity. This singularity becomes locally or globally naked, depending 
on the parameters ($\rho_0,r_1,r_2$). However, if the integer 
$n$ is greater than $2$, the final state of the dust cloud is a black
hole for any values of the parameters. Then, we consider three different
density profiles that correspond to three types of final states of the
dust cloud, globally and locally naked singularities and a black
hole. The outgoing null coordinate $u$ is chosen so that it agrees
with the proper time at the symmetric center. Therefore, even if a 
black hole background is considered, we can analyze the inside of the 
event horizon.
The corresponding parameter values are given in Table \ref{tab:parameter}.
Using this density profile, we numerically calculate the total 
gravitational mass of 
the dust cloud $M$. In our calculation we adopt the total
mass $M$ as the unit of the variables.

 \begin{table}
 \caption{Parameter values for initial density profiles, 
power law indices and damped oscillation frequencies.}
 \label{tab:parameter}
 \begin{tabular}{cccccccc} \hline \hline
  & final state & $\rho_0$ & $r_1$ & $r_2$ & $n$ &power index&
 damped oscillation frequency\\ \hline
 (a) & globally naked & $1 \times 10^{-2}$ & 0.25 & 0.5 & 2 & 5/3 & --- \\ 
 (b) & locally naked & $1 \times 10^{-1}$ & 0.25 & 0.5 & 2 & 5/3 & 0.37+0.089$i$
\\ 
 (c) & black hole & $2 \times 10^{-2}$ & 2 & 0.4 & 4 & --- &
 0.37+0.089$i$\\ 
\hline
 \end{tabular}
 \end{table}

The source term of Eq.\ (\ref{dphis/dlambda}),
\begin{equation}
  \label{souce}
  S(t,r)=-\frac{8\pi \alpha}{R}\left(
   \frac{r^2 \rho (r) U(r)}{R'R^2}\right)',
\end{equation}
is determined by $U(r)$. As mentioned above, the constraints on the 
functional form of
$U(r)$ are given by the regularity condition of $L_0$. From
Eq.\ (\ref{dJdr-c}), $U(r)$ should be
proportional to $r^{l+1}$ toward the center. 
We localize the matter perturbation near the center to diminish the
effects of the initial ingoing waves. Therefore we define $U(r)$ such that
\begin{equation}
  \label{dJdr}
  r^2 \rho(r) U(r) = \left\{ 
     \begin{array}{ccc}
      U_0 \left( \displaystyle \frac{r}{r_b} \right)^5 \left(1-\left
  ( \displaystyle \frac{r}{r_b} \right)^2 \right)^5  &{\mbox{for}} & 0\leq r\leq r_b,\\
        0 &{\mbox{for}} & r>r_b,
      \end{array}
    \right.
\end{equation}
where $U_0$ and $r_b$ are arbitrary constants. In our numerical calculation
we chose $r_b$ to be $r_{\mbox{\scriptsize core}}/2$. This choice of $r_b$ has no
special meaning, and the results of our numerical calculations are not
sensitive to it.

\begin{figure}[tbp]
  \begin{center}
    \leavevmode
    \epsfxsize=66mm\epsfbox{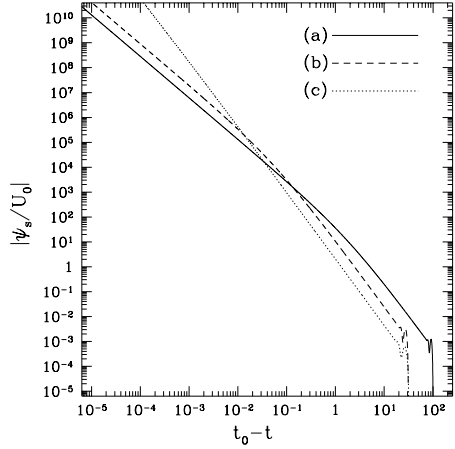}    
 \caption{Plots of $\psi_s$ at the center as a function of the time coordinate
   $t$. The solid curve represents the globally naked case (a), the dashed
   curve represents the locally naked case (b), and the dotted curve
   represents the black hole case (c).}
  \label{fig:center}
  \end{center}
 \end{figure}

\begin{figure}[tbp]
  \begin{center}
    \leavevmode
    \epsfxsize=66mm\epsfbox{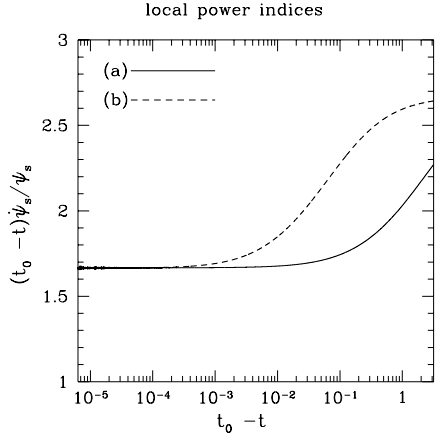}    
  \caption{Plots of the local power indices
    $d\ln \psi_s/d\ln (t_0-t)=(t_{0}-t)\dot{\psi_{s}}/\psi_{s}$. 
    The solid curve corresponds to the globally
    naked case (a), and the dashed curve corresponds to the 
    locally naked case
    (b). The two curves approach the same value near 5/3.}
 \label{fig:index}
  \end{center}
 \end{figure}

First we observe the behavior of $\psi_s$ at the center. The results are 
plotted in Fig.\ \ref{fig:center}. The initial oscillations correspond to
the initial ingoing waves. After these oscillations, $\psi_s$  grows
proportionally to $(t_0-t)^{- \delta}$ for the naked singularity cases
near the formation epoch of the naked singularity. 
For the case of black hole formation, $\psi_s$ exhibits power-law
growth in the early part. Later, its slope gradually changes,  
but it grows faster than in the case of a naked singularity. 
For naked singularity cases, the power-law indices $\delta$ are
determined by $d\ln \psi_s/d\ln (t_0-t)=(t_{0}-t)\dot{\psi_{s}}/\psi_{s}$ 
locally. The results are
shown in Fig.\ \ref{fig:index}. From this figure, we see that the final
indices are 5/3 for both naked cases. 
Therefore, the metric perturbations diverge at the central naked singularity.

\begin{figure}[tbp]
  \begin{center}
    \leavevmode
    \epsfxsize=66mm\epsfbox{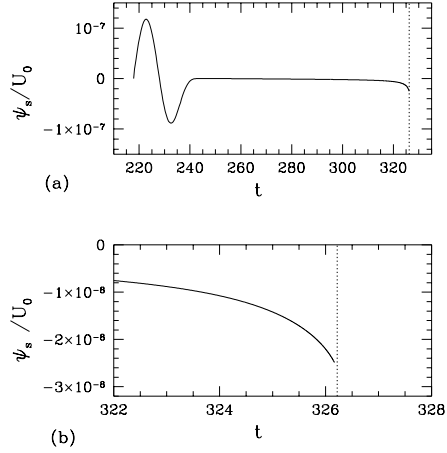}    
  \caption{Plots of $\psi_s$ for the globally naked case
    (a) at $R=100$. In (a), the first
    oscillation originates from the initial ingoing wave. In (b), we
    magnify the the right-hand edge, which is just before the Cauchy 
    horizon. The dotted lines represent the time at which the observer
    at $R=100$ 
    intersects the Cauchy horizon, which is determined by numerical
    integration of the null geodesic equation from the naked singularity.}
 \label{fig:outsideg}
  \end{center}
 \end{figure}

\begin{figure}[tbp]
  \begin{center}
    \leavevmode
    \epsfxsize=66mm\epsfbox{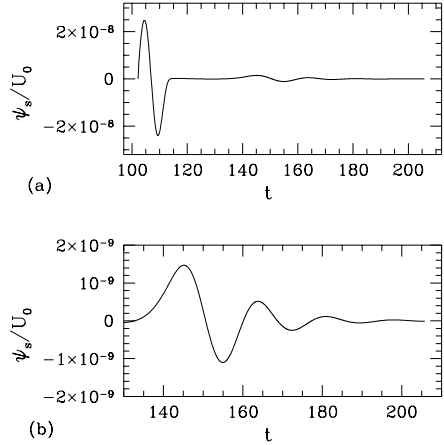}    
  \caption{Plots of $\psi_s$ for the locally naked case (b) at $R=100$. 
    In (a), the first
    oscillation originates from the initial ingoing wave. After this
    oscillation, the damped oscillation dominates, and this part of
    the wave form  is magnified in (b). }
 \label{fig:outsidel}
  \end{center}
 \end{figure}

\begin{figure}[tbp]
  \begin{center}
    \leavevmode
    \epsfxsize=66mm\epsfbox{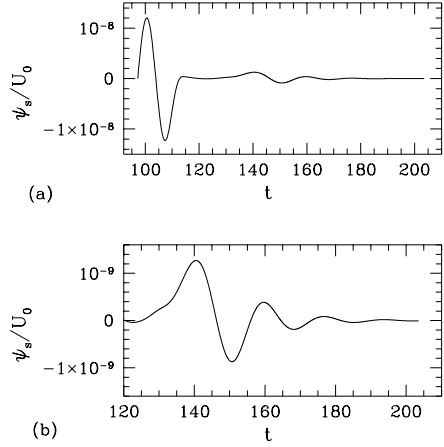}    
  \caption{Plots of $\psi_s$ for the black hole case (c) at $R=100$. 
    In (a), the first
    oscillation originates from the initial ingoing wave. After this
    oscillation, the damped oscillation dominates, as depicted
    in (b). }
 \label{fig:outsideb}
  \end{center}
 \end{figure}

We also observe the wave form of $\psi_s$ along the line of 
constant circumferential radius outside the dust cloud. The results are
shown in Figs.\ \ref{fig:outsideg}--\ref{fig:outsideb}. Figure 
\ref{fig:outsideg} displays the wave form of the globally naked case
(a), Fig.\ \ref{fig:outsidel} displays the wave form of the
locally naked case (b), and Fig.\ \ref{fig:outsideb} displays the
wave form of the black hole case (c).
The initial oscillations
correspond to the initial ingoing waves. In the cases of locally naked
singularity and black hole formation, damped oscillations dominate the 
gravitational waves. We read the frequencies and 
damping rates of these damped oscillations from Figs.\
\ref{fig:outsidel} and \ref{fig:outsideb}, and express them as the  
complex frequency $0.37+0.089i$ for the locally naked singularity
and black hole cases. These agree well with the
fundamental quasi-normal frequency of the quadrupole mode
$(2M\omega = 0.74734 + 0.17792 i)$ of a Schwarzschild black hole given
by Chandrasekhar and Detweiler.\cite{Chandrasekhar:1975} 
In the globally naked singularity case (a), 
we did not see this damped oscillation because of the
existence of the Cauchy horizon. In all cases, the gravitational waves 
generated by matter perturbations are at most quasi-normal modes of a
black hole that is generated outside the dust cloud. 
Therefore intense odd-parity gravitational waves would not be
produced by inhomogeneous dust cloud collapse. 
It is thus not expected that the central extremely high 
density region can be observed using this mode of gravitational waves.

\begin{figure}[tbp]
   \begin{center}
     \leavevmode
    \epsfxsize=66mm\epsfbox{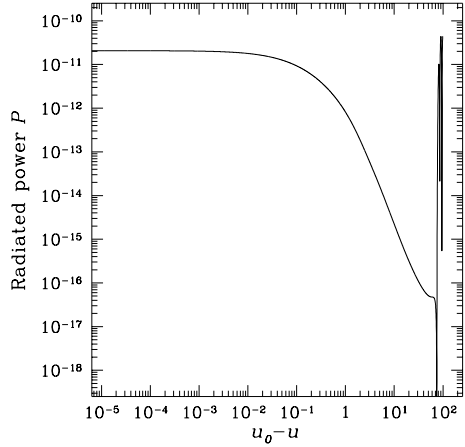}    
     \caption{Plots of the radiated power $P$ for the globally naked case
       (a) at $R=100$. The horizontal axis is the outgoing null
       coordinate $u$. At the Cauchy horizon, this coordinate has the
       value $u_0$.}
     \label{fig:power}
   \end{center}
 \end{figure}

We can calculate the radiated power of the gravitational waves and thereby
gain an understanding of the physical meaning of the gauge-invariant
quantities.\cite{CPM1978,CPM1979,CPM1980} \ 
The radiated power $P$ of the quadrupole mode 
is given by
\begin{equation}
  P = \frac{3}{32\pi}R'^2\left[\partial_{\tau}\left(R^3 \psi_s\right)\right]^2.
\end{equation}
Figure \ref{fig:power} displays the time evolution of the radiated power
$P$. The radiated power also has a finite value at the
Cauchy horizon. The total energy radiated by odd-parity quadrupole
gravitational waves during the dust collapse should not diverge.

\subsubsection{Even-parity perturbations}
\label{sec:even}
The behavior of the even-parity perturbations of the LTB spacetime is 
investigated in Ref.~\citen{Iguchi:2000jn}.
There are four gauge-invariant metric variables, $k_{ab}$ and $k$, 
and seven matter variables, $T_{ab}$, $T_a$, $T^2$, and $T^3$, where
$a$ and $b$ run over 0 and 1. 
The energy density $\bar{\rho}$ is perturbed by adding the scalar term $\delta
\rho Y$, while the four-velocity $\bar{u}_\mu$ is perturbed by adding the term
\begin{equation}
  \delta u_\mu =(V_0(x^d)Y,V_1(x^d)Y,V_2(x^d)Y_{,A}),
\end{equation}
where $Y$ is a scalar harmonic function.
The normalization for the four-velocity yields 
the relation $\bar{u}^\mu  \delta
u_\mu =0$. This relation implies that  $V_0$ vanishes. 
Then, there are only three matter perturbation variables, 
\begin{eqnarray}
  T_{00} &=& \delta \rho(t,r), \\
  T_{01} &=& \bar{\rho} V_1(t,r), \\
  T_0 &=& \bar{\rho} V_2(t,r), 
\end{eqnarray}
as the others vanish:
\begin{equation}
  T_{11}=T_1=T^3=T^2=0.
\end{equation}
Now we can write
down the perturbed Einstein field equations for the background LTB
spacetime. The resultant linearized Einstein equations are given in
Appendix \ref{chap:eqn}. 

We have derived seven differential equations, (\ref{00})--(\ref{z22}),
for seven variables (four metric and three matter). The right-hand sides
of four of these equations vanish. We can obtain the
behavior of the metric variables through the integration of these 
equations alone.
We transform these equations into more convenient forms.
  From Eq.\ (\ref{le4}), we have 
\begin{equation}
  k_{00} = \frac{1}{R'^2}k_{11}.
\end{equation}
Using this relation and the equations whose right-hand sides 
vanish, we obtain evolution 
equations for the gauge-invariant metric variables as
\begin{eqnarray}
\label{evQ}
  -\ddot{q}+\frac{1}{R'^2}q''&=& {\frac{4}{{{R}^2}}} q 
                             + \left(\frac{2}{RR'} + \frac{R''}{R'^3}\right)q'
                             + 3\frac{\dot{R'}}{R'}\dot{q} 
                             + 4\left({\frac{\dot{R}}{R}}
                             - \frac{\dot{R'}}{R'}\right)\dot{k} \nonumber\\
                         & & + \frac{2}{R'^3}\left(- \dot{R''}
                             - {\frac{2{{R'}^2}\dot{R}}{{{R}^2}}} 
                             - {\frac{R''\dot{R}}{R}} 
                             + {\frac{2R'\dot{R'}}{R}} 
                             + {\frac{2R''\dot{R'}}{R'}} 
                            \right)k_{01}\nonumber \\
                         & & + \frac{2}{R'^3}\left(- \dot{R'} 
                             + {\frac{R'\dot{R}}{R}}\right){k_{01}}' ,
\end{eqnarray}
\begin{eqnarray}
\label{evK}
 \ddot{k}&=&    - {\frac{2}{R^2}}q
                - \frac{q'}{RR'}+ \frac{\dot{R}}{R}\dot{q}
                - 4\frac{\dot{R}}{R}\dot{k}
                + \frac{2}{RR'}\left(- \frac{\dot{R'}}{R'}
                + \frac{\dot{R}}{R}\right) {k_{01}},
\end{eqnarray}
\begin{equation}
\label{evk01}
  \dot{k_{01}} = -\frac{\dot{R'}}{R'}k_{01}-q',
\end{equation}
where $q \equiv k-k_{00}$. If we solve these three equations with some 
initial data under appropriate boundary conditions, we can follow 
the full evolution of the metric perturbations. When we substitute
these metric perturbations into Eqs.\ (\ref{00}), (\ref{01}) and
(\ref{02}), the matter perturbation variables $\delta
\rho, V_1$ and $V_2$, respectively, are obtained.

We can also investigate the evolution of the matter perturbations from the 
linearized conservation equations $\delta (T^{\mu \nu}_{~~;\nu})=0$.
They reduce to
\begin{eqnarray}
 \label{rhodot}
  \left(\frac{\delta \rho}{\bar{\rho}}\right)^{.} 
    &=&\frac{1}{\bar{\rho}R^2R'}\left(\frac{R^2\bar{\rho}}{R'}
       \left(k_{01}+V_1\right)\right)' -\frac{6}{R^2}V_2
       -\dot{k}-\frac{3}{2}\left(\dot{k}-\dot{q}\right),\\
 \label{V1dot}
  \dot{V_1} &=& -\frac{1}{2}\left(k'-q'\right), \\
 \label{V2dot}
  \dot{V_2} &=& -\frac{1}{2}\left(k-q\right) .
\end{eqnarray}
Integration of these equations gives us the time evolution of the matter 
perturbations. 
We can check the consistency of the numerical calculation by 
comparison of these variables and those obtained from 
Eqs.\ (\ref{00}), (\ref{01}) and (\ref{02}).

To constrain the boundary conditions in our numerical calculation, we
should consider the regularity conditions at the center. These
conditions are obtained by requiring that all tensor quantities
be expandable in nonnegative integer powers of locally Cartesian
coordinates near the center.\cite{Bardeen:1983}\ 
 We simply quote the results. 
The regularity conditions for the metric perturbations 
are 
\begin{equation}
  k \approx k_0(t)r^2,~~ q \approx q_0(t)r^4,~~ k_{01} \approx k_0(t)r^3.
\end{equation}
For the matter perturbations, the regularity conditions at the center
are 
\begin{equation}
  \delta \rho \approx \delta \rho_0(t) r^2,~~ V_1 \approx V_{10}(t)r,
  ~~ V_2 \approx V_{20}(t)r^2.
\end{equation}
Therefore, all the variables we need to calculate vanish at the
center.

\begin{figure}[tbp]
  \begin{center}
    \leavevmode
    \epsfxsize=10cm\epsfbox{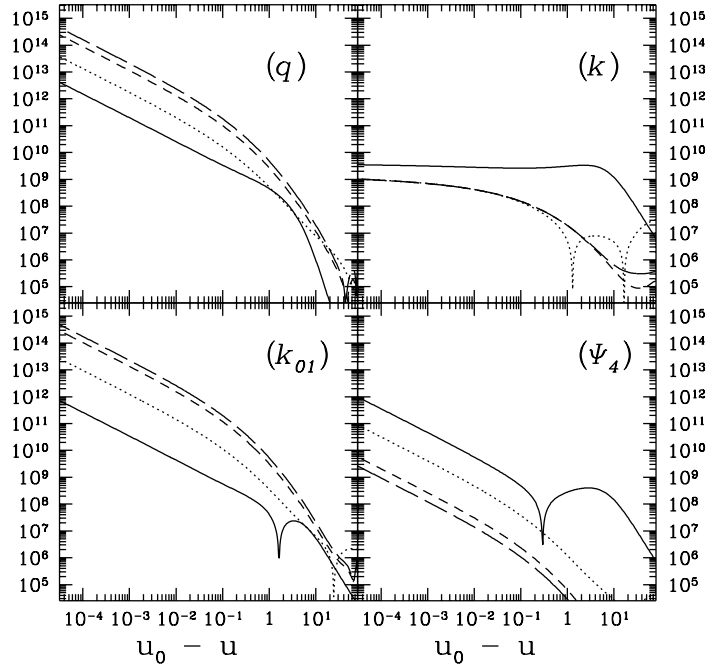}    
  \caption[Even-parity metric perturbation variables at constant
    circumferential radius]{Plots of perturbed variables $q$, $k$ and 
    $k_{01}$ and the Weyl scalar $\Psi_4$ 
   at constant circumferential radius $R$.
    The results for $R=1$,  $R=10$, $R=100$, and $R=200$ are
    plotted. The solid curves represent the results for $R=1$, the dotted
    curves for $R=10$, the dashed curves for $R=100$, and the long dashed
   curves for $R=200$. $u=u_0$ corresponds to the Cauchy horizon. } 
 \label{fig:out}
  \end{center}
 \end{figure}

First, we observe the behavior of the metric variables $q$, $k$ 
and $k_{01}$ and 
the Weyl scalar, which corresponds to outgoing waves 
\begin{eqnarray}
   \Psi_4 &\equiv& C_{\mu\nu\rho\sigma}n^{\mu}\bar{m}^{\nu}n^{\rho}\bar{m}^{\sigma}\\
     &=& -\frac{3}{32}\sqrt{\frac{5}{\pi}}\sin^2\theta\frac{k_{01}-\left(k-q\right)R'}{R^2 R'},
\end{eqnarray}
where
\begin{eqnarray}
  n^{\mu} &=& \left(\frac{1}{2},-\frac{1}{2R'},0,0\right),\\
  \bar{m}^{\nu} &=& \left(0,0,\frac{1}{\sqrt{2}R},-\frac{i}{\sqrt{2}R\sin\theta}\right),
\end{eqnarray}
outside the dust cloud. The results are plotted in 
Fig.\ \ref{fig:out}. We can see that the metric variables $q$ and 
$k_{01}$ and the Weyl scalar $\Psi_4$ 
diverge when they approach the Cauchy horizon. The
asymptotic power indices of these quantities are about 0.88. On the
other hand the metric quantity $k$ does not diverge when the
Cauchy horizon is approached. The energy flux is computed  by constructing the
Landau-Lifshitz pseudo-tensor. We can calculate the radiated power of
gravitational waves from
this. The result is given in Appendix \ref{chap:power}. For the
quadrupole mode, the total radiated power becomes 
\begin{equation}
  P = \frac{3}{8\pi} k^2.
\end{equation}
The radiated power of the gravitational waves is proportional to the
square of $k$. 
Therefore it seems unlikely that a system of
spherical dust collapse with linear perturbations 
can be a strong source of gravitational waves
within the linear perturbation scheme. However, we should also note that the 
divergence of the linear perturbation variables $q$, $k_{01}$ and $\Psi_4$
implies the breakdown of the linear perturbation scheme.

 \begin{figure}[tbp]
  \begin{center}
    \leavevmode
    \epsfxsize=10cm\epsfbox{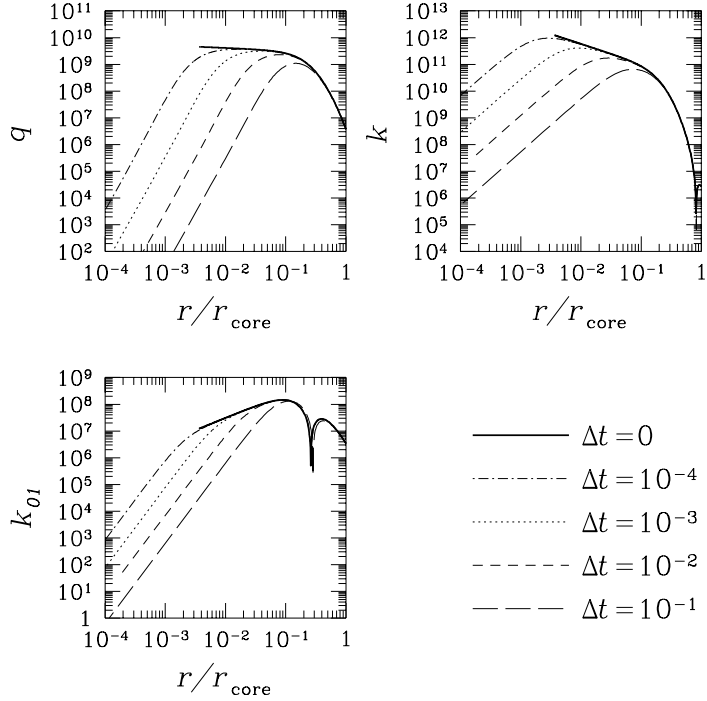}    
  \caption[Even-parity metric perturbation variables near the
    center]{Plots of perturbed variables  $q, ~k$ and $k_{01}$ near the
    center. The values for $\Delta t=t_0 -t=10^{-1}, 10^{-2}, 10^{-3},
    10^{-4}, 0$ are plotted. The solid curves represent the results for
    $\Delta t =0$, the long dashed curves for $\Delta t =10^{-1}$, the dashed 
    curves for $\Delta t =10^{-2}$, the dotted curves for $\Delta t =10^{-3}$,
    and the dotted dashed curves for $\Delta t =10^{-4}$.}
 \label{fig:center-e}
  \end{center}
 \end{figure}

\begin{figure}[tbp]
  \begin{center}
    \leavevmode
    \epsfxsize=10cm\epsfbox{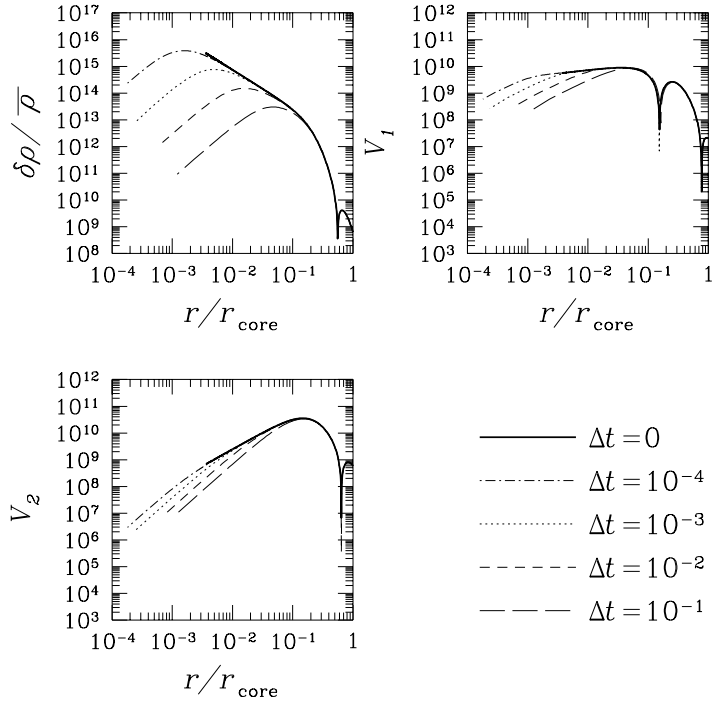}    
  \caption[Even-parity matter perturbation variables near the
    center]{Plots of perturbed variables  $\delta \rho,
    V_1,$ and $V_2 $ near the center. The values for $\Delta t =10^{-1},
    10^{-2}, 10^{-3}, 10^{-4},$ and  $0$ are
    plotted. The solid curves represent the results for
    $\Delta t =0$, the long dashed curves for $\Delta t =10^{-1}$, the dashed 
    curves for $\Delta t =10^{-2}$, the dotted curves for $ \Delta t =10^{-3}$,
    and the dotted dashed curves for $\Delta t =10^{-4}$.}
 \label{fig:centerm}
  \end{center}
 \end{figure}

Second, we observe the perturbations near the center. The results 
are plotted in Figs.\ \ref{fig:center-e} and \ref{fig:centerm}. In these
figures we plot the perturbations at $t-t_0 =$ $- 10^{-1},$ $ -10^{-2},$
$-10^{-3},$ $ -10^{-4},$ and $ 0$. 
Before the formation of the naked singularity, the perturbations obey
the regularity conditions at the center. Each curve in these figures 
displays this dependence if the radial coordinate is sufficiently small.
In this region, we can also see that all the variables grow according to 
power laws as functions of 
the time coordinate along the lines constant $r$.  
The asymptotic behavior of perturbations near the central naked
singularity is summarized as follows:
\begin{eqnarray}
  q \propto \Delta t ^{-2.1} r^4,~~ &k \propto \Delta t ^{-1.4} r^2, &~~k_{01}
    \propto \Delta t ^{-1.0} r^3, \nonumber \\ 
  \frac{\delta \rho}{\bar{\rho}} \propto \Delta t ^{-1.6} r^2, 
   ~~&V_1 \propto \Delta t ^{-0.4} r, &~~
   V_2 \propto \Delta t ^{-0.4} r^2.  \nonumber
\end{eqnarray}
Here $\Delta t = t_0 - t$. 
On the time slice at $\Delta t =0$, the perturbations behave as
\begin{eqnarray}
  q \propto r^{-0.09},~~ &k \propto r^{-0.74}, &~~k_{01} \propto
  r^{0.92}, \nonumber \\ 
  \frac{\delta \rho}{\bar{\rho}} \propto r^{-1.4}, 
   ~~&V_1 \propto r^{0.25}, &~~ V_2 \propto r^{1.3}.  \nonumber
\end{eqnarray}
On this slice $q$, $k$ and ${\delta \rho}/{\bar{\rho}}$ diverge
when they approach the central singularity. 
On the other hand, $k_{01}$, $V_1$ and $V_2$ go to zero.

\begin{figure}[tbp] 
  \begin{center}
    \leavevmode
    \epsfxsize=66mm\epsfbox{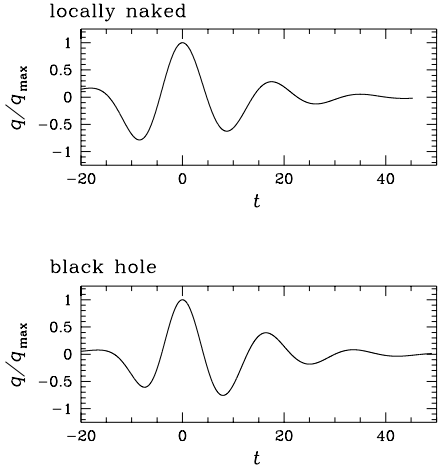}    
  \caption[Even-parity metric perturbation variables for locally naked
    singularity and black hole cases]{Plots of perturbed 
    variables  $q$  at constant
    circumferential radius $R=100$ in the locally naked singularity 
    and black hole cases.
    $q$ is normalized with respect to its maximum value, and the  
    origin of the time variable is adjusted to coincide with the time
    when $q$ is maximal.} 
 \label{fig:damp}
  \end{center}
 \end{figure}

In the cases of locally naked singularity and black hole formation, 
we expect to observe damped oscillation in the asymptotic region outside
the dust cloud, as in the odd-parity case.  
The results are plotted in Fig.\ \ref{fig:damp}. These
figures show that damped oscillations are dominant. 
We read the frequencies and 
damping rates of these damped oscillations from Fig.\
\ref{fig:damp} and express them as the  
complex frequencies $0.36+0.096i$ and $0.36+0.093i$ for locally naked
and black hole cases, respectively. These results agree well with the
fundamental quasi-normal frequency of the quadrupole mode
$(2M\omega = 0.74734 + 0.17792 i)$.\cite{Chandrasekhar:1975} \

\subsubsection{Summary of relativistic perturbations of spherical 
dust collapse}
We investigated odd-parity perturbations in 
\S \ref{sec:odd}. We concluded there that 
the Cauchy horizon is not destroyed by gravitational waves, while the 
properties of a shell-focusing naked central singularity 
may change, for
example, the divergence of the magnetic part of the Weyl curvature tensor.

In \S \ref{sec:even}, we investigated the behavior of the
even-parity perturbations in the LTB spacetime. In contrast to the
results for the odd-parity mode, the numerical analysis for the even-parity
perturbations shows that the Cauchy horizon should be destroyed by 
even-parity gravitational radiation. The energy flux of this radiation, 
however, is finite for an observer at constant circumferential radius
outside of the dust cloud. Therefore inhomogeneous aspherical dust
collapse appears unlikely as a strong source of 
gravitational wave bursts.

The difference between odd and even modes seems to originate from the
properties of matter perturbations. Odd-parity matter perturbations
are produced by the rotational motion of the dust cloud, and their 
evolution decouples from the evolution of metric perturbations. 
On the other hand, 
the even-parity matter perturbations contain the radial motion
of dust fluid and their evolution couples to the metric perturbations.
These two modes of odd and even parity, couple to each other when we
consider second-order perturbations.

\subsection{Estimate of the gravitational radiation from a
homogeneous spheroid}
Intuitively, at the formation of a singularity, 
disturbances of spacetime with short wavelength 
will be created. If there is no event horizon, 
these disturbances may propagate as gravitational radiation, so that a
naked singularity may be a strong source of short wavelength
gravitational radiation.
Nakamura, Shibata and Nakao\cite{Nakamura:1993xr} have suggested that a 
naked singularity may emit considerable gravitational wave radiation.
First, they considered Newtonian prolate dust collapse and
estimated the amount of the gravitational radiation 
using a quadrupole formula.

The shape of the homogeneous prolate spheroid is given by
\begin{equation}
\frac{x^2+y^2}{a^2}+\frac{z^2}{c^2}=1.
\end{equation}
The quantities $a$ and $c$ obey 
\begin{eqnarray}
 \ddot{a} &=& -\frac{3M}{2} \alpha(e) \frac{1}{ac}, \\
 \ddot{c} &=& -\frac{3M}{2} \gamma(e) \frac{1}{a^2}, \\
\alpha(e)&=&\frac{1}{e^2}\left
( 1-\frac{1-e^2}{2e}\ln\frac{1+e}{1-e}\right), \\
\gamma(e)&=&\frac{2(1-e^2)}{e^2}\left
( -1+\frac{1}{2e}\ln\frac{1+e}{1-e}\right). 
\end{eqnarray}
The luminosity of the gravitational waves in the frame of the quadrupole
formula is given by 
\begin{equation}
L=\frac{1}{45}(D^{\cdot\cdot\cdot})^2_{\alpha\beta}=\frac{2M^2}{375}
((a^{\cdot\cdot\cdot})^2-(c^{\cdot\cdot\cdot})^2)^2.
\end{equation}
The total amount of energy $\Delta E$ is proportional to
\begin{equation}
I = \int^{t_{c}-\epsilon}_{0} ((a^{\cdot\cdot\cdot})^2-
(c^{\cdot\cdot\cdot})^2)^2dt,
\end{equation}
where $t_c$ is the time at which $a$ becomes zero, and $\epsilon>0$.
It is easily found that the integral $I$ diverges as $\epsilon\to 0$.
Therefore,
under Newtonian gravity with the quadrupole formula, an
infinite amount of energy is 
radiated before the formation of a spindle-like singularity.
Of course, the collapse of a homogeneous
spheroid to a singularity cannot be properly described with 
Newtonian gravity, but this result is suggestive.

To extend this result to a relativistic analysis,
they modeled spindle-like 
naked singularity formation in gravitational collapse
using a sequence of general relativistic, momentarily static initial data
for the prolate spheroid.
It should be noted that their conclusion is the subject of debate.

\subsection{Newtonian analysis on the linear perturbation of
spherical dust collapse}
The dynamics of perturbations of the LTB spacetime has been re-analyzed 
in the framework of the Newtonian approximation.\cite{Nakao:2001as}
In order for a singularity of a spherically symmetric spacetime to be naked, 
``the gravitational potential'' $2m/R$ must be smaller than unity in 
the neighborhood of the singularity, where $m$ is the Misner-Sharp mass 
function and $R$ is the circumferential radius. A central shell-focusing 
naked singularity 
of the LTB spacetime satisfies this condition, and further,  
the gravitational potential vanishes even at this singularity. 
The speed of the dust fluid is also much smaller than the speed of light 
both before and at the time of the 
central shell-focusing naked singularity formation. 
Therefore, the Newtonian approximation seems to be applicable, 
even though the spacetime curvature diverges near the singularity. 
The advantage of the Newtonian approximation scheme is that 
the dynamics of perturbations of the dust fluid and 
gravitational waves generated by the motion of the dust fluid 
are estimated separately; the evolution of the perturbations 
of the dust fluid is obtained using Newtonian dynamics, and 
the gravitational radiation is obtained using the quadrupole formula. 
Hence, it is possible to make a 
semi-analytic estimate of the gravitational radiation 
due to the matter perturbation of the LTB spacetime 
if we adopt the Newtonian approximation.  
This suggests that the Newtonian analysis may be a powerful tool 
in the analysis of some category of naked singularities. 
However, we should stress that the neighborhood 
of a naked singularity is not Newtonian in an ordinary sense, 
because there is an indefinitely strong tidal force.   
Thus the Newtonian approximation scheme can be used to describe the dynamics 
of the neighborhood of the naked singularity, but the situation 
as a whole is not Newtonian. 

In this subsection, we consider the gravitational collapse 
of a spherically symmetric dust fluid in the framework of the 
Newtonian approximation and show that the Newtonian approximation 
is valid even at the moment of the formation of a central shell-focusing 
naked singularity if the initial conditions are 
appropriate, as in the case of the example in 
the previous subsection. 

\subsubsection{Eulerian coordinates}
 
In the Newtonian approximation, the maximal time slicing condition 
and Eulerian coordinates (for example, 
the minimal distortion gauge condition) are usually adopted.  
The line element is expressed in the form
\begin{equation}
ds_{\rm E}^{2}=-\left(1+2\Phi_{\rm N}\right)dT^{2}+dR^{2}
+R^{2}d\Omega^{2},
\end{equation}
where $\Phi_{\rm N}$ is the Newtonian gravitational potential, and 
we have adopted a polar coordinate system as the spatial coordinates. 
The equations for a spherically symmetric dust fluid 
and Newtonian gravitational potential $\Phi_{\rm N}$
are 
\begin{eqnarray}
\partial_{T}{\bar\rho}+{1\over R^{2}}\partial_{R}(R^{2}{\bar\rho} V)&=&0, \\
\partial_{T}V+V\partial_{R}V&=&-\partial_{R}\Phi_{\rm N},\\
{1\over R^{2}}\partial_{R}(R^{2}\partial_{R}\Phi_{\rm N})&=&4\pi{\bar\rho},
\end{eqnarray}
where $V$ is the velocity of the dust fluid element. 
The assumptions in the Newtonian approximation are 
\begin{equation}
|V|\ll1~~~~{\rm and}~~~~
|\Phi_{\rm N}|\ll1, \label{eq:N-condition}
\end{equation}
and further
\begin{equation}
|\partial_{T}V|\ll|\partial_{R}V|,~~~~
|\partial_{T}\Phi_{\rm N}|\ll|\partial_{R}\Phi_{\rm N}|~~~~
{\rm and}~~~~
|\partial_{T}{\bar\rho}|\ll|\partial_{R}{\bar \rho}|.
\end{equation}

\subsubsection{Lagrangian coordinates}

For the purpose of following the motion of a dust ball, 
Lagrangian coordinates are more suitable than Eulerian coordinates. 
The transformation 
matrix between Eulerian and Lagrangian coordinate systems 
is given by
\begin{eqnarray}
dT&=&d\tau, \label{eq:L-matrix1}\\
dR&=&{\dot R}d\tau+R'dx,
\label{eq:L-matrix2}
\end{eqnarray}
where $\tau$ and $x$ are regarded as independent variables,   
the dot represents a partial derivative with respect to $\tau$, 
and the prime represents a partial derivative with respect to $x$. 
Then the line element in the Lagrangian coordinate system 
is obtained as
\begin{equation}
ds_{\rm L}^{2}=-\left(1+2\Phi_{\rm N}-{\dot R}^{2}\right)d\tau^{2}
+2{\dot R}R'd\tau dx
+{R'}^{2}dx^{2}+R^{2}d\Omega^{2}.
\end{equation}
The equations for the dust fluid and Newtonian gravitational 
potential are 
\begin{eqnarray}
F(x)&=&8\pi\int_{0}^{x}{\bar\rho}R'R^{2}dx, \\
V^{2}&=&{\dot R}^{2}=f(x)+{F(x)\over R}, \\
\Phi_{\rm N}'&=&{R'\over 2R^{2}}
F(x),\label{eq:L-poisson}
\end{eqnarray}
where $f(x)$ and $F(x)$ are regarded as 
arbitrary functions.  
Since the equation for the circumferential radius $R$ is the same as that 
in the LTB spacetime, its solution in the case of  
the marginally bound collapse, $f(x)=0$, has the 
same functional form as Eq.~(\ref{eq:radius}),
\begin{equation}
R=\left({9F\over4}\right)^{1\over3}[\tau_{R}(x)-\tau]^{2\over3},
\label{eq:N-radius}
\end{equation}
where $\tau_{R}(x)$ is an arbitrary function that determines 
the time of singularity formation.

Here we consider the Newtonian approximation of the example 
given in \S \ref{sec:gr-pertub}.  
Therefore we choose the time of the singularity formation as 
\begin{equation}
\tau_{R}(x)={x^{3/2}\over 3}\sqrt{4\over F}, \label{eq:N-s-time}
\end{equation}
so that $R$ is equal to $x$ at $\tau=0$. 
For the initial density configuration, we adopt 
the same functional form as Eq.~(\ref{density}),
\begin{equation}
{\bar \rho}(0,x)={F'\over 8\pi x^{2}}
={1\over6\pi}
\left\{1+\exp\left(-{x_{1}\over2x_{2}}\right)\right\}
\left\{1+\exp\left({{x^{2}-x_{1}^{2}\over
2x_{1}x_{2}}}\right)\right\}^{-1}, \label{eq:N-initial-density}
\end{equation}
where $x_{1}$ and $x_{2}$ are positive constants.
The above choice guarantees the regularity of all the variables 
before singularity formation and that a central shell-focusing 
singularity is formed at $\tau=1$. 

Imposing the boundary condition $\Phi_{\rm N}\rightarrow0$ for 
$x\rightarrow\infty$, the solution of Eq.~(\ref{eq:L-poisson}) 
can be formally expressed as
\begin{equation}
\Phi_{\rm N}
=\Phi_{\rm N1}(\tau,x)+\Phi_{\rm N2}(\tau),
\end{equation}
where
\begin{eqnarray}
\Phi_{\rm N1}(\tau,x)&\equiv&
\int_{0}^{x}{R'\over 2R^{2}}Fdx,\\
\Phi_{\rm N2}(\tau)&\equiv& -\int_{0}^{\infty}{R'
\over 2R^{2}}Fdx
=-{1\over2}\int_{0}^{\infty}{F'\over R}dx. \label{eq:Phi2-def}
\end{eqnarray}
Here it is worthwhile noting that the right hand side of 
Eq.~(\ref{eq:L-poisson}) at $x=0$ diverges at the
time of the central shell-focusing naked singularity formation,  
\begin{equation}
\Phi_{\rm N}'\longrightarrow {14\over 27\tau_{R(1)}^{2/3}}
~x^{-1/3}~~~~~~{\rm for}~~x\longrightarrow0~~~~{\rm at}~~\tau=1,
\end{equation}
where 
\begin{equation}
\tau_{R(1)}\equiv{1\over2}{d^{2}\tau_{R}(x)\over dx^{2}}\biggr|_{x=0}.
\end{equation}
However, since the power index of $x$ is larger than $-1$, $\Phi_{\rm N}$ 
itself is finite at $x=0$, even at the time of the central 
shell-focusing singularity formation, $\tau=1$. 

In order for the Newtonian approximation to be successful, 
temporal derivatives of all the quantities should be always 
smaller than their radial derivatives. Here we focus on 
the neighborhood of the central shell-focusing naked singularity only.  
For this purpose, we introduce a new variable $w$ defined by
\begin{equation}
w\equiv \delta\tau^{-1/2}x, \label{eq:w-definition}
\end{equation}
where $\delta\tau\equiv (1-\tau)/\tau_{R(1)}$. 
Then, we consider the limit $\tau\rightarrow 1$ 
as $w$ is kept constant.
It should be noted that $x$ also goes to zero in this limit. 
The mass function $F$, rest-mass density ${\bar \rho}$, and 
circumferential radius $R$ behave as
\begin{eqnarray}
F &\longrightarrow&{{4\over 9}} w^{3} \delta \tau^{3/2}, \label{eq:F-ap}\\
{\bar\rho}&\longrightarrow&{1\over2\pi}\tau_{R(1)}^{-2}(3+7w^{2})^{-1}
(1+w^{2})^{-1}\delta\tau^{-2}, \label{eq:rho-ap} \\
R   &\longrightarrow&\tau_{R(1)}^{2/3}w(1+w^{2})^{2/3}\delta\tau^{7/6}. 
\label{eq:R-ap}
\end{eqnarray}
All these variables are proportional to powers of $\delta\tau$,  
and the coefficients are functions of $w$. 
It is easy to see that the derivatives of these quantities 
with respect to $\tau$ or  
$x$ also have the same basic functional structure 
with respect to $\delta\tau$ and 
$w$. Thus the $x$ dependent part, $\Phi_{\rm N1}$, of  
the Newtonian gravitational potential also behaves in the manner
\begin{equation}
\Phi_{\rm N1}\longrightarrow \phi_{\rm N1}(w)\delta\tau^{i},
\end{equation}
where $i$ is a constant and $\phi_{\rm N1}$ is 
a function of $w$. Substituting the above equation 
into Eq.~(\ref{eq:L-poisson}) and using the asymptotic behavior 
 described by Eqs.(\ref{eq:F-ap}) and (\ref{eq:R-ap}), we obtain
\begin{equation}
{d\phi_{\rm N1}(w)\over dw}\delta\tau^{i-1/2}
={2w(3+7w^{2})\over 27\tau_{R(1)}^{2/3}(1+w^{2})^{5/3}}~\delta\tau^{-1/6}.
\label{eq:Phi-ap-equation}
\end{equation}
In order for the dependences on $\delta \tau$ of two sides of the above
equation to agree, $i$ must be equal to $1/3$. 
Then, integration of Eq.~(\ref{eq:Phi-ap-equation}) leads to
\begin{equation}
\phi_{\rm N1}(w)={1\over 9\tau_{R(1)}^{2/3}(1+w^{2})^{2/3}}
\left\{7(1+w^{2})-9(1+w^{2})^{2/3}+2\right\}.
\end{equation}

In order to see the asymptotic dependence of $\Phi_{\rm N2}(\tau)$ on $\tau$, 
we differentiate Eq.~(\ref{eq:Phi2-def}), obtaining 
\begin{equation}
{\dot \Phi_{\rm N2}}=-{1\over2}\int_{0}^{\infty}{F'\over R^{2}}
\sqrt{F\over R}dx.\label{eq:dt-Phi}
\end{equation}
The integrand on the right hand side of the above equation 
behaves near the origin at the time of the central shell-focusing 
singularity formation as
\begin{equation}
{F'\over R^{2}}\sqrt{F\over R}\longrightarrow 
{8\over 9\tau_{R(1)}^{5/3}}x^{-7/3}~~~~~{\rm for}~~
x\longrightarrow 0~~~~~{\rm at}~~\tau=1.
\end{equation}
Therefore, the integral in Eq.~(\ref{eq:dt-Phi}) 
does not have a finite value at $\tau=1$. Since, as shown above, 
this divergence comes from the irregularity of the integrand 
at the origin, $x=0$, we estimate the contribution to the integral 
near the origin in Eq.~(\ref{eq:dt-Phi}). We again consider the limit  
$\delta\tau\rightarrow0$ as $w$ is kept constant and obtain
\begin{equation}
\int_{0}^{\infty}{F'\over R^{2}}\sqrt{F\over R}dx
\longrightarrow {8\over 9\tau_{R(1)}^{5/3}}\delta\tau^{-2/3}
\int_{0}^{\infty}{wdw \over (1+w^{2})^{5/3}}
={2\over3\tau_{R(1)}^{5/3}}\delta\tau^{-2/3}.
\end{equation}
Substituting the above equation into 
Eq.~(\ref{eq:dt-Phi}) and integrating it with respect to $\tau$, 
we obtain
\begin{equation}
\Phi_{\rm N2}\longrightarrow {\delta\tau^{1/3}\over \tau_{R(1)}^{2/3}}
+\Phi_{\rm N2}(1). \label{eq:PhiN2-ap}
\end{equation}
Therefore, in the limit $\delta\tau\rightarrow0$, 
with $w$ constant, $\Phi_{\rm N}$ can be expressed as 
\begin{equation}
\Phi_{\rm N}\longrightarrow{9+7w^{2}\over 9\tau_{R(1)}^{2/3}(1+w^{2})^{2/3}}
~\delta\tau^{1/3}+\Phi_{\rm N2}(1). \label{eq:N-Phi-ap}
\end{equation}

We now know that in the limit $\delta\tau\rightarrow 0$, with 
$w$ constant, all the variables behave as
\begin{equation}
Z(\tau,x)\longrightarrow z(w)\delta\tau^{j}+{\rm constant}, \label{eq:Z-ap}
\end{equation}
where $z(w)$ is some function of $w$, and $j$ is a constant. 
The derivatives of $Z$ with respect to $T$ and $R$ can be 
expressed in terms of derivatives with respect to $\tau$ and $x$ as
\begin{eqnarray}
(\partial_{T}Z)_{R}&=&(\partial_{T}\tau)_{R}{\dot Z}
+(\partial_{T}x)_{R}Z', \\
(\partial_{R}Z)_{T}&=&(\partial_{R}\tau)_{T}{\dot Z}
+(\partial_{R}x)_{T}Z'.
\end{eqnarray}
 From Eqs.~(\ref{eq:L-matrix1}) and (\ref{eq:L-matrix2}), we find 
\begin{equation}
(\partial_{T}\tau)_{R}=1,~~(\partial_{R}\tau)_{T}=0,~~
(\partial_{T}x)_{R}=-{{\dot R}\over R'}
~~{\rm and}~~(\partial_{R}x)_{T}={1\over R'}.
\end{equation}
Then, we derive the relation 
\begin{equation}
{(\partial_{T}Z)_{R}\over (\partial_{R}Z)_{T}}
=R'{{\dot Z}\over Z'}-{\dot R}.
\end{equation}
Inserting Eq.~(\ref{eq:Z-ap}) into the above equation, we obtain
\begin{equation}
{(\partial_{T}Z)_{R}\over (\partial_{R}Z)_{T}}
\longrightarrow {1\over3\tau_{R(1)}^{1/3}(1+w^{2})^{1/3}}
\left\{{1\over2}w(1+7w^{2})-jz(3+7w^{2})\left({dz\over dw}\right)^{-1}
\right\}~\delta\tau^{1\over6}.
\end{equation}
This equation implies that in the limit $\delta\tau\rightarrow0$ 
with $w$ constant, the inequality 
\begin{equation}
|(\partial_{T}Z)_{R}| \ll |(\partial_{R}Z)_{T}|
\end{equation}
holds. Therefore, the validity of the 
order counting of the Newtonian approximation 
is guaranteed even in the neighborhood of the central 
naked singularity. 

Here it is worth noting that in the limit 
$\delta\tau\rightarrow0$ with $w$ fixed, $|{\dot Z}|$ is much larger 
than $|Z'|$ as seen from  
\begin{equation}
{{\dot Z}\over Z'}\longrightarrow {1\over \tau_{R(1)}}
\left({1\over2}w-jz{dw\over dz}\right)~\delta\tau^{-1/2}\longrightarrow\infty.
\end{equation}
Hence, the vicinity of a central shell-focusing naked singularity 
is not Newtonian in the ordinary sense.

\subsubsection{Basic equations of even mode perturbations}

We now consider nonspherical linear perturbations in the system of a 
spherically symmetric dust ball.  
First, we consider perturbations in the Eulerian coordinate system. 
Here, the line element is written as
\begin{equation}
ds_{\rm E}^{2}=-\left(1+2\Phi_{\rm N}+2\delta\Phi_{\rm N}\right)dT^{2}
+dR^{2}+R^{2}d\Omega^{2},
\end{equation}
where $\delta\Phi_{\rm N}$ is a perturbation of the Newtonian 
gravitational potential.
Using the transformation matrix given by Eqs.(\ref{eq:L-matrix1}) 
and (\ref{eq:L-matrix2}), we obtain the perturbed line element 
in the background Lagrangian coordinate system as
\begin{equation}
ds^{2}_{\rm L}=-\left(1+2\Phi_{\rm N}+2\delta\Phi_{\rm N}
 -{\dot R}^{2}\right)d\tau^{2}+2{\dot R}R'd\tau dx
+{R'}^{2}dx^{2}+R^{2}d\Omega^{2}.
\end{equation}
Hereafter we study the behavior of perturbations in this coordinate
system. 

The density $\rho$ and four-velocity $u^{\mu}$ 
are written in the forms
\begin{eqnarray}
\rho&=&{\bar \rho}(1+\delta_{\rho}), \\
u^{\mu}&=&{\bar u}^{\mu}+\delta u^{\mu}.
\end{eqnarray}
By definition of the Lagrangian coordinate system, 
the components of the background four-velocity are 
given by 
\begin{equation}
\left({\bar u}^{\mu}\right)=\left({\bar u}^{0},0,0,0\right).
\end{equation}
 From the normalization of the four-velocity, we find 
\begin{equation}
\delta u^{0}=-\delta\Phi_{\rm N}+{\dot R}R'\delta u^{1}.
\end{equation}

The order-counting with respect to the expansion parameter 
$\varepsilon$ of the Newtonian approximation is given by
\begin{equation}
\delta u^{0}=O(\varepsilon^{2}),~~~~\delta u^{\ell}=O(\varepsilon),
~~~~\delta_{\rho}=O(\varepsilon^{0})~~~~{\rm and}~~~~
\delta\Phi_{N}=O(\varepsilon^{2}). 
\end{equation}
Then, the equations for the perturbations are given by
\begin{eqnarray}
\partial_{\tau}\delta_{\rho}
+{1\over {\bar \rho}\sqrt{\bar \gamma}}
\partial_{\ell}\left({\bar \rho}\sqrt{\bar\gamma} 
\delta u^{\ell}\right)&=&0, \label{eq:rho-eq1}\\
\partial_{\tau}\delta u_{\ell}+\partial_{\ell}\delta\Phi_{N}
&=&0, \label{eq:v-eq1}\\
{1\over \sqrt{{\bar \gamma}}}\partial_{\ell}
\left(\sqrt{\bar \gamma}{\bar \gamma}^{\ell m}\partial_{m}
\delta\Phi_{N}\right)-4\pi{\bar\rho}\delta_{\rho}&=&0,\label{eq:p-eq1}
\end{eqnarray}
where 
\begin{equation}
\sqrt{\bar \gamma}\equiv R'R^{2}\sin\theta, 
\end{equation}
and ${\bar \gamma}^{\ell m}$ are the contravariant components 
of the background three-metric. 

Here we focus on axisymmetric even mode of perturbations. Hence 
the perturbations we consider are expressed in the forms
\begin{eqnarray}
\delta_{\rho}&=&\sum_{l}\Delta_{\rho(l)}(\tau,x)
 P_{l}(\cos\theta), \\
\delta\Phi_{N}&=&\sum_{l}\Delta_{\Phi(l)}(\tau,x)
 P_{l}(\cos\theta), \\
\delta u_{1}&=&\sum_{l}U_{x(l)}(\tau,x)
 P_{l}(\cos\theta), \\
\delta u_{2}&=&\sum_{l}U_{\theta(l)}(\tau,x){d\over d\theta}
 P_{l}(\cos\theta), \\
\delta u_{3}&=&0.
\end{eqnarray}
 From Eqs.~(\ref{eq:rho-eq1}), (\ref{eq:v-eq1}) and (\ref{eq:p-eq1})  
we obtain
\begin{eqnarray}
{\dot\Delta}_{\rho(l)}
+{1\over F'}\left({F'\over R'^{2}}U_{x(l)}
\right)'-l(l+1){U_{\theta(l)}\over R^{2}}&=&0, \label{eq:rho-eq}\\
{\dot U}_{x(l)}+\Delta_{\Phi(l)}'&=&0, \label{eq:Ux-eq}\\
{\dot U}_{\theta(l)}+\Delta_{\Phi(l)}&=&0, \label{eq:Ut-eq}\\
{1\over R'R^{2}}
\left({R^{2}\over R'}\Delta_{\Phi(l)}'\right)'-
l(l+1){\Delta_{\Phi(l)}\over R^{2}}
-4\pi{\bar\rho}\Delta_{\rho(l)}&=&0.\label{eq:P-eq}
\end{eqnarray}
Comparing the basic equations for the relativistic perturbations 
with $l=2$ in \S \ref{sec:even} to the above equations, 
we find the following correspondence between the Newtonian and relativistic 
variables: $\Delta_{\Phi(2)}=-2K$, 
$U_{x(2)}=-V_{1}$ and $U_{\theta(2)}=-V_{2}$.

\subsubsection{Mass-quadrupole formula}

Hereafter we focus on the quadrupole mode, $l=2$, and therefore 
omit the subscript $(l)$ specifying the multipole component 
of the perturbation variables. 
The mass-quadrupole moment $Q_{\ell m}$ is given by
\begin{equation}
Q_{\ell m}\equiv \int\rho
\left(X_{\ell}X_{m}-{1\over3}R^{2}\delta_{\ell m}\right)d^{3}X
={4\pi\over 15}Q(T){\rm diag}(-1,-1,2),
\end{equation}
where 
\begin{equation}
Q(T)\equiv\int_{0}^{\infty}{\bar\rho}\Delta_{\rho}R^{4}dR.
\label{eq:QB-relation}
\end{equation}
For a function $g(\tau,x)$ with sufficiently rapid 
decay as $x\rightarrow \infty$, we find that
\begin{eqnarray}
{d\over dT}\int_{0}^{\infty}gdR
&=&\int_{0}^{\infty}(\partial_{T} g)_{R}dR
=\int_{0}^{\infty}\left({\dot g}-{{\dot R}\over R'}g'\right)R'dx \\
&=&\int_{0}^{\infty}\left(R'{\dot g}
 +{\dot R}'g\right)dx-\left[{\dot R}g\right]_{0}^{\infty}
=\int_{0}^{\infty}\partial_{\tau}(R'g)dx.
\end{eqnarray}
Using the above formula, we obtain
\begin{equation}
{d^{m}Q\over dT^{m}}
=\int_{0}^{\infty}
{\partial^{m}\over\partial\tau^{m}} 
\left({\bar\rho }\Delta_{\rho}R'R^{4}\right)dx
={1\over8\pi}\int_{0}^{\infty}F'
{\partial^{m}\over\partial\tau^{m}} 
\left(\Delta_{\rho}R^{2}\right)dx. \label{eq:delt-Q}
\end{equation}
The power $L_{\rm GW}$ carried by the gravitational radiation 
at the future null infinity $T+R\rightarrow\infty$ is 
given by
\begin{equation}
L_{\rm GW}={32\pi^{2}\over375}\left({d^{3}Q(u)\over du^{3}}\right)^{2},
\label{eq:G-power}
\end{equation}
where $u\equiv T-R$ is the retarded time. 
The Weyl scalar $\Psi_{4}$ carried by outgoing gravitational 
waves at the future null infinity is estimated as 
\begin{equation}
\Psi_{4}=-C_{abcd}n^{a}{\bar m}^{b}n^{c}{\bar m}^{d}
=-{3\pi\over5}{d^{4}Q(u)\over du^{4}}\sin^{2}\theta,
\label{eq:curvature}
\end{equation}
where $C_{abcd}$ is the Weyl tensor, and $n^{a}$ and ${\bar m}^{a}$ 
are two of the null tetrad basis vectors, whose components in  
spherical polar coordinates are given by
\begin{eqnarray}
(n_{\mu})&=&{1\over \sqrt{2}}(1,-1,0,0), \\
({\bar m}_{\mu})&=&{1\over\sqrt{2}}(0,0,R,-iR\sin\theta).
\end{eqnarray}
It should be noted that the power $L_{\rm GW}$ is proportional 
to the square of the third-order derivative of $Q(u)$, while 
the Weyl scalar $\Psi_{4}$ is  
proportional to the fourth-order derivatives of $Q(u)$. 

\subsubsection{Asymptotic analysis of the perturbations}
We assume that all of the perturbation variables are regular before the 
central shell-focusing singularity formation and hence
can be written as 
\begin{eqnarray}
\Delta_{\rho}&=&x^{2}\Delta{\rho}^{*}, \\
U_{x}&=&xU_{x}^{*}, \\
U_{\theta}&=&x^{2}U_{\theta}^{*}. \\
\Delta_{\Phi}&=&x^{2}\Delta_{\Phi}^{*}, 
\end{eqnarray}
where each variable with an asterisk is given in the form 
of a Taylor series with respect to $x^{2}$.

In order to obtain information concerning the asymptotic behavior of the 
mass-quadrupole moment, we should carefully examine the asymptotic behavior 
of the perturbation variables near the origin. 
For this purpose, we introduce $w$ defined in Eq.~(\ref{eq:w-definition}) 
and then consider the limit $\tau\rightarrow 1$ with fixed $w$. 
Since all the background variables appearing in the 
equations of the perturbations are proportional to some powers 
of $\delta\tau$ and their coefficients are functions of $w$, 
as given in Eqs. (\ref{eq:F-ap})--(\ref{eq:R-ap}) and (\ref{eq:N-Phi-ap}),  
we expect that the perturbation variables 
also behave in the same manner as the background variables and 
hence we assume 
\begin{eqnarray}
{\Delta}^{*}_{\rho}&=&{\delta}^{*}_{\rho}(w)\delta\tau^{-p}, 
\label{eq:rho-asp}\\
{U}^{*}_{x}&=&{1\over x}\partial_{x}(x^{2}{U}^{*}_{\theta})
=\tau_{R(1)}^{1/3}
\left(w{d{u}^{*}_{\theta}\over dw}+2{u}^{*}_{\theta}\right)
\delta\tau^{-q}, \\
{U}^{*}_{\theta}&=&\tau_{R(1)}^{1/3}{u}^{*}_{\theta}(w)\delta\tau^{-q}, 
\label{eq:Ut-asp}\\
{\Delta}^{*}_{\Phi}&=&\tau_{R(1)}^{-2/3}
{\delta}^{*}_{\Phi}(w)\delta\tau^{-r}.\label{eq:P-asp}
\end{eqnarray}
 
Now, by virtue of our knowledge about the asymptotic forms 
(\ref{eq:rho-asp})--(\ref{eq:P-asp}), a rigorous 
analysis about the evolution of the mass-quadrupole moment 
is possible. Substituting 
Eqs.~(\ref{eq:rho-asp})--(\ref{eq:P-asp}) into
Eqs.~(\ref{eq:rho-eq})--(\ref{eq:P-eq}), 
and using the asymptotic behavior of the 
background variables (\ref{eq:F-ap})--(\ref{eq:R-ap}), 
we obtain 
\begin{eqnarray}
\left(w{d{\delta}^{*}_{\rho}\over dw}
+2p{\delta}^{*}_{\rho}\right)\delta\tau^{-p-1}
&+&\left[{18\over w^{4}}{d\over dw}\left\{{w^{2}(1+w^{2})^{2/3}
\over (3+7w^{2})^{2}}{d\over
dw}(w^{2}{u}^{*}_{\theta})\right\}\right. \nonumber \\
&-& \left.{12{u}^{*}_{\theta}\over w^{2}(1+w^{2})^{4/3}}\right]
\delta\tau^{-q-7/3}=0,\label{eq:conti-ap}
\end{eqnarray}
\begin{equation}
\left(w{d{u}^{*}_{\theta}\over dw}+2q{u}^{*}_{\theta}\right)
\delta\tau^{-q-1}+2{\delta}^{*}_{\Phi}\delta\tau^{-r}
=0, \label{Euler-ap}
\end{equation}
and
\begin{equation}
\left[{d\over dw}\left\{{w^{2}(1+w^{2})^{5/3}\over 3+7w^{2}}
{d\over dw}(w^{2}{\delta}^{*}_{\Phi})\right\}
-{2w^{2}(3+7w^{2})\over3(1+w^{2})^{1/3}}{\delta}^{*}_{\Phi}\right]
\delta\tau^{-r+5/3}
-{2\over9}w^{4}{\delta}^{*}_{\rho}\delta\tau^{-p+2}=0.\label{eq:P-ap}
\end{equation}
Since the powers of $\delta\tau$ should be balanced 
in each equation, we obtain
\begin{equation}
q=p-{4\over3}~~~~~~{\rm and}~~~~~~r=p-{1\over3}. \label{eq:power}
\end{equation}
Equations (\ref{eq:conti-ap})--(\ref{eq:P-ap}) constitute a closed system 
of ordinary differential equations. 

Through an appropriate manipulation, we obtain a single decoupled 
equation for ${u}^{*}_{\theta}$ as
\begin{equation}
{d^{4}{u}^{*}_{\theta}\over dy^{4}}
+c_{3}{d^{3}{u}^{*}_{\theta}\over dy^{3}}
+c_{2}{d^{2}{u}^{*}_{\theta}\over dy^{2}}
+c_{1}{d{u}^{*}_{\theta}\over dy}
+c_{0}{u}^{*}_{\theta}=0, \label{eq:u-eq}
\end{equation}
where $y\equiv w^2$ and 
\begin{eqnarray}
c_{0}&=&-2\{-24(9 + 26y + 21y^2)  
+3q^2(63 + 414y + 1016y^2 + 1106y^3 + 441y^4) \nonumber \\
&+& q(252 + 1323y + 3149y^2 + 3465y^3 + 1323y^4)\} 
\left\{9y^3(1 + y)^3 (3 + 7y)^2\right\}^{-1}, \nonumber \\
&& \nonumber \\
c_{1}&=&\{378 + 2196y + 4758y^2 + 2006y^3 - 4424y^4 - 3234y^5 
\nonumber \\
&+&3q^2(1 + y)^2 (189 + 1035y + 1911y^2 + 1225y^3) \nonumber \\
&+& q(1323 + 7893y + 18966y^2 + 21202y^3 + 9247y^4 + 
    441y^5)\} \nonumber \\
&\times&\{18y^3(1 + y)^3(3 + 7y)^2\}^{-1}, \nonumber \\
&& \nonumber \\
c_{2}&=&\{1269 + 7731y + 18453y^2 + 19565y^3 + 7350y^4 
+9q^2(3 + 10y + 7y^2)^2 \nonumber \\
&+& 6q(153 + 993y + 2387y^2 + 2527y^3 + 980y^4)\} 
\{9y^2(1 + y)^2(3 + 7y)^2\}^{-1}, \nonumber \\
&& \nonumber \\
c_{3}&=&\{159 + 506y + 427y^2 
+ 12q(3 + 10y + 7y^2)\} \{6y(1 + y)(3 + 7y)\}^{-1}. \nonumber 
\end{eqnarray}
With appropriate boundary conditions, we can numerically solve 
Eqs.~(\ref{eq:conti-ap})--(\ref{eq:P-ap}) as a kind of the 
eigenvalue problem to obtain $q$ and the solution for 
${u}^{*}_{\theta}$. 

The boundary condition at $y=0$ for Eq.~(\ref{eq:u-eq}) is 
uniquely determined by the Taylor expandability with respect to $y$
in terms of the unknown parameter $q$
and the normalization condition $u_{\theta}^{*}|_{y=0}=1$.
We numerically integrate 
Eq.~(\ref{eq:u-eq}) outward from $y=0$ using the 
fourth order Runge-Kutta method. 
The behavior of ${u}^{*}_{\theta}$ depends on the value 
of $q$. 

Considering the behavior of Eq.~(\ref{eq:u-eq}) in the 
limit $y\rightarrow \infty$ and imposing the 
condition that $U_{\theta}^{*}$ is nonzero and finite for 
$0<x<\epsilon$ at $\delta\tau=0$, 
where $\epsilon$ is a positive infinitesimal number,
we obtain the outer numerical boundary as
\begin{equation}
{u}^{*}_{\theta}\longrightarrow {\rm const}\times y^{-q}
~~~~~~{\rm as}~~~~~~y\longrightarrow\infty. 
\end{equation}
The numerical calculation reveals that the above behavior is realized when 
\begin{equation}
q=0.3672~.
\end{equation}
 From Eq.~(\ref{eq:power}), we obtain
\begin{equation}
p=1.701~~~~~{\rm and}~~~~~~r=1.367~. \label{eq:ap-results}
\end{equation}
The above values agree with the results obtained from 
the direct numerical simulation of partial differential equations
(\ref{eq:rho-eq})--(\ref{eq:P-eq})
quite well.\cite{Nakao:2001as}

Now we examine the mass-quadrupole moment $Q(T)$ and its time derivatives 
$d^{m}Q/dT^{m}$. 
In order to find the contribution of the central singularity 
to $d^{m}Q/dT^{m}$, we consider the integrand on the right hand side of 
Eq.~(\ref{eq:delt-Q}). 
Using Eqs.~(\ref{eq:F-ap}), (\ref{eq:R-ap}) and (\ref{eq:rho-asp}), 
we obtain
\begin{eqnarray}
F'\Delta_{\rho}R^2&\longrightarrow&
{4\over3}\tau_{R(1)}^{4/3}w^{6}(1+w^{2})^{4/3}{\delta}^{*}_{\rho}(w) 
~\delta\tau^{13/3-p}\nonumber \\
&=&{4\over3}\tau_{R(1)}^{4/3}x^{2(13/3-p)}w^{2(p-4/3)}
(1+w^{2})^{4/3}{\delta}^{*}_{\rho}(w).
\end{eqnarray}
>From the above equation, we obtain
\begin{eqnarray} 
I^{(m)}(\tau,x)&\equiv&
{\partial^{m}\over\partial\tau^{m}}\left(F'\Delta_{\rho}R^{2}\right)
\longrightarrow
{2^{2-m}\over3}\tau_{R(1)}^{4/3-m}
\delta\tau^{13/3-p-m}w^{26/3-2p-2m} \nonumber \\
&\times&\left(w^{3}{d\over dw}\right)^{m}
\left\{w^{2(p-4/3)}(1+w^{2})^{4/3}{\delta}^{*}_{\rho}(w)\right\}.
\label{eq:Im-def}
\end{eqnarray}

We consider the integral of $I^{(m)}$ from $x=0$ to $x=x_{\rm o}$ 
to determine the contribution of the central 
shell-focusing naked singularity to 
the time derivatives of the mass-quadrupole moment. 
Here we take the limit $\delta\tau\rightarrow0$ with 
$w_{o}\equiv x_{\rm o}\delta\tau^{-1/2}$ constant, and 
then consider the limit $w_{o}\rightarrow\infty$. 
In this way, we obtain
\begin{eqnarray}
\int_{0}^{x_{\rm o}}I^{(m)}(\tau,x)dx
&=&\delta\tau^{1/2}\int_{0}^{w_{o}}
I^{(m)}(\tau,\delta\tau^{1/2} w)dw  \nonumber \\
&\longrightarrow&{2^{2-m}\over3}\tau_{R(1)}^{4/3-m}
\delta\tau^{29/6-p-m} \nonumber \\
&\times&\int_{0}^{\infty}
w^{26/3-2p-2m}
\left(w^{3}{d\over dw}\right)^{m}
\left\{w^{2(p-4/3)}(1+w^{2})^{4/3}{\delta}^{*}_{\rho}(w)\right\}dw. \nonumber
\end{eqnarray}
The above equation and Eq.~(\ref{eq:ap-results}) 
show that the contribution of the 
central singularity to $d^{m}Q/dT^{m}$ diverges for $\tau\rightarrow1$ 
if and only if $m$ is larger than or equal to four. 
This result and the quadrupole formula imply that the metric perturbation 
corresponding to the gravitational radiation and its first-order 
temporal derivative are finite, but the second-order temporal derivative 
diverges. Hence the power $L_{\rm GW}$ of the gravitational radiation 
is finite, but the curvature $\Psi_{4}$ carried by the gravitational waves 
from the central naked singularity diverges (see Eqs.~(\ref{eq:G-power}) 
and (\ref{eq:curvature})). This conclusion agrees with 
the relativistic perturbation analysis. Further, we find that 
in the limit of $\tau\rightarrow1$,
\begin{equation}
\Psi_{4}=-{3\pi\over5}{d^{4}Q\over dT^{4}}
\propto\delta\tau^{5/6-p}\propto(1-\tau)^{-0.867}.
\end{equation}
This result is also consistent with the relativistic 
perturbation analysis in \S \ref{sec:even}. 

\subsubsection{Summary of Newtonian perturbations of spherical dust collapse}
We analyzed the even-mode perturbations of $l=2$ for  
spherically symmetric dust collapse in the framework 
of the Newtonian approximation and estimated the gravitational 
radiation generated by these perturbations using the quadrupole 
formula. 
Since we treat separately the dynamics of the matter perturbations 
and the gravitational 
waves in the wave zone, we can estimate 
the asymptotic behavior semi-analytically, and  
we obtain the results by solving 
gentle ordinary differential equations.
This is the great advantage of the Newtonian approximation. 

>From this analysis, we found that the power carried 
by the gravitational waves from the neighborhood of a  
naked singularity at the symmetric center is finite. 
However, the spacetime curvature 
associated with the gravitational waves becomes infinite, 
in accordance with the power law. 
This result is consistent with the
relativistic perturbation analysis in \S \ref{sec:even}. 
Furthermore, the power index obtained from the Newtonian analysis 
also agrees with that obtained from the relativistic perturbation analysis 
quite well. 

The agreement between the results of the Newtonian and 
relativistic analyses suggests that the perturbations 
themselves are always confined within the range to 
which the Newtonian approximation is applicable. 
Here we focus on the metric perturbation, ${\Delta}^{*}_{\Phi}$. 
Since the asymptotic solution of ${\Delta}^{*}_{\Phi}$ has the same form 
as Eq.~(\ref{eq:Z-ap}), we immediately find that 
in the limit $\delta\tau\rightarrow0$ with $w$ constant, 
\begin{equation}
{\partial_{T}{\Delta}^{*}_{\Phi}\over \partial_{R}{\Delta}^{*}_{\Phi}}\propto 
\delta\tau^{1/6},
\end{equation}
and hence the assumption 
$|\partial_{T}{\Delta}^{*}_{\Phi}|\ll|\partial_{R}{\Delta}^{*}_{\Phi}|$
of the Newtonian approximation is valid in the Eulerian coordinate
system. We can also find the second order derivatives.
In the same limit, we find 
\begin{equation}
{\partial_{T}\partial_{R}{\Delta}^{*}_{\Phi}
\over \partial_{R}^{2}{\Delta}^{*}_{\Phi}}\propto
\delta\tau^{1/6}~~~~{\rm and}~~~~
{\partial_{T}^{2}{\Delta}^{*}_{\Phi}
\over \partial_{R}\partial_{T}{\Delta}^{*}_{\Phi}}\propto 
\delta\tau^{1/6}.
\end{equation}
 From the above equations, we obtain 
\begin{equation}
{\partial_{T}^{2}{\Delta}^{*}_{\Phi}
\over \partial_{R}^{2}{\Delta}^{*}_{\Phi}}\propto 
\delta\tau^{1/3}.
\end{equation}
The above equation implies that in the limit of $\delta\tau\rightarrow0$ 
with fixed $w$, the inequality 
\begin{equation}
|\partial_{T}^{2}{\Delta}^{*}_{\Phi}|\ll
|\partial_{R}^{2}{\Delta}^{*}_{\Phi}|
\end{equation}
is also satisfied. This inequality implies 
that the wave equation for the metric perturbation 
$\Delta_{\Phi}$ is approximated well by a Poisson type equation 
if we adopt the Eulerian coordinate system. 

However, the gravitational collapse producing 
the shell-focusing globally naked singularity is not Newtonian 
in the ordinary sense. The same is true for the 
perturbation variables, because 
$|{\dot\Delta_{\Phi}}/\Delta_{\Phi}'|\gg1$ 
in the limit $\delta\tau\rightarrow0$ with fixed $w$.  
Even though the Newtonian approximation is valid in the 
Eulerian coordinate system, 
Newtonian order counting breaks down if we adopt the
Lagrangian coordinate system as the spatial coordinates. 

\subsection{Cylindrical collapse and gravitational radiation}
It has long been known that collapsing cylindrically symmetric fluids
form naked singularities.\cite{Thorne:1972ji}
Such examples are not considered as direct
counterexamples to the cosmic censorship conjecture,  
because these spacetimes are not asymptotically
flat. However, there is an  expectation 
in which the local behavior of prolate collapse
to a spindle singularity is very similar to that of infinite
cylindrical collapse. For this reason, properties 
of cylindrical collapse have been
studied in this context. Apostolatos and Thorne \cite{Apostolatos:1992}
investigated the collapse of a
counterrotating dust shell cylinder and showed that rotation, even if
it is infinitesimally small, can halt the gravitational collapse of the
cylinder.  Echeveria\cite{Echeverria:1993wf} studied the evolution 
of a cylindrical dust shell
analytically at late times and numerically for all
times. It was found
that the shell collapses to form a strong  singularity in finite proper
time. The
numerical results showed that a sharp burst of gravitational waves is
emitted by the shell just before the singularity forms.
Chiba \cite{Chiba:1996} showed that the maximal time slicing 
never possesses the singularity
avoidance property in cylindrically symmetric spacetimes 
and proposed a new time slicing that may be
suitable to investigate the formation of cylindrical singularities. 
He numerically investigated cylindrical dust collapse to elucidate 
the role of gravitational waves and found that a negligible amount of 
gravitational waves is emitted during the free fall time.  
There seems to be a discrepancy in the results of Echeveria and Chiba
with regard to the emission of gravitational waves.
The origin of this discrepancy may be the difference between the 
matter fields: a thin, massive dust shell in the former case and 
a dust cloud in the latter.
However, further careful analysis is necessary.

\section{Quantum particle creation from a forming naked singularity}
\label{sec:quantum}
As we have seen in \S \ref{sec:intro}, naked singularities 
are formed in some models of gravitational collapse.
If this is the case for more realistic situations,
then what happens?
The existence of naked singularities implies 
that the high-curvature region due to strong gravity
is exposed to us.
If naked singularities emit anything, we
may obtain information of some features of quantum gravity.
In this context, particle creation 
due to effects of quantum fields in curved space
will be one of the interesting possibilities.
Using a semi-classical theory of quantum fields in curved space,
Hawking~\cite{hawking1975} derived black body radiation
from a black hole formed in complete gravitational collapse.
Ford and Parker~\cite{fp1978} calculated quantum
emission from a shell-crossing naked singularity
and obtained a finite amount of flux.  
Hiscock, Williams and Eardley~\cite{hwe1982} considered a shell-focusing
naked singularity which results from a self-similar implosion
of null dust and obtained diverging flux.
Here, we will apply a semi-classical theory of quantum fields in curved
space to the collapse of a dust ball, i.e., the LTB solution. 

\subsection{Particle creation by a collapsing body}
\subsubsection{Power, energy and spectrum}
We consider both minimally and conformally coupled 
massless scalar fields in a four-dimensional spacetime
which is spherically symmetric and asymptotically flat. 
Let $T,R,\theta,\phi$ denote the usual 
quasi-Minkowskian time and spherical coordinates,
which are asymptotically related to null coordinates 
$u$ and $v$ through $u\approx T-R$ and $v\approx T+R$.
If the exterior region is vacuum and spherically symmetric,
it is described by the Schwarzschild metric,
which is given by
\begin{equation}
ds^2=-\left(1-\frac{2M}{R}\right)dT^2+\left(1-\frac{2M}{R}\right)^{-1}
dR^2+R^2d\Omega^{2},
\end{equation}
where $d\Omega^{2}\equiv d\theta^{2}+\sin^{2}\theta d\phi^{2}$.
In this case, $u$ and $v$ are naturally given by 
the Eddington-Finkelstein null coordinates $u$ and $v$, which 
are given by 
\begin{eqnarray}
u&\equiv& T-R_{*}, 
\label{u}\\
v&\equiv& T+R_{*},
\label{v}
\end{eqnarray}
with $R_{*}\equiv R+2M\ln[(R/2M)-1]$.

An incoming null ray $v=\mbox{const}$, originating from 
past null infinity ${\cal I^{-}}$,
propagates through the center becoming an outgoing null ray 
$u=\mbox{const}$, and arriving on future null infinity ${\cal I^{+}}$ 
at a value $u\equiv F(v)$.
Conversely, we can trace a null ray from $u$ on ${\cal I^{+}}$ to
$v\equiv G(u)$ on ${\cal I^{-}}$, where $G$ is the inverse of $F$.
Here, we assume that the geometrical optics approximation is valid.
The geometrical optics approximation 
implies that the trajectories of the null rays
give surfaces of constant phase.
Then, in the asymptotic region, the mode function 
which contains an ingoing mode of the 
standard form on ${\cal I^{-}}$ is the following
\begin{equation}
u^{\rm in}_{\omega l m}\approx \frac{1}
{\sqrt{4\pi \omega} R}(e^{-i\omega v}-e^{-i
\omega G(u)})Y_{lm}(\theta,\phi),
\end{equation}
where we have imposed the 
reflection symmetry condition at the center.
In the asymptotic region, the mode function 
which contains an outgoing mode of the 
standard form on ${\cal I^{+}}$ is the following
\begin{equation}
u^{{\rm out}}_{\omega l m}\approx \frac{1}
{\sqrt{4\pi \omega} R}(e^{-i\omega 
F(v)}-e^{-i\omega u})Y_{lm}(\theta,\phi).
\end{equation}
Note that in the above we have normalized the mode functions as
\begin{equation}
(u_{\omega l m},u_{\omega^{\prime} l^{\prime} m^{\prime}})
=\delta(\omega-\omega^{\prime})\delta_{ll^{\prime}}\delta_{mm^{\prime}},
\end{equation}
where the inner product is defined by integration on
the space-like hypersurface $\Sigma$ as
\begin{equation}
(f,g)=-i\int_{\Sigma}(fg^{*}_{,\mu}-f_{,\mu}g^{*})d\Sigma^{\mu}.
\end{equation}

Using the above mode functions we can express the scalar field $\phi$ as
\begin{eqnarray}
\phi&=&\sum_{l,m}\int d\omega^{\prime}
(\mbox{\boldmath{$a$}}^{{\rm in}}_{\omega^{\prime}lm}
u^{{\rm in}}_{\omega^{\prime}lm}+
\mbox{\boldmath{$a$}}^{{\rm in}\dagger}_{\omega^{\prime}lm}
u^{{\rm in}*}_{\omega^{\prime}lm}), \\
\phi&=&\sum_{l,m}\int 
d\omega^{\prime}(\mbox{\boldmath{$a$}}^{{\rm out}}_{\omega^{\prime}lm}
u^{{\rm out}}_{\omega^{\prime}lm}+
\mbox{\boldmath{$a$}}^{{\rm out}\dagger}_{\omega^{\prime}lm}
u^{{\rm out}*}_{\omega^{\prime}lm}).
\end{eqnarray}
According to the usual procedure of canonical quantization, 
we obtain the following commutation relations
\begin{eqnarray}
\left[\mbox{\boldmath{$a$}}_{\omega lm}^{{\rm in}}, 
\mbox{\boldmath{$a$}}_{\omega^{\prime}l^{\prime}m^{\prime}}
^{{\rm in}\dagger}\right]&=&
\delta(\omega-\omega^{\prime})
\delta_{l l^{\prime}}\delta_{m m^{\prime}}, \\
\left[
\mbox{\boldmath{$a$}}_{\omega lm}^{{\rm out}}, 
\mbox{\boldmath{$a$}}_{\omega^{\prime}l^{\prime}m^{\prime}}
^{{\rm out} \dagger}\right]&=&
\delta(\omega-\omega^{\prime})
\delta_{l l^{\prime}}\delta_{m m^{\prime}},
\end{eqnarray}
where it is noted that the Lagrangian in the Minkowski spacetime
is common for both minimally 
and conformally coupled scalar fields.
Here, $\mbox{\boldmath{$a$}}_{\omega lm}^{{\rm in}}$ 
and $\mbox{\boldmath{$a$}}_{\omega lm}^{{\rm out}}$ 
are interpreted as annihilation operators 
corresponding to in and out modes, respectively.
Then we set the initial quantum state to in vacuum, i.e.,
\begin{equation}
  \mbox{\boldmath{$a$}}_{\omega lm}^{{\rm in}}|0\rangle=0.
\end{equation}

The radiated power for fixed $l$ and $m$ is given by
estimating the expectation value of stress-energy tensor
through the point-splitting regularization in a flat spacetime
as~\cite{fp1978}
\begin{equation}        
P_{lm}\equiv \int \langle T_{T}^{~R}\rangle R^2d\Omega
=\frac{1}{24\pi}\left[\frac{3}{2}\left(
\frac{G^{\prime\prime}}{G^{\prime}}\right)^2-
\frac{G^{\prime\prime\prime}}{G^{\prime}}\right]
=\frac{1}{48\pi}
\left(\frac{G^{\prime\prime}}{G^{\prime}}\right)^2
-\frac{1}{24\pi}\left(\frac{G^{\prime\prime}}
{G^{\prime}}\right)^{\prime}
\label{llm}
\end{equation}
for a minimally coupled scalar field, and
\begin{equation}        
\hat{P}_{lm}\equiv \int \langle \hat{T}_{T}^{~R}\rangle 
R^2d\Omega=\frac{1}{48\pi}
\left(\frac{G^{\prime\prime}}{G^{\prime}}\right)^2
\label{hatllm}
\end{equation}  
for a conformally coupled scalar field.
Here the prime denotes the differentiation 
with respect to the argument of the function.
It implies that the amount of the power depends on
the way of coupling of the scalar field with 
gravity.
However, if 
\begin{equation}
\left.\frac{G^{\prime\prime}}{G^{\prime}}\right|_{a}
=\left.\frac{G^{\prime\prime}}{G^{\prime}}\right|_{b}
\end{equation}
holds,
the radiated energy from $u=a$ to
$u=b$ of a minimally coupled field
\begin{equation}
E_{lm}\equiv \int_{a}^{b}P_{lm}du
\end{equation}
and that of a conformally coupled field
\begin{equation}
\hat{E}_{lm}\equiv \int_{a}^{b}\hat{P}_{lm}du
\end{equation}
coincide exactly. 
The actual power is given by summation of all $(l,m)$.
The simple summation diverges.
This is because we have neglected the back scattering effect by
the curvature potential which will reduce the radiated flux
considerably for larger $l$.
Therefore we should recognize that the above expressions for 
the power,
(\ref{llm}) and (\ref{hatllm}),
are a good approximation only for smaller $l$.
Hereafter we omit the suffixes $l$ and $m$.

The spectrum of radiation is derived 
from the Bogoliubov coefficients
which relate in and out modes given as:~\cite{bd1982}
\begin{eqnarray}
\alpha_{\omega^{\prime}\omega}&=&(u^{{\rm in}}_{\omega^{\prime}},
u^{{\rm out}}_{\omega})=\frac{1}{2\pi}
\sqrt{\frac{\omega^{\prime}}{\omega}}\int^{\infty}_{-\infty}
dv e^{i\omega F(v)-i\omega^{\prime}v}, 
\label{alpha}\\
\beta_{\omega^{\prime}\omega}&=&
-(u^{{\rm in}}_{\omega^{\prime}},u^{{\rm out}*}_{\omega})=
-\frac{1}{2\pi}
\sqrt{\frac{\omega^{\prime}}{\omega}}\int^{\infty}_{-\infty}
dv e^{-i\omega F(v)-i\omega^{\prime}v}.
\label{beta}
\end{eqnarray}
The expectation value $N(\omega)$ of the 
particle number of a frequency $\omega$
on ${\cal I^{+}}$ is obtained by
\begin{equation}
N(\omega)
=\int^{\infty}_{0}d\omega^{\prime}|\beta_{\omega^{\prime}
\omega}|^2.
\label{nomega}
\end{equation}

It is noted that these results are free of 
ambiguity coming from local curvature
because the regularization is done
only in a flat spacetime.

\subsubsection{Quantum stress-energy tensor in a two-dimensional spacetime}
We can estimate the vacuum expectation value of 
stress-energy tensor in a two-dimensional spacetime
without ambiguity which may come from local curvature in contrast
to a four-dimensional case.
Unfortunately, this is not the case for 
four-dimensional spacetime.
For simplicity, we consider a minimally coupled scalar field
as a quantum field, although the situation would not be changed for 
other massless fields.

It is known that any two-dimensional spacetime is conformally flat. 
Then its metric can be expressed by double null coordinates 
$(\hat{u},\hat{v})$ as
\begin{equation}
 ds^2 = -C^2(\hat{u},\hat{v})d\hat{u}d\hat{v}.
\end{equation}
If the initial quantum state is set to the vacuum state in the 
Minkowski spacetime $ds^{2}=-d\hat{u}d\hat{v}$, then
the vacuum expectation value of the stress-energy tensor of the scalar field is
given by \cite{dfu1976}
\begin{eqnarray}
 \langle T_{\hat{u}\hat{u}}\rangle 
  &=& -\frac{1}{12\pi}C\left(C^{-1}\right)_{,\hat{u}\hat{u}},\label{hatuu}\\
 \langle T_{\hat{v}\hat{v}}\rangle 
  &=& -\frac{1}{12\pi}C\left(C^{-1}\right)_{,\hat{v}\hat{v}}\label{hatvv}\\
 \langle T_{\hat{u}\hat{v}}\rangle &=& \frac{{\cal{R}}C^2}{96\pi},
 \label{hatuv}
\end{eqnarray}
where $\cal R$ is the two-dimensional scalar curvature. 

As usual, we require that the regular center is given by 
$\hat{u}=\hat{v}$ and that
$\hat{v}$ coincides with the standard Eddington-Finkelstein
advanced time coordinate $v$.
We introduce the internal double null coordinates $U$ and $V$.
Assuming the relation between internal and external null 
coordinates, 
\begin{eqnarray}
 U&=&\alpha(u), \\
 v&=&\beta(V),
\end{eqnarray}
we can obtain all the coordinate relations as
\begin{eqnarray}
 \hat{v}&=&v=\beta(V), \\
 \hat{u}&=&\beta(U)=\beta(\alpha(u)). \label{relation}
\end{eqnarray}
In terms of the interior double null coordinates,
the line element in the interior is given by
\begin{equation}
ds^{2}=-A^{2}(U,V)dUdV,
\end{equation}
and we further require $U=V$ at the regular center.
Then we can transform the quantum stress-energy tensor given by
$(\hat{u},\hat{v})$ coordinates to the ones in the interior and the
exterior double null coordinates.

Using the coordinate relation equation (\ref{relation}) we can transform
Eqs.~(\ref{hatuu})--(\ref{hatuv}) to the components expressed in 
the interior coordinates and in the exterior coordinates.
The components in the interior coordinates are given by
\begin{eqnarray}
 \langle T_{UU}\rangle &=& F_{U}(\beta ')- F_{U}(A^2) \label{TUU}\\
 \langle T_{VV}\rangle &=& F_{V}(\beta ')- F_{V}(A^2) \label{TVV}\\
 \langle T_{UV}\rangle &=& -\frac{1}{24 \pi}(\ln A^2)_{,UV} 
\label{TUV}
\end{eqnarray}
where 
\begin{equation}
 F_x(y) \equiv \frac{1}{12\pi}\sqrt{y}\left(\frac{1}{\sqrt{y}}\right)_{,xx}.
\end{equation}
The $\beta '$ in Eq.~(\ref{TUU}) should be considered as
$\beta '(U)$.

In the exterior vacuum region, the line element is given by
\begin{equation}
ds^{2}=-\left(1-\frac{2M}{R}\right)dudv,
\end{equation}
where $u$ and $v$ are the Eddington-Finkelstein null coordinates.
Then, we obtain
\begin{eqnarray}
 \langle T_{uu}\rangle &=& - F_{u}(D^2)+\alpha'^2 F_{U}(\beta ')+ F_{u}(\alpha') \label{Tuu}\\
 \langle T_{vv}\rangle &=& - F_{v}(D^2) \label{Tvv}\\
 \langle T_{uv}\rangle &=& -\frac{1}{24 \pi}(\ln D^2)_{,uv},
\label{Tuv}
\end{eqnarray}
where $D^{2}= 1-2M/R$.
The $\beta '$ in Eq.~(\ref{Tuu}) should be also considered as
$\beta '(U)$. 
In this region we obtain
\begin{eqnarray}
(\ln D^2)_{,uv}&=&-\left(2\frac{M^2}{R^4}-\frac{M}{R^3}\right), \\
 F_u(D^2)&=&F_v(D^2)=-\frac{1}{24\pi}\left(\frac{3}{2}\frac{M^2}{R^4}
                    -\frac{M}{R^3}\right).
 \label{F_u=F_v}
\end{eqnarray}
It is found that the outgoing part of flux $\langle T_{uu}\rangle$
has terms which are dependent on the internal structure of 
the collapsing body. 

\subsection{Particle creation in analytic dust collapse}

\subsubsection{Mapping}
We can find the function $G$ or $F$ by solving the trajectories 
of outgoing and ingoing null rays in the dust cloud and
determining the retarded time $u$ and the advanced time $v$
through Eqs.~(\ref{u}) and (\ref{v})
at the time when the outgoing and ingoing null rays reach 
the surface boundary, respectively.
The LTB solution for marginally bound collapse 
is given by Eq.~(\ref{eq:radius}).
The trajectories of null rays in the dust cloud are given by
the following ordinary differential equation
\begin{equation}
\frac{dt}{dr}=\pm R_{,r},
\label{dtdr}
\end{equation} 
where the upper and lower signs denote
outgoing and ingoing null rays, respectively. 

We have solved the ordinary differential equation (\ref{dtdr})
numerically.~\cite{hin2000a,hin2000b}
We have used the Runge-Kutta method of the fourth order.
We have executed the quadruple precision calculations. 

\begin{figure}[tbp]
        \centerline{\epsfxsize 10cm \epsfbox{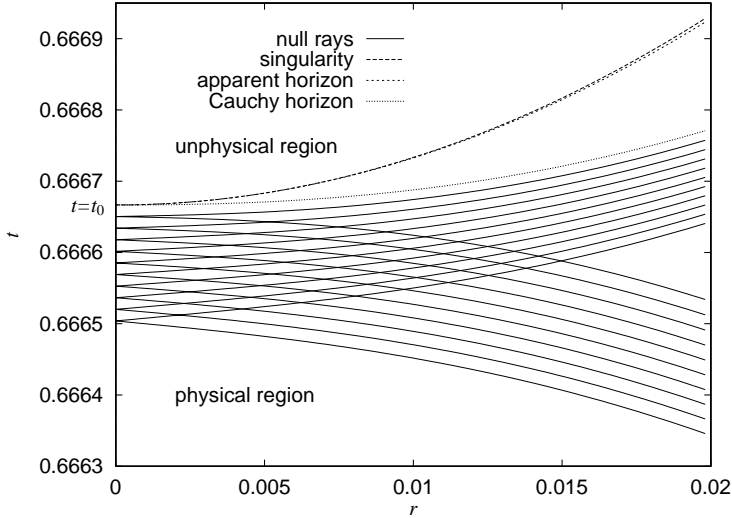}}
        \caption{Null rays interior of the 
dust cloud in the analytic LTB spacetime. 
Parameters are chosen as
$F_{3}=1$, $F_{5}=-2$ and $r_{b}=0.02$.
See text for definitions of these parameters.}
        \label{fg:rays_ltb}
\end{figure}

We have chosen the mass function as
\begin{equation}  
        F(r)=F_{3}r^3+F_{5}r^5.
\end{equation}
We have found that the central singularity has been globally naked for very 
small $r_{b}$ if we have fixed the value of $F_{3}$ and $F_{5}$. 
Although we have calculated several models,
we only display the numerical results for the model with
$F_{3}=1$, $F_{5}=-2$ and $r_{b}=0.02$ in an arbitrary unit
because the features have been the same if there has been globally
naked singularity in the model.
The total gravitational mass $M$ is given by $M=3.9968\times 10^{-6}$
for this model. 
See Fig.~\ref{fg:rays_ltb} for trajectories of null geodesics.
We also indicate the location of the singularity, apparent horizon
and Cauchy horizon in this figure.

\begin{figure}[tbp]
\begin{center}
\leavevmode
\begin{tabular}{ll}
        \subfigure[$G$]{\epsfxsize=66mm \epsfbox{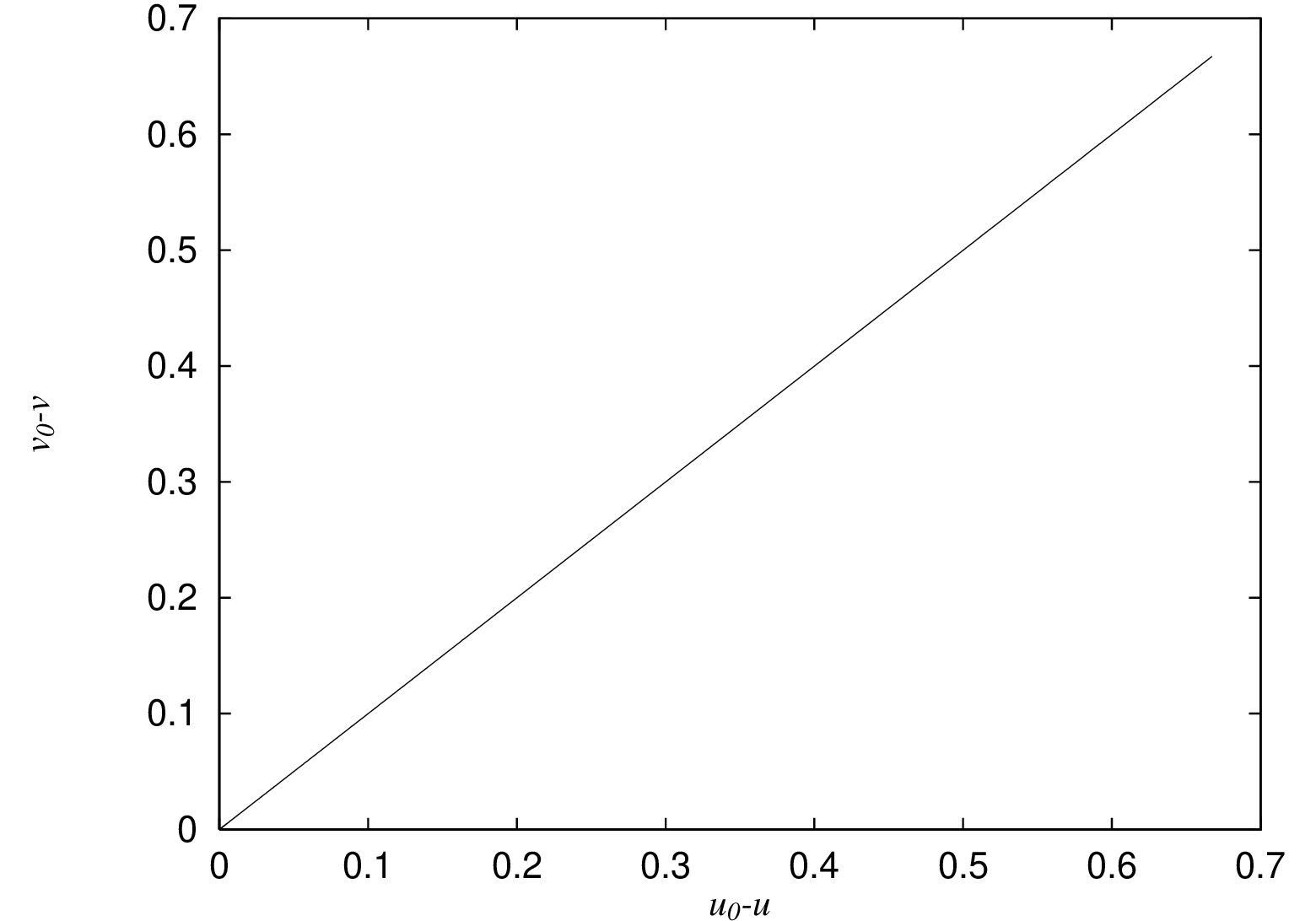}}
        \subfigure[$G'$]{\epsfxsize=66mm \epsfbox{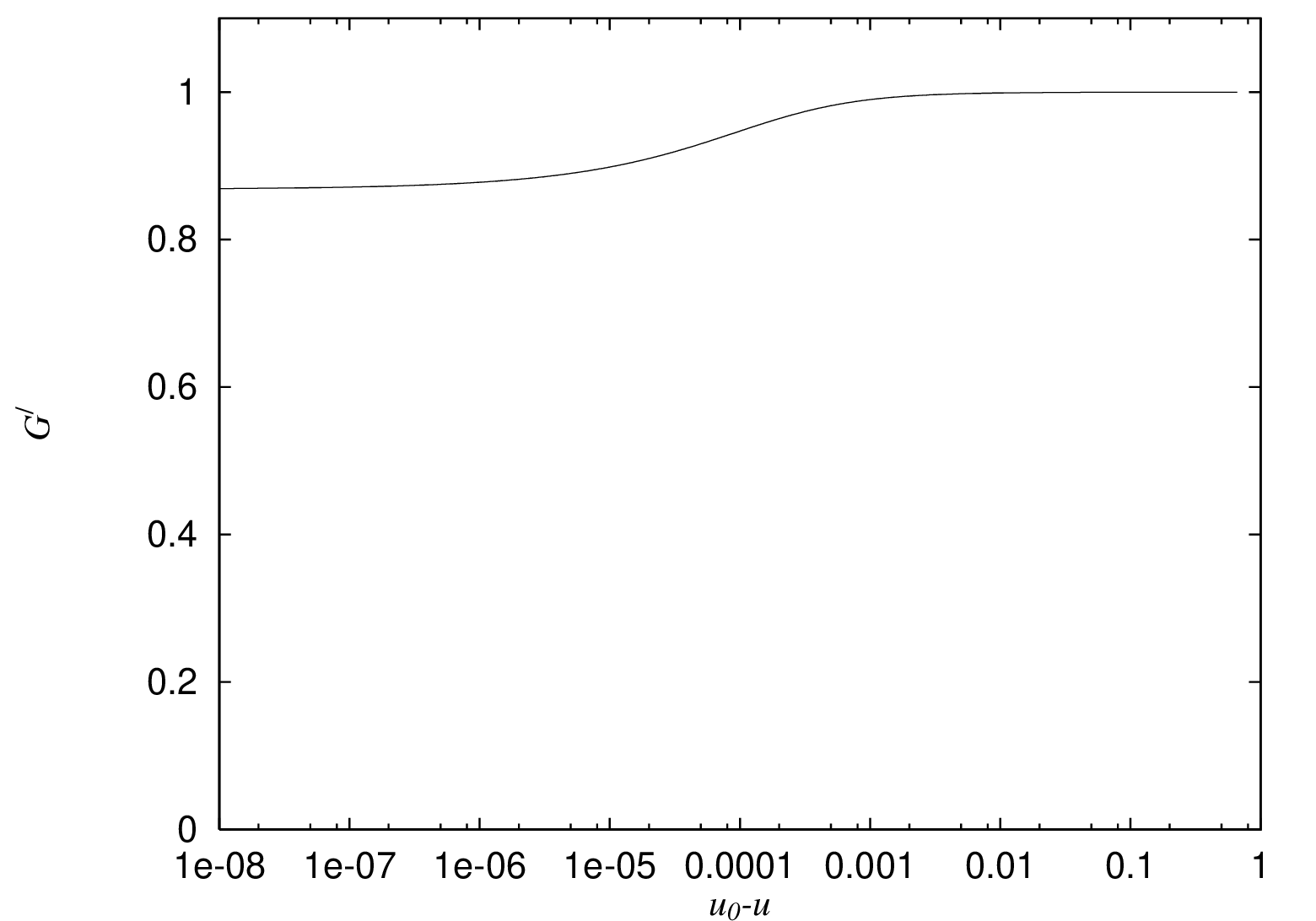}}\\
\end{tabular}
        \subfigure[$G''$]{\epsfxsize=10cm \epsfbox{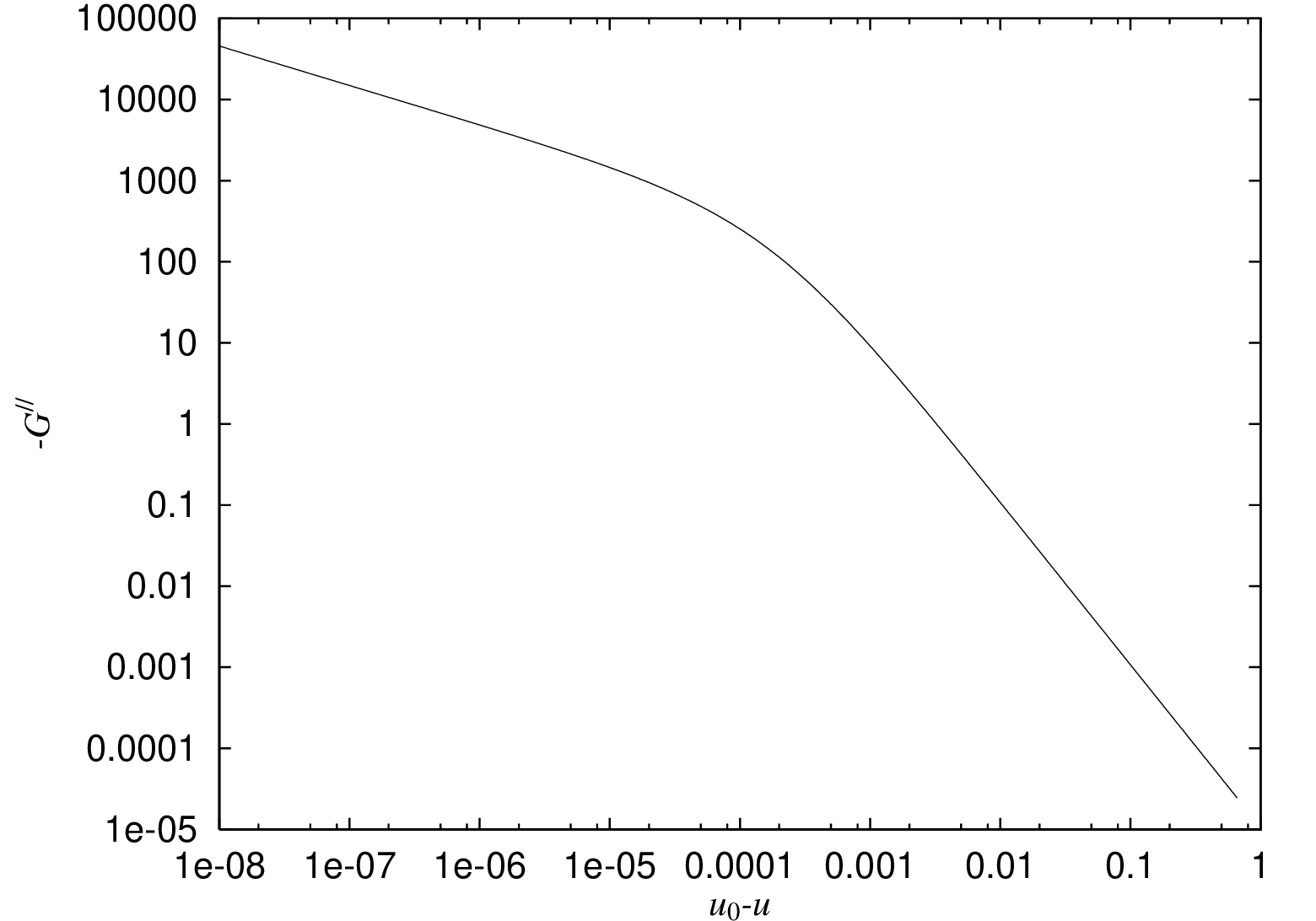}}
\end{center}
	\caption{Plots of (a) $G(u)$, (b) $G'(u)$ and (c) $G''(u)$ 
for the analytic LTB spacetime. 
Parameters are the same as those in 
Fig.~\ref{fg:rays_ltb}}
	\label{fg:g_ltb}
\end{figure}

See Fig.~\ref{fg:g_ltb}. In Fig.~\ref{fg:g_ltb}(a), 
the relation between $u$ and $v$, i.e., the function $G(u)$
is shown, where 
$u_{0}$ ($v_{0}$) is defined as the retarded (advanced) 
time of the earliest 
light ray which originates from (terminates at) the singularity.
In Fig.~\ref{fg:g_ltb}(b), the first derivative $G^{\prime}(u)$ is shown.
This implies that $G^{\prime}(u)$ does not diverge
but converges to some positive constant $A$ with $0<A<1$.
In Fig.~\ref{fg:g_ltb}(c), it is found that
the second derivative $G^{\prime\prime}(u)$ does diverge 
as $u\to u_{0}$.
This figure shows that the behaviors of the 
growth of $G^{\prime\prime}(u)$ 
are different from each other during
early times ($10^{-4}\lesssim u_{0}-u $) and during late times 
($0 < u_{0}-u \lesssim 10^{-4}$).
During late times, the dependence is written as
\begin{equation}
        G^{\prime\prime}\propto -(u_{0}-u)^{-1/2}.
        \label{G2late}
\end{equation}
It is noted that this singular behavior was re-derived 
analytically under several assumptions.~\cite{ts2001}

Here we determine the amplitude of the power by physical 
considerations.
We assume that the particle creation during late times
is due to the growth of the central curvature.
Then, we consider a constant of proportion which will appear
in Eq.~(\ref{G2late}).
We should note that the coefficient must be
written by initial data because it must not depend on time.
Equation~(\ref{G2late})
demands a dimensionful constant of proportion.
For late times, 
the only possible quantity
is $(t_{0}^7 /l_{0}^{6})$, where $t_{0}$ the free-fall time given by
\begin{equation}
t_{0}=\frac{2}{3}F_{3}^{-1/2}
\end{equation}
and $l_{0}$ denotes the
scale of inhomogeneity defined as
\begin{equation}
l_{0}\equiv\left(\frac{-F_{5}}{F_{3}}\right)^{-1/2}.
\end{equation}
This is because only $(t_{0}^{7}/l_{0}^{6})$ does not 
depend on the origin of time coordinate and can be formed
by local quantities at the center.
See Appendix~\ref{sec:conserved_quantity} for the derivation 
and the physical meaning of this constant.
Thus we can determine the coefficient up a numerical factor as
\begin{eqnarray}
G'&\approx& A, \\
G^{\prime\prime}&\approx& -fA\left(\frac{t_{0}^{7}}{l_{0}^{6}}
\right)^{-1/2}(u_{0}-u)^{-1/2},
\end{eqnarray}
for late times,
where $0<A<1$ and $f$ is a dimensionless 
positive constant of order unity.
It implies that there is characteristic frequency of singularity
which is defined as
\begin{equation}
        \omega_{s}\equiv\frac{l_{0}^{6}}{t_{0}^7}.
        \label{omegas}
\end{equation}
The late-time behavior is good for $0<u_{0}-u\ll \omega_{s}^{-1}$.
The above estimate shows a good agreement with numerical results. 
        
\subsubsection{Power and energy}
\begin{figure}[tbp]
        \centerline{\epsfxsize 10cm \epsfbox{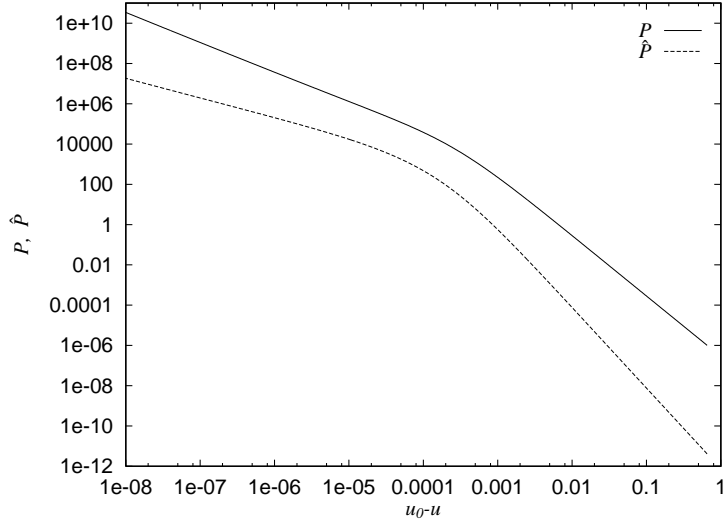}}
        \caption{Power for minimally and conformally
        coupled scalar fields in the analytic LTB spacetime.
Parameters are the same as those in 
Fig.~\ref{fg:rays_ltb}}

        \label{fg:p_ltb}
\end{figure}

Since $G^{\prime\prime}$ and therefore $G^{\prime\prime\prime}$ diverge 
as $u\to u_{0}$, the power of radiation diverges for 
both minimally and conformally coupled scalar fields.
For this model, we have numerically calculated 
the power of radiation by 
Eqs.~(\ref{llm}) and (\ref{hatllm}).
The results are displayed in Fig.~\ref{fg:p_ltb}.

Based on the above physical considerations, we can obtain
the formula for the power by the particle creation
using Eqs.~(\ref{llm}) and (\ref{hatllm}).
As seen in Fig.~\ref{fg:g_ltb}(b), we find  
\begin{equation}
        G^{\prime}\approx A,
\end{equation}
during late times.
The power during late times is obtained as
\begin{eqnarray}
P&\approx& 
\frac{1}{48\pi}f
\omega_{s}^{1/2}(u_{0}-u)^{-3/2}, \\
\hat{P}&\approx&
\frac{1}{48\pi}f^2 
\omega_{s}(u_{0}-u)^{-1}.
\label{hatl}
\end{eqnarray}
Therefore the power diverges to positive infinity 
for both minimally and 
conformally coupled scalar fields as $u\to u_{0}$.
The radiated energy is estimated by integration.
During late times, the totally radiated energy is estimated as
\begin{eqnarray}
E&\approx&\frac{1}{24\pi}f\omega_{s}^{1/2}
(u_{0}-u)^{-1/2}, \\
\hat{E}&\approx&\frac{1}{48\pi}f^2
\omega_{s}\ln\frac{1}{\omega_{s}(u_{0}-u)}.
\end{eqnarray}
Therefore, the totally radiated energy diverges to positive infinity
for both minimally and conformally 
coupled scalar fields as $u\to u_{0}$.
However, in realistic situations,
we may assume that the naked singularity formation 
is prevented by some mechanism and that the quantum particle creation
is ceased at the time $u_{0}-u\approx \Delta t$.
In other words, $G^{\prime\prime}/G^{\prime}$ 
tends to vanish for $u_{0}-u\lesssim 
\Delta t$.
In such situations, the total derivative term in the 
expression of the power
of a minimally coupled scalar field gives no contribution to the
totally radiated energy.
Therefore, the total energy for a minimally coupled 
scalar field and that for a conformally coupled one 
coincide exactly, i.e.,
\begin{equation}
E=\hat{E}\approx \frac{1}{48\pi}f^{2}
\omega_{s}\ln \frac{1}{\omega_{s}\Delta t}.
\end{equation}

\subsubsection{Spectrum}
From the above numerical results and physical 
considerations, we obtain the 
following formula for the function $G(u)$ or $F(v)$: 
\begin{equation}
G(u)\approx A (u-u_{0}) -\frac{4}{3}
A f\omega_{s}^{1/2}
(u_{0}-u)^{3/2}+v_{0}
\label{GlateLTB}
\end{equation}
or
\begin{equation}
F(v)\approx A^{-1}(v-v_{0})+\frac{4}{3}f
\omega_{s}^{1/2}[A^{-1}(v_{0}-v)]^{3/2}+u_{0}
\label{FlateLTB}
\end{equation}
for late times.

Now that we have obtained the function $F(v)$, we can calculate the 
spectrum of radiation by Eqs.~(\ref{alpha}),
(\ref{beta}) and (\ref{nomega}).
See Harada, Iguchi and Nakao~\cite{hin2000b} for details.
In order to determine the spectrum numerically,
we introduce the Gaussian window function in the Fourier 
transformation as usual.

Although we could use the calculated data for the function $F(v)$,
we have used the analytic formula which have been derived based on
the numerical results for 
convenience of retaining accuracy.
Since the diverging power
is associated with the late-time behavior, 
we concentrate on the late-time radiation.
In order to obtain the totally integrated spectrum, we extrapolate the 
function $F$ linearly as
\begin{equation}
F(v)=\left\{
\begin{array}{ll}
A^{-1}(v-v_{0})+u_{0} & \qquad (v_{0}<v) \\
A^{-1}(v-v_{0})+\displaystyle{\frac{4}{3}}
f\omega_{s}^{1/2}[A^{-1}(v_{0}-v)]^{3/2}
+u_{0} &\qquad (v_{1}<v\le v_{0})
\end{array}\right. ,
\label{extrapolate_linear}
\end{equation}
where 
$v_{1}$ is given as 
\begin{equation}
v_{0}-v_{1}\equiv\left[\frac{1}{2}(1-A)f^{-1}\right]
^{2}A\omega_{s}^{-1}.
\end{equation}
We should note that Eq.~(\ref{extrapolate_linear}) can be used
for $u_{0}-u\lesssim  \omega_{s}^{-1}$.
In fact, it turns out that we only have to pay attention to
the spectrum above $\omega_{s}$.
We should note that there is no radiation for $u>u_{0}$ in this 
extrapolation.
Because the second derivative of $F$ is divergent at $v=v_{0}$, 
we can only require that the first derivative of $F$ should be
continuous at $v=v_{0}$.
If we allow discontinuity of the first derivative,
radiation due to this discontinuity dominates the 
radiated energy flux, which is out of our concern.

\begin{figure}[tbp]
\begin{center}
\begin{tabular}{cc}
 \subfigure[Totally integrated spectrum]{\epsfxsize 66mm \epsfysize 66mm 
	\epsfbox{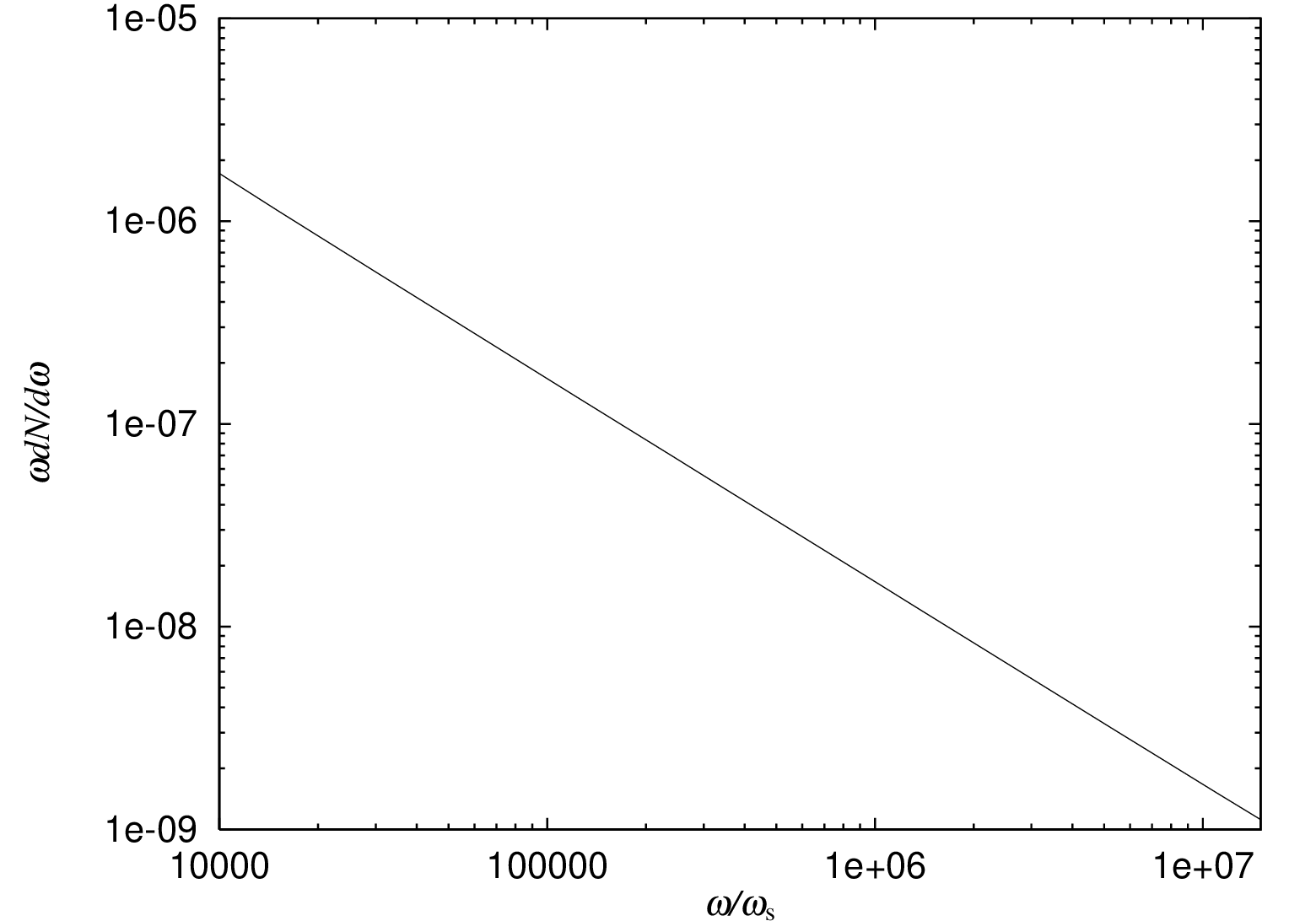}}
 \subfigure[Momentary spectrum]
	{\epsfxsize 66mm \epsfysize 66mm \epsfbox{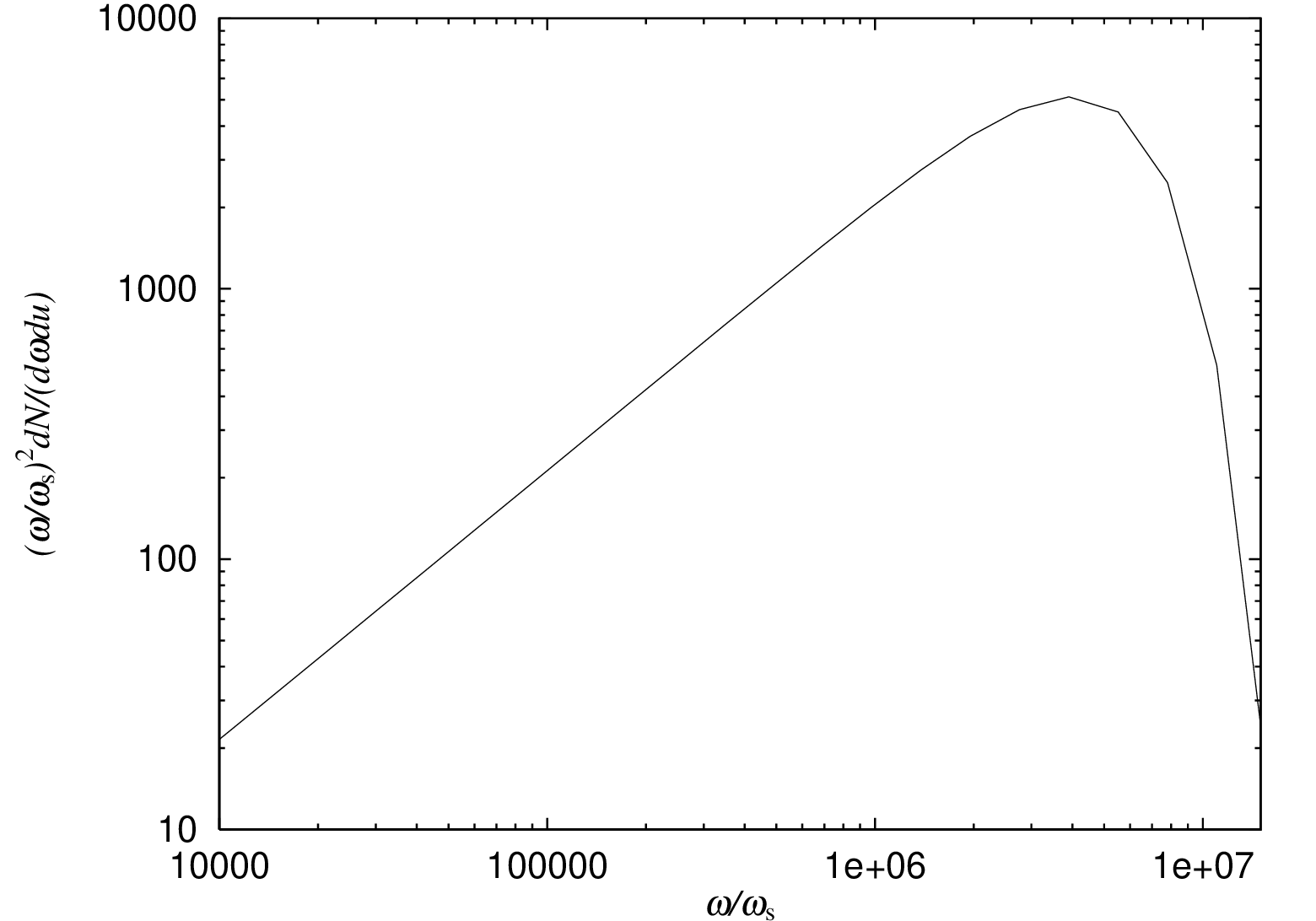}}
\end{tabular}
 \caption{(a) Totally integrated and (b) 
 momentary spectra for $A=0.8$ and $f=1$. 
	The latter is for 
	the moment $u_{0}-u=10^{-6}\omega_{s}^{-1}$.}
\label{fg:spec}
\end{center}
\end{figure}

The obtained spectrum is shown in Fig.~\ref{fg:spec}(a).
The parameters are fixed as $A=0.8$ and $f=1$.
We should note that the contribution to the total 
radiated energy mainly comes from $\omega\gtrsim \omega_{s}$. 
In this figure, we find 
\begin{equation}
\omega\frac{dN}{d\omega}\propto \omega^{-1}.
\end{equation}
Therefore, the total energy, which will be obtained by
\begin{equation}
E=\int^{\infty}_{0}d\omega \omega \frac{dN}{d\omega},
\end{equation}
is logarithmically divergent, which is 
consistent with the divergent radiated energy obtained
based on
the point-splitting regularization.

It is important in estimating the validity of the geometrical optics
approximation to examine which frequency band dominates the power
at each moment.
For this purpose, we determine the momentary spectrum
by a wavelet analysis.
We use the Gabor wavelet
in place of $e^{i x}$
because it has a very clear physical meaning 
as the Gaussian wave packet.
The Gabor wavelet is written as
\begin{equation}
\psi(x)=\frac{1}{\sqrt{4\pi}\sigma}e^{ix}e^{-x^{2}/\sigma^2},
\end{equation} 
where we have set $\sigma=8$.
The result below is not so sensitive to the value of $\sigma$.

We have calculated numerically the momentary spectrum.
The result is shown in Fig.~\ref{fg:spec}(b).
This figure displays the contribution of each logarithmic bin  
of frequency to the power at the moment.
The parameters are fixed as $A=0.8$ and $f=1$.
We have chosen the time of observation $u_{0}-u=10^{-6}\omega_{s}^{-1}$.
In this figure, it is found that
the contribution to the power
is dominated by the frequency $\omega\sim 2\pi (u_{0}-u)^{-1}$. 

\subsubsection{Null geodesics and redshifts}
Here we analytically examine null geodesics and redshifts 
in the LTB spacetime
and derive some useful results to comprehend the problem.
We assume the function $F(r)$ of the form Eq.~(\ref{eq:Fodd})
with $F_{3}>0$ and $F_{5}<0$.
We define $\eta$ as
\begin{equation}
\eta\equiv t-t_{0}.
\label{eta}
\end{equation}
Moreover, $\eta_{s}(r)\equiv t_{s}(r)-t_{0}$ and 
$\eta_{ah}(r)\equiv t_{ah}(r)-t_{0}$ are 
approximated for $r/l_{0}\ll 1$ as
\begin{eqnarray}
\frac{\eta_{s}(r)}{t_{0}}&\approx &
\frac{1}{2}\left(\frac{r}{l_{0}}\right)^2,
\label{etas}\\
\frac{\eta_{ah}(r)}{t_{0}}&\approx& 
\frac{1}{2}\left(\frac{r}{l_{0}}\right)^2
-\left(\frac{2}{3}\right)^3 
\left(\frac{l_{0}}{t_{0}}\right)^3
\left(\frac{r}{l_{0}}\right)^{3}.
\label{etaah}
\end{eqnarray}

For a while, we assume
\begin{equation}
   \frac{r_{b}}{l_{0}}\ll 1.
\end{equation}
This implies that we can safely expand in powers of $r/l_{0}$
and take the leading order term in the whole of the cloud. 

Then we examine trajectories of null geodesics in 
this spacetime.
First we consider the region where 
\begin{equation}
\left(\frac{r}{l_{0}}\right)^2\ll \left|\frac{\eta}{t_{0}}\right|
\label{A}
\end{equation}
is satisfied.
This region is approximately recognized as the Friedmann universe.
From Eq.~(\ref{etas}), it is found that this condition can 
be satisfied only for $\eta<0$.
In this case, $R_{,r}$ is approximated as
\begin{equation}
R_{,r}\approx\left(\frac{-\eta}{t_{0}}
\right)^{2/3}.
\end{equation}
The ordinary differential equation (\ref{dtdr})
can be easily integrated to 
\begin{equation}
\left(\frac{-\eta}{t_{0}}\right)^{1/3}
\approx \mp \frac{1}{3}\left(\frac{l_{0}}
{t_{0}}\right)\left(\frac{r}{l_{0}}
\right)+C_{A\pm},
\label{nullA}
\end{equation}
where $C_{A\pm}$ is a constant of integration.
It is easily found that each null ray can be drawn
by a parallel transport of the $C_{A\pm}=0$ 
curve along the $r$-axis direction.

Next we consider the region where
\begin{equation}
\left(\frac{r}{l_{0}}\right)^2\gg \left|\frac{\eta}{t_{0}}\right|
\label{B}
\end{equation}
is satisfied.
This region is not at all approximated by the Friedmann model.
In this case, $R_{,r}$ is approximated as
\begin{equation}
R_{,r}\approx \frac{7}{2^{2/3}3}
\left(\frac{r}{l_{0}}\right)^{4/3}.
\end{equation}
Then, Eq.~(\ref{dtdr}) is integrated to 
\begin{equation}
\frac{\eta}{t_{0}} \approx \pm 2^{-2/3}
\left(\frac{l_{0}}{t_{0}}\right)
\left(\frac{r}{l_{0}}
\right)^{7/3}-C_{B\pm},
\label{nullB}
\end{equation}
where $C_{B\pm}$ is a constant of integration.
It is easily found that each null ray can be drawn
by a parallel transport of the $C_{B\pm}=0$ curve along 
the $\eta$-axis direction.

\begin{figure}[tbp]            
 \centerline{\epsfxsize 10cm \epsfbox{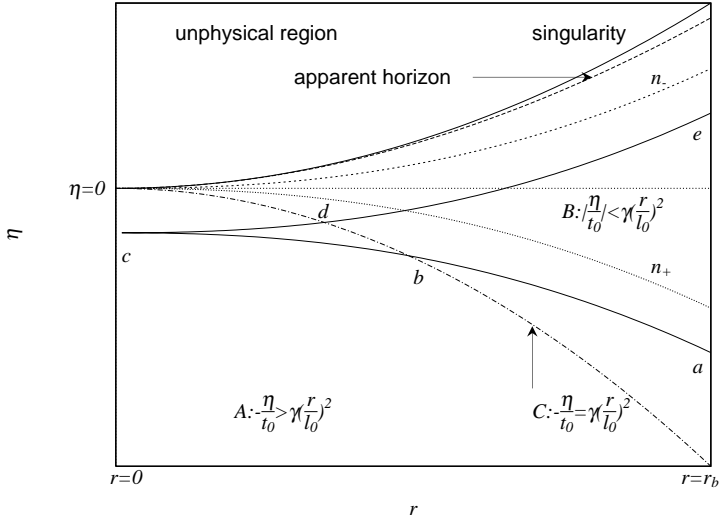}}
 \caption{Schematic figure of null geodesics 
around naked singularity in the analytic 
LTB spacetime}
 \label{fg:sketch}
\end{figure}

See Fig.~\ref{fg:sketch}, which illustrates the spacetime 
around $(\eta,r)=(0,0)$.
Condition (\ref{A}) is satisfied in region $A$, while
condition (\ref{B}) is satisfied in region $B$.
The boundary of regions $A$ and $B$ will be described by
\begin{equation}
\frac{-\eta}{t_{0}}= \gamma \left(\frac{r}{l_{0}}\right)^2,
\label{C}
\end{equation}
where $\gamma$ is a constant of order unity.
We denote this boundary curve as $C$.
It may be kept on mind that this treatment 
is rather simple.
However we believe that the present approximation
that we divide the spacetime to two regions
will be enough to comprehend an essence of the problem.

Then, we concentrate on the behavior of null geodesics
around the naked singularity.
See Fig.~\ref{fg:sketch} for the trajectory of null geodesics.
From the regular center $\eta<0$ and $r=0$, 
the outgoing null ray runs region $A$
which is described by the upper sign of Eq.~(\ref{nullA})
with $C_{A+}>0$.
Then this null ray goes through boundary $C$.
After that the null ray goes into region $B$
and the trajectory is described by the upper sign of 
Eq.~(\ref{nullB}) with  $C_{B+}>0$.
Since all null rays with $C_{B+}>0$ in region $B$
emanate from the regular center, the outgoing null geodesic
with $C_{B+}=0$ in region $B$ is the Cauchy horizon, i.e.,
the earliest outgoing null ray from the singularity at $r=0$.
We denote this outgoing null geodesic as $n_{-}$.
There are infinitely many outgoing 
null geodesics later than $n_{-}$ which emanates from the
singularity at $r=0$.
In fact, these null geodesics form one parameter family.
These null rays cannot asymptote to $n_{-}$ as $r\to 0$ because 
these null rays are obtained by a parallel transport 
of $n_{-}$ in $\eta$-axis direction in region $B$.
Instead, these null rays asymptote to the 
location of the apparent horizon in 
an approach to the singularity. 

It is clear that the ingoing null ray with $C_{B-}=0$ 
terminates at the singularity $r=0$.
We denote this ingoing null ray as $n_{+}$.
For small positive $C_{B-}$, 
the ingoing null ray in region $B$ crosses boundary $C$
and the null ray in region $A$ becomes described by the
lower sign of Eq.~(\ref{nullA}) with $C_{A-}>0$,
and then terminates at the regular center.
For small negative $C_{B-}$, 
the ingoing null ray in region $B$ 
terminates at the space-like singularity $r>0$.

Using the above results, we can derive the condition
for globally naked singularity.
If $n_{-}$ reach ${\cal I^{+}}$, then the singularity is globally naked,
otherwise the singularity is locally naked.
Noting that the intersection of the apparent horizon 
with the cloud surface 
is on the event horizon, we find that the 
condition for the singularity to 
be globally naked is given as
\begin{equation}
\left[\left(\frac{l_{0}}{t_{0}}
\right)\left(\frac{r_{b}}{l_{0}}\right)^{1/3}\right]^{-1}
-\frac{16}{27}\left[\left(\frac{l_{0}}{t_{0}}\right)
\left(\frac{r_{b}}{l_{0}}\right)^{1/3}
\right]^{2}\gtrsim 2^{1/3}.
\end{equation}
Since $x^{-1}-(16/27)x^{2}$ is a decreasing 
function of $x>0$, we can find the condition
for being globally naked as
\begin{equation}
\frac{r_{b}}{l_{0}}\lesssim 0.66\left(\frac{t_{0}}{l_{0}}\right)^3.
\end{equation}
This implies that the singularity is globally naked for sufficiently
small $r_{b}$ when $t_{0}$ and $l_{0}$ are fixed.
Thus, it turns out that 
the present assumption $r_{b}/ l_{0}\ll 1$ is relevant 
for the globally naked singularity.
We can translate the above condition to the condition 
for $M$ and $\omega_{s}$ as
\begin{equation}
M\lesssim 6.4\times 10^{-2} \omega_{s}^{-1}.
\label{globallynaked}
\end{equation}
It implies that the mass of the dust cloud with globally 
naked singularity is bounded from above.

Since it is necessary for later discussions, 
we turn our attention to estimate of redshifts.
Let $k^{\mu}$ be the tangent vector of a radial null geodesic.
The frequency $\hat{\omega}$ observed by an observer comoving 
with a fluid element is
calculated from $k^{\mu}$ as
\begin{equation}
\hat{\omega}=-k^{\mu}u_{\mu}=-k^{t}.
\end{equation}

In region $A$, null geodesic equations 
can be integrated as
\begin{equation}
k^{t}(-\eta)^{2/3}\approx \mbox{const}, 
\end{equation}
in the lowest order.
Therefore a particle is blueshifted for the comoving observer
in region $A$.
In region $B$, null geodesic equations
can be integrated as
\begin{equation}
k^{t}\approx\mbox{const}, 
\end{equation}
in the lowest order.
Therefore a particle is neither redshifted nor blueshifted
in region $B$.

In the external Schwarzschild spacetime,
in a similar way, we obtain the tangent vector of
a radial null geodesic as
\begin{equation}
k^{T}\left(1-\frac{2M}{R}\right)=\mbox{const}, \\
\end{equation}
A static distant observer observes the frequency
\begin{equation}
\omega=-k^{T}(T,\infty).
\end{equation}
Using the matching condition~(\ref{matching}),
the observed frequency $\hat{\omega}=-k^{\mu}u_{\mu}$ by 
the comoving observer
at the surface is written using the above $\omega$ as
\begin{equation}
\hat{\omega}=\frac{\omega}{1\mp\alpha},
\label{redshiftfactor}
\end{equation}
where $\alpha$ is defined by
\begin{equation}
\alpha\equiv\sqrt{\frac{2M}{R}},
\end{equation}
and $R$ is the circumferential radius of the dust surface 
when the null ray crosses the surface.

Let $\alpha_{\pm}$ be the value of $\alpha$ for $n_{\pm}$.
Let $\eta_{\pm}$ be the value of $\eta$ 
when $n_{\pm}$ crosses the dust surface.
Then, from Eq.~(\ref{matching}), 
it is easy to see that the Taylor expansion is valid
around $\eta=\eta_{+}$ for $v$ as a function of $\eta$ as
\begin{equation}
v=v_{0}+\frac{1}
{1+\alpha_{+}}(\eta-\eta_{+})+O((\eta-\eta_{+})^2).
\label{taylorv}
\end{equation}
For $\alpha_{-}<1$, from Eq.~(\ref{matching}), it is also 
easy to see that the Taylor expansion is valid
around $\eta=\eta_{-}$ for $u$ as a function of $\eta$ as
\begin{equation}
u=u_{0}+\frac{1}
{1-\alpha_{-}}(\eta-\eta_{-})+O((\eta-\eta_{-})^2),
\label{tayloru}
\end{equation}
or equivalently,
\begin{equation}
\eta_{-}-\eta=(1-\alpha_{-})
(u_{0}-u)+O((u_{0}-u)^2).
\label{tayloreta}
\end{equation}

See Fig.~\ref{fg:sketch} again.
The light ray which is close to $n_{\pm}$ enters the dust 
cloud in region $B$ ($a$),
enters region $A$ ($b$), 
reaches the center ($c$),
enters again region $B$ ($d$) and
leaves the dust cloud ($e$). 
In order to determine $G$ or $F$, we 
need to relate the time coordinate
$\eta_{a}$ at which the incoming 
light ray enters the dust cloud with
$\eta_{e}$ at which the outgoing 
light ray leaves the dust cloud.
Within the present approximation, we find
\begin{equation}
\eta_{-}-\eta_{e}\approx \eta_{+}-\eta_{a},
\end{equation}
which can be derived using the solutions of 
null geodesic equations obtained above.

Then, we find for $\alpha_{-}<1$
\begin{equation}
v=G(u)\approx v_{0}-\frac{1-\alpha_{-}}{1+\alpha_{+}}
(u_{0}-u),
\end{equation}
or
\begin{equation}
u=F(v)\approx u_{0}-\frac{1+\alpha_{+}}{1-\alpha_{-}}(v_{0}-v).
\end{equation}
From the above discussions, we can determine 
$A=\lim_{u\to u_{0}}G^{\prime}$ as
\begin{equation}
A=\frac{1-\alpha_{-}}{1+\alpha_{+}}.
\end{equation}

\subsubsection{Validity of geometrical optics approximation}

The geometrical optics approximation is exact for 
a two-dimensional spacetime,
where the line element is given by setting $d\Omega^{2}=0$
in the four-dimensional line element (\ref{eq:lineelement}).
Therefore, the obtained power, energy and spectrum
are exact in this sense for the two-dimensional version of 
the LTB spacetime.

In a four-dimensional case, the geometrical 
optics approximation is only an
approximation because of the existence of curvature potential.
The geometrical optics approximation is valid for waves 
with the wave length
shorter than the curvature radius of the spacetime geometry.
In other words, the geometrical optics approximation is good if 
condition \begin{equation}
\hat{\omega}\gtrsim 2\pi 
|R_{\hat{\alpha}\hat{\beta}\hat{\gamma}\hat{\delta}}|^{1/2}
\end{equation}
is satisfied,
where the hat denotes the components 
in the local inertial tetrad frame.

In region $A$, 
nonvanishing components of the Riemann tensor
are approximated as
\begin{equation}
|R_{\hat{t}\hat{r}\hat{t}\hat{r}}
|\approx|R_{\hat{t}\hat{\theta}\hat{t}\hat{\theta}}|
\approx|R_{\hat{r}\hat{\theta}\hat{r}\hat{\theta}}
|\approx|R_{\hat{\theta}\hat{\phi}\hat{\theta}\hat{\phi}}|
\approx(-\eta)^{-2}.
\end{equation}
In region $B$, they are approximated as
\begin{equation}
|R_{\hat{t}\hat{r}\hat{t}\hat{r}}|\approx
|R_{\hat{t}\hat{\theta}\hat{t}\hat{\theta}}|
\approx|R_{\hat{r}\hat{\theta}\hat{r}\hat{\theta}}|
\approx|R_{\hat{\theta}\hat{\phi}\hat{\theta}\hat{\phi}}|\approx
\frac{1}{t_{0}^2}\left(\frac{r}{l_{0}}\right)^{-4}.
\end{equation}
We have already seen that, 
in region $B$, $\hat{\omega}$ is kept constant approximately.
In region $A$, 
$\hat{\omega}$ may be considerably blueshifted. 
However, we can see
that the null ray which goes into the dust cloud
enters region $A$ and get out of region $A$
at the same time in the lowest order, i.e,
$\eta_{b}\approx \eta_{c}\approx \eta_{d}$.
This implies that the frequency of such a particle is kept
almost constant all over the dust cloud.
The Riemann tensor along the 
light ray reaches the maximum 
when the light ray goes through region $A$. 
Now we can write down the condition for the geometric
optics approximation to be valid as
\begin{equation}
\hat{\omega}\gtrsim 2\pi(-\eta_{c})^{-1}.
\end{equation}
This condition can be rewritten by the quantities on ${\cal I^{+}}$
using Eq.~(\ref{redshiftfactor}) and (\ref{tayloru}).
The result is
\begin{equation}
\omega\gtrsim \omega_{cr}(u)\equiv 2\pi (u_{0}-u)^{-1}
\end{equation}
for $\alpha_{-}<1$.

As we have seen,
the power at each moment is mainly carried
by particles of which frequency is about $2\pi (u_{0}-u)^{-1}$.
Since the transmission coefficient $\Gamma_{\omega}$
will be a function of $(\omega/\omega_{cr})$,
it is natural to estimate $\Gamma_{\omega}(\omega\approx\omega_{cr})$ 
as of order unity.
It implies that the calculations based on
the geometrical optics approximation will be 
valid for a rough order estimate.

\subsection{Particle creation in self-similar dust collapse}
\subsubsection{Null coordinates}
As we have seen in \S \ref{sec:dust},
the self-similar dust collapse corresponds to 
marginally bound collapse with 
$F(r)=\lambda r$, where $\lambda$ is a constant.
The self-similar collapse solution is then
\begin{equation}
 \label{R}
 R=\left(\frac{9}{4}\lambda\right)^{1/3}r (1-z)^{2/3},
\end{equation}
and the partial derivative $R$ with respect to $r$ is
\begin{equation}
 R_{,r}=\left(\frac{9}{4}\lambda\right)^{1/3} \frac{1-z/3}{(1-z)^{1/3}},
\end{equation}
where $z\equiv t/r$.

In the following, we derive the mapping function
for the self-similar dust collapse
acceding to Barve et al.~\cite{bsvw1998b}
For self-similar dust collapse, 
we can obtain exact null coordinates.
For this purpose, we restrict our attention to
the two-dimensional part of metric tensor. 
In the interior region, we introduce null coordinates $(\eta, \zeta)$
\begin{equation}
 \eta=\left\{\begin{array}{ll} r e^{\int dz/(z-R_{,r})} & (z-R_{,r} > 0)\\
                         - r e^{\int dz/(z-R_{,r})} & (z-R_{,r} < 0)
           \end{array}\right. 
\label{eq:eta}
\end{equation}
and
\begin{equation}
 \zeta=\left\{\begin{array}{ll} r e^{\int dz/(z+R_{,r})} & (z+R_{,r} > 0)\\
                         - r e^{\int dz/(z+R_{,r})} & (z+R_{,r} < 0).
           \end{array}\right. 
\label{eq:zeta}
\end{equation}
We introduce another null coordinates $(U,V)$ as
\begin{eqnarray}
 U &=&\left\{\begin{array}{ll}
       \ln \eta &  (z-R_{,r} > 0)\\
        -\ln |\eta| & (z-R_{,r} < 0)\\
       \end{array}\right. \\
 V &=& \left\{\begin{array}{ll}
           \ln \zeta &  (z+R_{,r} > 0)\\
           - \ln |\zeta| & (z+R_{,r} < 0).
           \end{array}\right. 
\end{eqnarray}
Hereafter we concentrate our attention to
the outside of the Cauchy horizon where $z-R_{,r}<0$. 
Then the two-dimensional metric is expressed as
\begin{equation}
\label{UVmetric}
 ds^2 = - A^2(U,V) dU dV
\end{equation}
where 
\begin{equation}
 A^2(U,V) = \left\{\begin{array}{ll}
             -r^2 ( z^2 -R_{,r}^2) &  (z+R_{,r} > 0)\\
              r^2 ( z^2 -R_{,r}^2) &  (z+R_{,r} < 0).\\
           \end{array}\right.       
\end{equation}
If we define
\begin{equation}
 f_{\pm}(y) \equiv y^4 \mp \frac{a}{3} y^3 - y \mp \frac{2}{3}a 
\end{equation}
where $y\equiv (1-z)^{1/3}$ and  
$a\equiv\left(\frac{9}{4}\lambda\right)^{1/3}$,
we can write
\begin{equation}
I_{\pm}\equiv\int\frac{dz}{z\pm R_{,r}}=\int\frac{3y^{3}dy}{f_{\pm}(y)}.
\label{eq:Ipm}
\end{equation}

In the exterior region, the metric can be expressed as 
\begin{equation}
 ds^2=-\left(1-\frac{2M}{R}\right)dudv
\end{equation}
using the Eddington-Finkelstein double null coordinates.
The matching surface is given by $r=r_{b}$. Hence, 
we have $2M=\lambda r_{b}$.
As in the analytic case,
the matching condition is given by Eq.~(\ref{matching}).

\subsubsection{Mapping}
The mapping $v=G(u)$ from ${\cal I^{+}}$ to ${\cal I^{-}}$ can be determined 
as follows.
First we consider the limit to the regular center $t<0$ and $r= 0$.
Since $y\to \infty$ in this limit, we have
\begin{eqnarray}
\eta &\to & -r y^{3} \to t, \\
\zeta &\to & -r y^{3} \to t.
\end{eqnarray}
Therefore, $\eta =\zeta $ holds at the regular center.
The singularity at $t=0$ and $r=0$ is mapped into $\eta=\zeta=0$.

Next we consider the earliest outgoing null geodesic $n_{-}$
which emanates from the center at $t=0$.
Since $\eta$ is outgoing null coordinate,
$\eta=0$ along $n_{-}$.
Since it is only possible when $I_{-}= -\infty$ along the 
null geodesic, we find that $y$ along $n_{-}$
satisfies an algebraic equation $f_{-}(y)=0$.
In the domain $0<y<1$, there exists two real roots for 
$\lambda < 6(26-15\sqrt{3})$, one degenerate root for 
$\lambda = 6(26-15\sqrt{3})$, and no real roots otherwise.
$\lambda \le 6(26-15\sqrt{3})$ is 
the condition for the naked singularity to exist
in this model.~\cite{jd1993}
It is clear that the largest positive root $y=y_{-}\in (0,1)$ 
corresponds to
$n_{-}$, i.e., the Cauchy horizon.
Hereafter we restrict our attention to the nondegenerate case.
For the first ingoing null ray $n_{+}$ 
which terminates at the singular center at $t=0$,
we encounter a similar situation.
In the domain $y>1$,
an algebraic equation $f_{+}(y)=0$ has always 
the only one real root $y=y_{+}$, which corresponds to $n_{+}$.

Then we consider the matching condition for null rays.
The matching condition at the surface is given in terms 
of $u$ and $v$ as
\begin{eqnarray}
u&=&u_{b}(y)=-\frac{r_{b}}{a}y^{3}-\frac{4}{3}ar_{b}y-r_{b}y^{2}
-\frac{8}{9}a^{2}r_{b}\ln\left|\frac{3y}{2a}-1\right|, \\
v&=&v_{b}(y)=-\frac{r_{b}}{a}y^{3}-\frac{4}{3}ar_{b}y+r_{b}y^{2}
+\frac{8}{9}a^{2}r_{b}\ln\left|\frac{3y}{2a}-1\right|.
\end{eqnarray}
It is noted that these functions are regular around $y=y_{\pm}$.

We shall consider a family of null rays which emanate from
${\cal I^{-}}$, cross the regular center, and reach ${\cal I^{+}}$.
For an ingoing null ray which is earlier than and close to $n_{+}$,
from Eqs.~(\ref{eq:zeta}) and (\ref{eq:Ipm}),
we can find
\begin{equation}
\zeta\approx - r (y_{in}-y_{+})^{\gamma_{+}},
\end{equation}
where $y_{in}$ is the value of $y$ with which the ingoing null
ray crosses the surface and
\begin{equation}
\gamma_{+}\equiv \frac{3y_{+}^{3}}{f'_{+}(y_{+})}.
\end{equation}
Similarly, for an outgoing null ray which is 
earlier than and close to $n_{-}$,
from Eqs.~(\ref{eq:eta}) and (\ref{eq:Ipm}),
we can find
\begin{equation}
\eta\approx - r (y_{out}-y_{-})^{\gamma_{-}},
\end{equation}
where $y_{out}$ is the value of $y$ with which the ingoing null
ray crosses the surface and
\begin{equation}
\gamma_{-}\equiv \frac{3y_{-}^{3}}{f'_{-}(y_{-})}.
\end{equation}
For an ingoing null ray which is earlier than and close to $n_{+}$,
we find
\begin{equation}
v\approx v_{0}+\left(\frac{dv_{b}}{dy}\right)_{y_{+}}(y_{in}-y_{+}),
\end{equation}
where $v_{0}\equiv v_{b}(y_{+})$.
Similarly, for an outgoing null ray which is 
earlier than and close to $n_{-}$,
we find
\begin{equation}
u\approx u_{0}+\left(\frac{du_{b}}{dy}\right)_{y_{-}}(y_{out}-y_{-}),
\end{equation}
where $u_{0}\equiv u_{b}(y_{-})$.
The coefficients
$(du_{b}/dy)_{y_{-}}$ and $(dv_{b}/dy)_{y_{+}}$
are both finite and negative.
Then we can find the following mapping from ${\cal I^{+}}$ to ${\cal I^{-}}$
using the fact that the regular center is given by $\eta=\zeta$:
\begin{equation}
v=G(u)\approx v_{0}+\Gamma(u_{0}-u)^{\gamma},
\end{equation}
where
\begin{eqnarray}
\gamma&\equiv& \frac{\gamma_{-}}{\gamma_{+}}, \\
\Gamma&\equiv&\left|\left(\frac{dv_{b}}{dy}\right)_{\alpha_+}
\right|
\left|\left(\frac{du_{b}}{dy}\right)_{\alpha_-}\right|^{-\gamma}.
\end{eqnarray}
It should be noted that we can show $\gamma> 1$.

\subsubsection{Power and energy}
Now that we have the map $G$ for late times, 
we can calculate the emitted power and energy for late times.
We have to assume that geometrical optics approximation
is valid for a four-dimensional problem.
The result is
\begin{eqnarray}
P&=&\frac{1}{48\pi}\frac{\gamma^{2}-1}{(u_{0}-u)^{2}}, \\
\hat{P}&=&\frac{1}{48\pi}\frac{(\gamma-1)^{2}}{(u_{0}-u)^{2}},
\end{eqnarray}
and
\begin{eqnarray}
E&=&\frac{1}{48\pi}\frac{\gamma^{2}-1}{u_{0}-u}, \\
\hat{E}&=&\frac{1}{48\pi}\frac{(\gamma-1)^{2}}{u_{0}-u}.
\end{eqnarray}
Therefore, the emitted power diverges to positive infinity
for both minimally and conformally coupled scalar fields.
The totally radiated energy also diverges to positive infinity 
for both minimally and conformally coupled scalar fields.
It should be noted that there appears no characteristic scale 
in the above expression for emitted power and energy in 
contrast to the analytic case.

We should note that no complete analysis for 
the spectrum of radiation has been done yet
for the particle creation from self-similar dust collapse,
although it was pointed out that the spectrum is
different from that of a black body.~\cite{vw1998}

\subsubsection{Quantum stress-energy tensor}
Since we have an exact expression for the metric tensor 
in double null coordinates,
we can estimate the vacuum expectation value of 
stress-energy tensor in a two-dimensional spacetime.~\cite{bsvw1998a}

The expectation value of stress-energy tensor
in the interior coordinates 
is given by Eqs.~(\ref{TUU})--(\ref{TUV}).
It is found that $\langle T_{UV}\rangle$ in the interior is given by
\begin{equation}
 \langle T_{UV}\rangle = \frac{1}{24\pi}
 \frac{|z^2-R_{,r}^2|}{2R_{,r}}\frac{d^2R_{,r}}{dz^2}. 
\end{equation}
After a long calculation, we obtain
\begin{eqnarray}
 \label{F_V}
 F_V(\beta') &=& \frac{1}{12\pi}\left\{\frac{3}{4}\left(\frac{(\beta')_{,V}}{\beta'}\right)^2 -\frac{1}{2}\frac{(\beta')_{,VV}}{\beta'}\right\}  \nonumber\\
             &=& \frac{1}{12\pi}\frac{1}{4(1+x)^2} 
             \left\{\left(1+x-\frac{x^2}{2}\right)^2 
	\right. \nonumber \\
	& &\left.
               +x^5\left(1+\frac{x}{2}\right)\left(\frac{1}{\lambda}
            -\frac{2}{3x^3}+\frac{x}{\lambda}+\frac{1}{3x^2}\right)\right\}
\end{eqnarray}
where $x$ is determined by the equation
\begin{equation}
 \label{(r_b-v)/2M}
 \frac{r_b-v}{2M} = \frac{2}{3x^3}+\frac{2}{x}-\frac{1}{x^2}
                    -2\ln\frac{1+x}{x} .
\end{equation}
$F_U(\beta')$ can be expressed by the same equation (\ref{F_V})
but with $x$ which 
is determined by the equation related to the retarded time as
\begin{equation}
 \label{(r_b-beta)/2M}
 \frac{r_b-\beta(\alpha(u))}{2M} = 
        \frac{2}{3x^3}+\frac{2}{x}-\frac{1}{x^2}-2\ln\frac{1+x}{x} .
\end{equation}
$F_U(A^2)$ and $F_V(A^2)$ become
\begin{eqnarray}
 \label{F_U(A^2)}
 F_U(A^2) &=& \frac{1}{12\pi}\left\{\frac{1}{4}\left
 ( \frac{dR_{,r}}{dz}-1\right)^2 + \frac{1}{2}\frac{z^2 -
 R_{,r}^2}{2R_{,r}}\frac{d^2R_{,r}}{dz^2}\right\} \\
 \label{F_V(A^2)}
 F_V(A^2) &=& \frac{1}{12\pi}\left\{\frac{1}{4}\left( \frac{dR_{,r}}{dz}+1\right)^2 + \frac{1}{2}\frac{z^2 - R_{,r}^2}{2R_{,r}}\frac{d^2R_{,r}}{dz^2}\right\}. 
\end{eqnarray}

In the exterior Schwarzschild region,
the vacuum expectation value is given by Eqs.~(\ref{Tuu})--(\ref{Tuv}).
We obtain
\begin{eqnarray}
\label{F_u}
 F_u(\alpha') &=&
  \frac{1}{12\pi}\left[-\frac{\alpha'^2}{4}\left(1-\frac{w^2}{2\lambda}
	+\frac{w}{3}\right)^2
	+\frac{\alpha'}{4\left(2M\right)}\left(\frac{w^4}{3}
	+\frac{2w^8}{\lambda}-\frac{2w^7}{\lambda}-\frac{w^5}{3}\right)
	\right. \nonumber \\
	& &\left.
  	+\frac{1}{16\left(2M\right)^2}
	\left(8w^7-7w^8\right)\right] \nonumber \\
  &=&\frac{1}{12\pi}\frac{1}{16(2M)^2(3w^3-2\lambda-3w^4-w\lambda)^2}
	\times 3w^6\lambda \nonumber \\
	& &\times\left[-(5w^4+12w^3-8w^2-24w+12)\lambda
               -6w^7+18w^6-12w^5\right].
\end{eqnarray}
where $w$ is determined by the equation
\begin{equation}
 \label{r_b-u/2M}
 \frac{r_b-u}{2M} = \frac{2}{3w^3}+\frac{2}{w}+\frac{1}{w^2}
                    +2\ln\frac{1-w}{w} .
\end{equation}
The relation between $u$ and $U$ is obtained as
\begin{equation}
 \frac{1}{\alpha'(u)}= - \frac{2M}{1-w}\left(\frac{1}{\lambda}
                       -\frac{2}{3w^3}-\frac{w}{\lambda}-\frac{1}{3w^2}\right).
\end{equation}

Here we interpret the outgoing flux emitted from the star.
For this sake, we look into the right hand side of Eq.~(\ref{Tuu}).
The first term clearly denotes the vacuum polarization,
which tends to vanish so rapidly as $R$ goes to infinity
that there is no net contribution to the flux at infinity.
The second term is originated in the interior of the star
as seen in Eqs.~(\ref{TUU}) and (\ref{TVV}).
First it appears as ingoing flux at the passage of the 
ingoing rays through the stellar surface, crosses the center and 
becomes outgoing flux.
Since this term depends on both $\alpha^{\prime}$
and $\beta^{\prime}$,
not only the outgoing but also ingoing null rays 
are relevant to this term.
It implies that this term strongly depends on the 
details of the spacetime geometry in the interior of the star.  
Since the third term depends only on $\alpha^{\prime}$,
only the outgoing null rays determine this contribution. 
Such a term is not seen in Eq.~(\ref{TUU}).
Therefore, this contribution seems to originate at the passage of
the outgoing rays through the stellar surface.
As will be discussed later, the third term corresponds to the
Hawking radiation, while the second term becomes important
in the naked singularity explosion.

\subsubsection{Physical mechanism of particle creation}

Using the above exact expressions,
we can now investigate the exact 
behaviors of the quantum stress-energy tensor for a massless scalar 
field.~\cite{ih2001}  

To determine what would be actually measured, the world line 
of an observer must be specified. 
For the observer with the velocity $u^\mu$, the energy density 
$\langle T_{\mu\nu}\rangle u^\mu u^\nu$ and energy 
flux $\langle T_{\mu\nu}\rangle u^\mu n^\nu$ are considered to be
measured, where $u^\mu n_\mu=0$ and $n^{\mu}n_{\mu}=1$. 

\begin{figure}[tbp]
  \begin{center}
    \leavevmode
    \begin{tabular}{c}
    \subfigure[Lines of $\rho_{\mbox{\scriptsize q}}=\mbox{const}$]
	{\epsfxsize=66mm \epsfbox{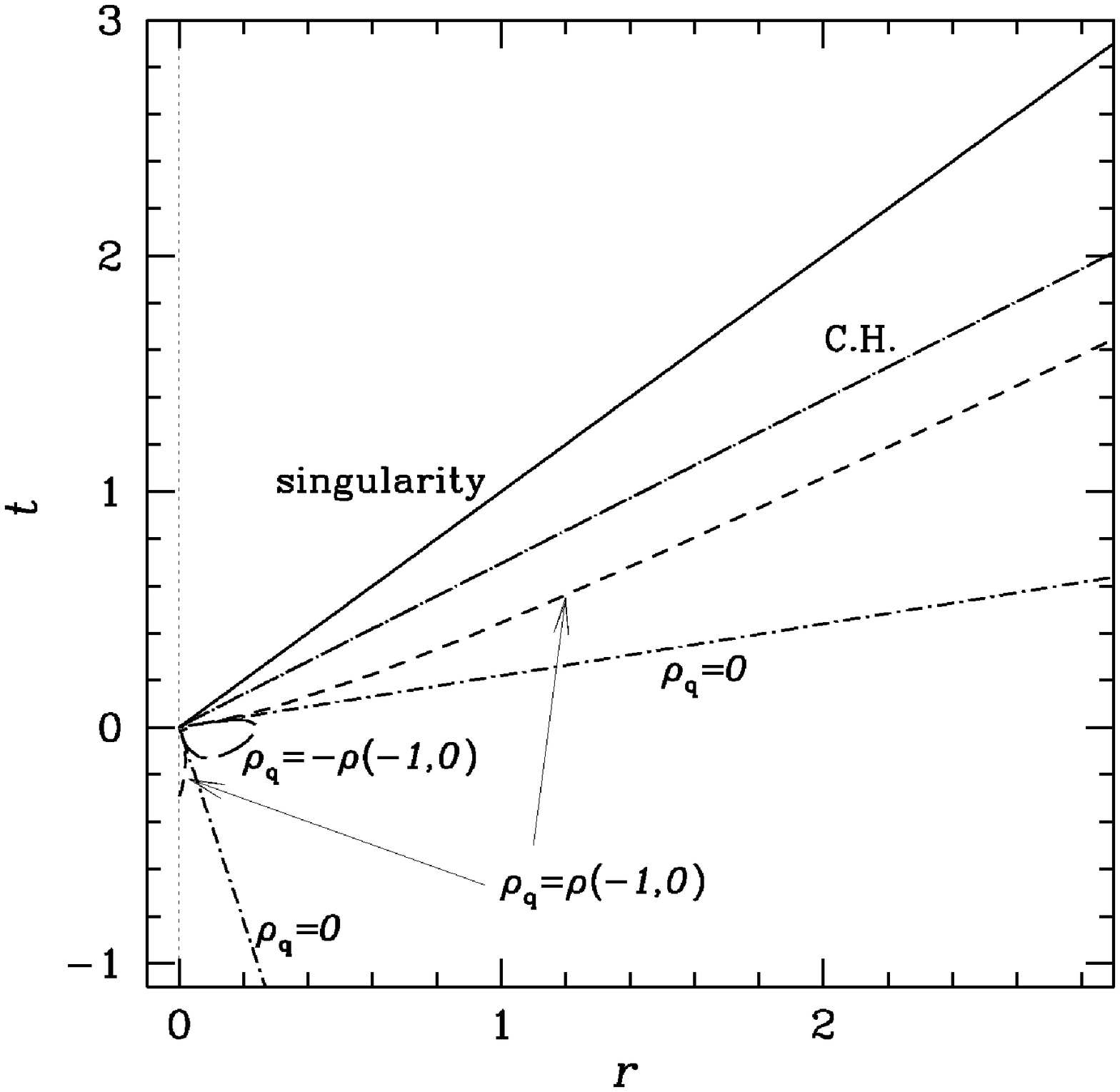}}  
    \subfigure[Lines of $\rho=\mbox{const}$ and $\rho_{\mbox{\scriptsize
  q}}=\rho$]{\epsfxsize=66mm \epsfbox{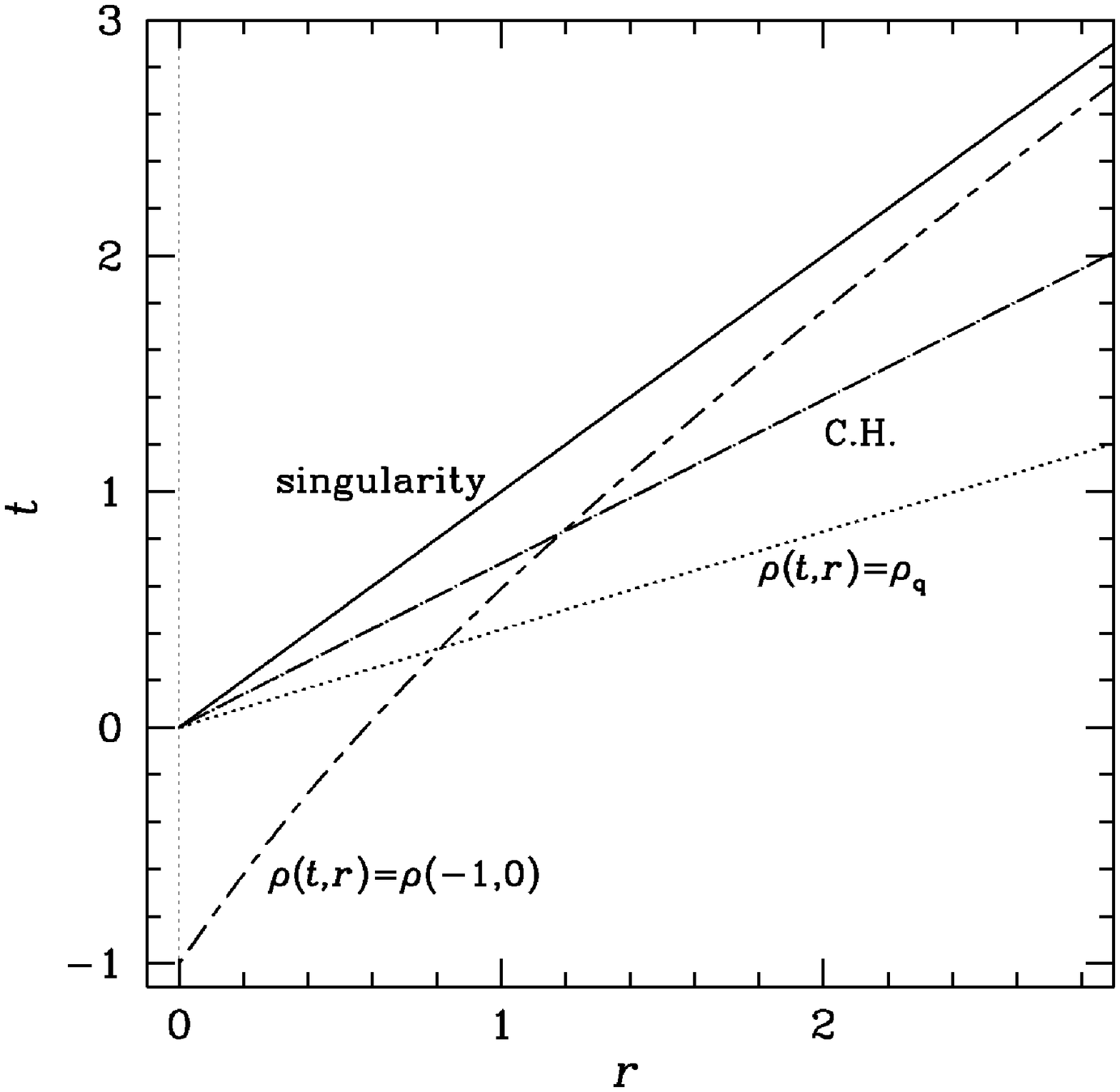}}  
    \end{tabular}  
  \caption{Plots of (a) lines $\rho_{\mbox{\scriptsize q}}=$ const
  and (b) $\rho=$ const and $\rho_{\mbox{\scriptsize
  q}}=\rho$ in the self-similar LTB spacetime.
  Solid, long-dashed-dotted, dashed, dashed-dotted, long-dashed,
  long-dashed-short-dashed and dotted lines denote
  singularity, Cauchy horizon,
  $\rho_{\mbox{\scriptsize q}}=\rho(-1,0)$, 
  $\rho_{\mbox{\scriptsize q}}=0$, 
  $\rho_{\mbox{\scriptsize q}}=-\rho(-1,0)$, 
  $\rho=\rho(-1,0)$ and $\rho_{\mbox{\scriptsize q}}=\rho$,
  respectively.
    }
 \label{fig:density}
  \end{center}
 \end{figure}

To investigate the importance of the back reaction for the central
singularity formation, we compare the energy density observed by the
comoving observer to the background energy density around the center.
The energy density observed by the comoving observer becomes
\begin{equation}
 \label{rho_{q}}
 \rho_{\mbox{\scriptsize {q}}}\equiv\langle T_{\mu\nu}\rangle u^\mu u^\nu 
         = \frac{\langle T_{UU}\rangle }{r^2(z-R_{,r})^2}
                        +\frac{\langle T_{VV}\rangle }{r^2(z+R_{,r})^2}
                        \mp 2 \frac{\langle T_{UV}\rangle}{r^2(z^2-R_{,r}^2)}.
\end{equation}
The results are shown in Fig.~\ref{fig:density}. 
Basically we use the background parameters
$\lambda=0.1$ and $r_b=10^3$. The lines of
$\rho_{\mbox{\scriptsize {q}}}=$ const are plotted in (a) and the line of
$\rho(t,r)=$ const is in (b).  These two values coincide with each other on
the dotted line in (b). Below this line the background density is
larger than the quantum energy density. 
We see especially that near the center the background density is larger than 
the  energy density of the scalar field until the central
singularity. Therefore we conclude that  the
back reaction does not become significant during the semi-classical
evolution. 

\begin{figure}[tbp]
  \begin{center}
    \leavevmode
    \epsfxsize=10cm \epsfbox{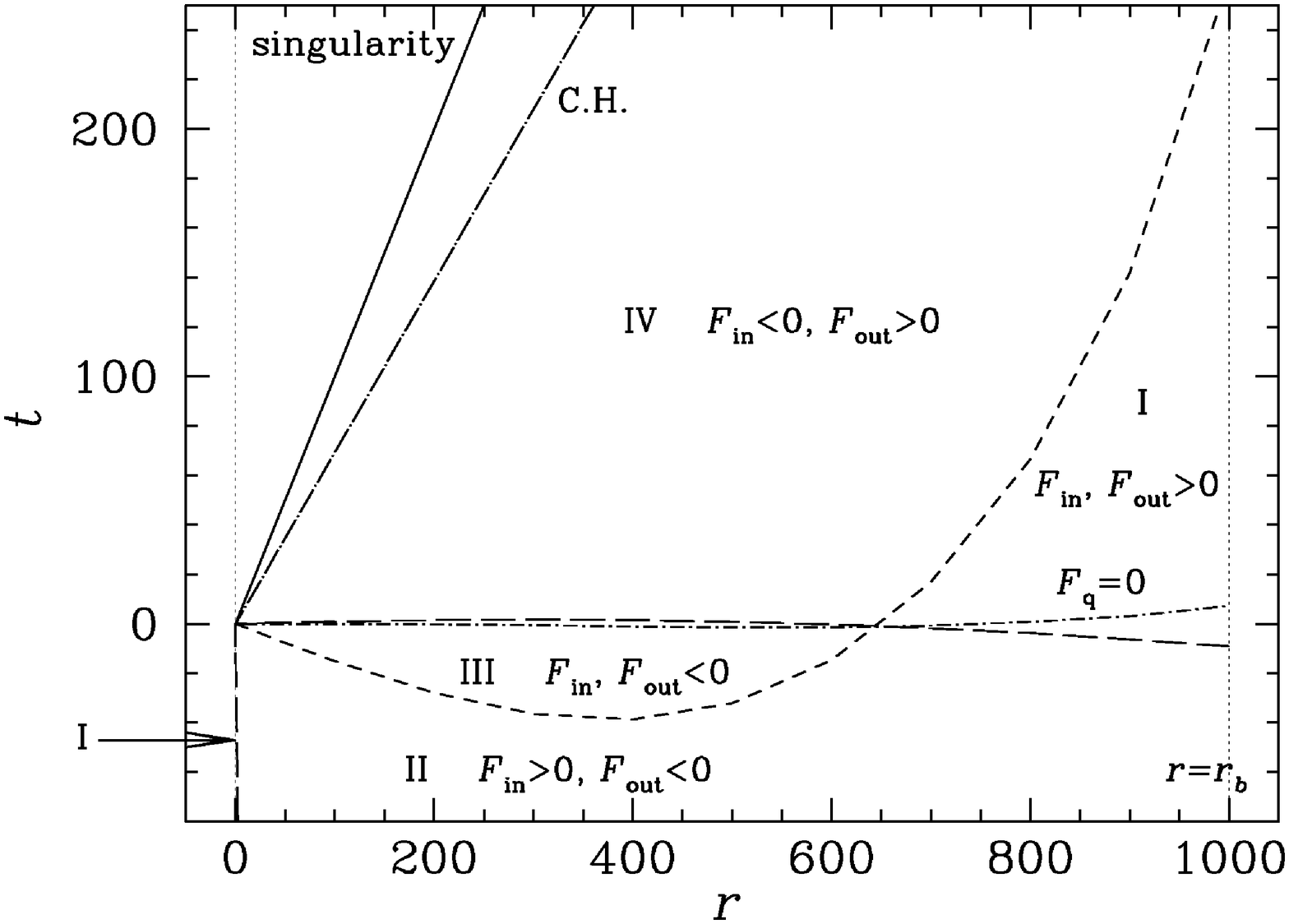}    
  \caption{
  Signs of ingoing and outgoing parts of flux 
  in the self-similar LTB spacetime.
   The solid and long-dashed-dotted lines correspond to the singularity 
  and the Cauchy horizon, respectively. The dashed line denotes the
  line on which the ingoing part of flux vanishes. On the long dashed lines
  the outgoing part vanishes. Regions I, II, III, and IV
  correspond to positive ingoing and outgoing parts, positive ingoing and
  negative outgoing parts, negative ingoing and outgoing parts, and negative 
  ingoing
  and positive outgoing parts, respectively. On the short-dashed-dotted line, 
  the observed net flux vanishes.
   }
 \label{fig:region}
  \end{center}
 \end{figure}

Next we consider the energy flux measured by the comoving observer.
For this observer, the energy flux becomes
\begin{equation}
 \label{flux}
 F_{\mbox{\scriptsize q}}\equiv \langle T_{\mu\nu} \rangle u^\mu n^\nu 
 = \frac{\langle T_{UU}\rangle}{r^2(z-R_{,r})^2}
                        -\frac{\langle T_{VV} \rangle}{r^2(z+R_{,r})^2}.
\end{equation} 
In the interior region, outgoing part of flux is proportional to 
$\langle T_{UU}\rangle$
and ingoing part of flux is proportional to $\langle T_{VV}\rangle$. 
Here we define the ingoing part $F_{\mbox{\scriptsize in}}$ and the outgoing
part $F_{\mbox{\scriptsize out}}$ of the flux as
\begin{equation}
 F_{\mbox{\scriptsize in}}\equiv \frac{\langle
 T_{VV}\rangle}{r^2(z+R_{,r})^2}, ~~~~~
 F_{\mbox{\scriptsize out}}\equiv
 \frac{\langle T_{UU}\rangle}{r^2(z-R_{,r})^2}.
\end{equation}
Following the sign of these two values we divide the 
interior region into four parts. Region I: positive ingoing and outgoing
parts, region II: negative outgoing and positive ingoing parts, region III:
negative ingoing and outgoing parts, and region IV: 
positive outgoing and negative ingoing 
parts. The results are shown in Fig.~\ref{fig:region}.
 Flux observed by a time-like observer may become zero in region I
and III. In fact, the observed flux vanishes at $r=0$ and 
on the dotted dashed
line in Fig.~\ref{fig:region}. 

\begin{figure}[tbp]
  \begin{center}
  \begin{tabular}{cc}
  \subfigure[Observed flux]{\epsfysize=66mm\epsfxsize=66mm\epsfbox{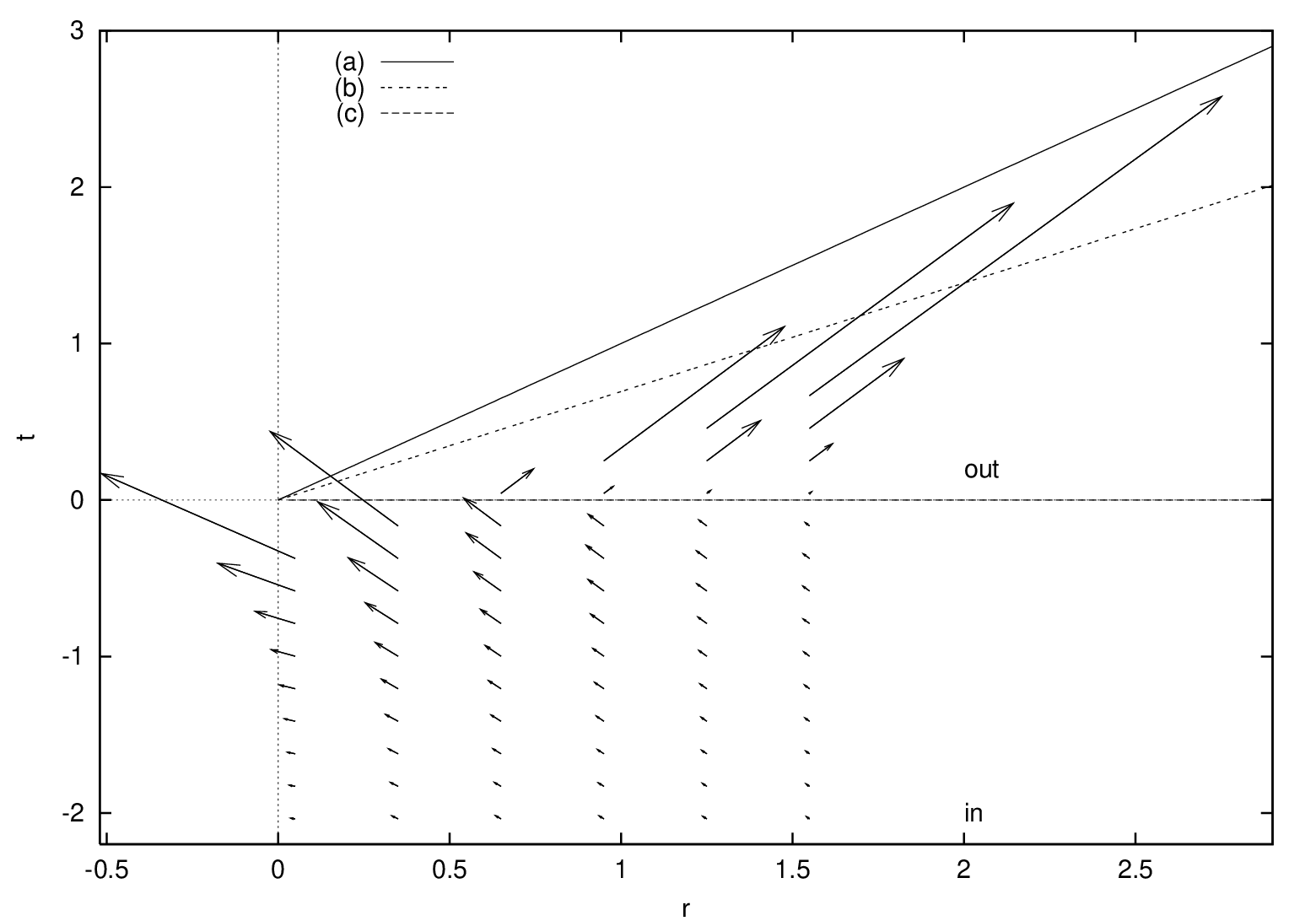}}    
  \subfigure[Ingoing and outgoing parts of flux]
	{\epsfysize=66mm\epsfxsize=66mm\epsfbox{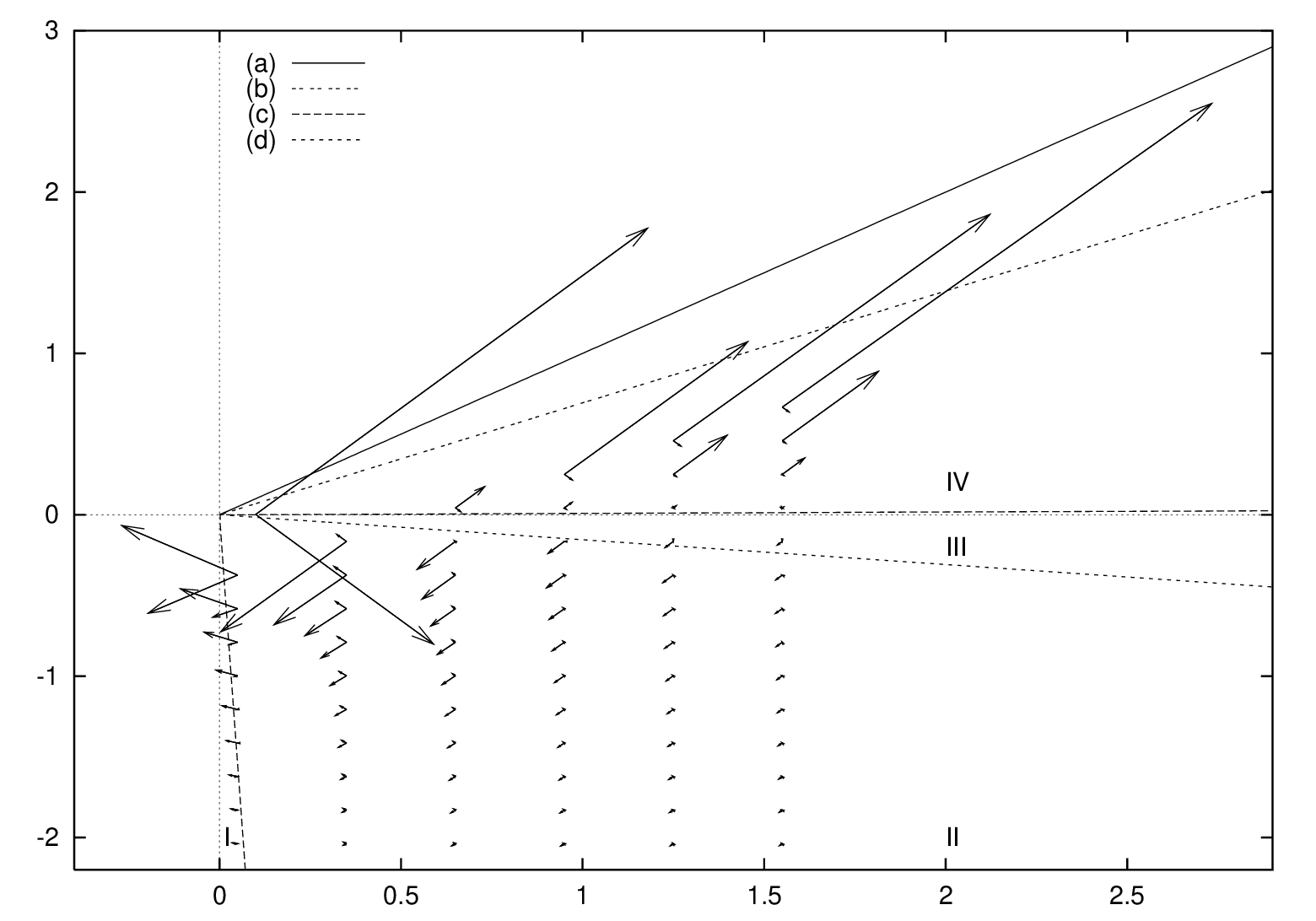}}    
  \end{tabular}	
  \caption{
   Schematic figures (a) for the flux observed by comoving observer and
(b) for the ingoing and outgoing parts of flux in the self-similar 
   LTB spacetime.
  Lines (a) and (b) correspond to the singularity and the Cauchy
  horizon, respectively. Line (c) denotes the line on which the observed
  flux vanishes. Regions I, II, III, and IV are similar to those in 
  Fig.~\ref{fig:region}.
   }
 \label{fig:flux}
  \end{center}
 \end{figure}

We show this flux schematically in Fig.~\ref{fig:flux}(a). The intensity 
is proportional to the length of arrows along the $r$-axis. The comoving 
observers receive inward flux first and then outward flux after the
crossing of the line (c). The
intensity grows as the Cauchy horizon is approached.

In Fig.~\ref{fig:flux}(b), we show ingoing and outgoing parts of the flux
schematically. The intensity 
is proportional to the length of arrow along the $r$-axis.
The future and past directed arrows correspond to the positive and
negative flux respectively. 
The outgoing part has large intensity near the Cauchy
horizon. The ingoing part has large intensity only near the central
singularity. 

Here we summarize the behavior of the quantum field in the interior of
the star. At first inward flux appears and then the positive energy 
is concentrated  near the center 
surrounded by a slightly negative energy region. 
As the collapse proceeds the central positive energy increases and
concentrates within a smaller region. 
At the naked singularity formation the
concentrated positive energy is converted to the outgoing positive diverging
flux.  The left negative energy goes down across the
Cauchy horizon.  It will be reach the space-like singularity.

\begin{figure}[tbp]      
  \begin{center}
  \begin{tabular}{cc}
    \subfigure[Power]{
    \epsfxsize=66mm \epsfysize=66mm\epsfbox{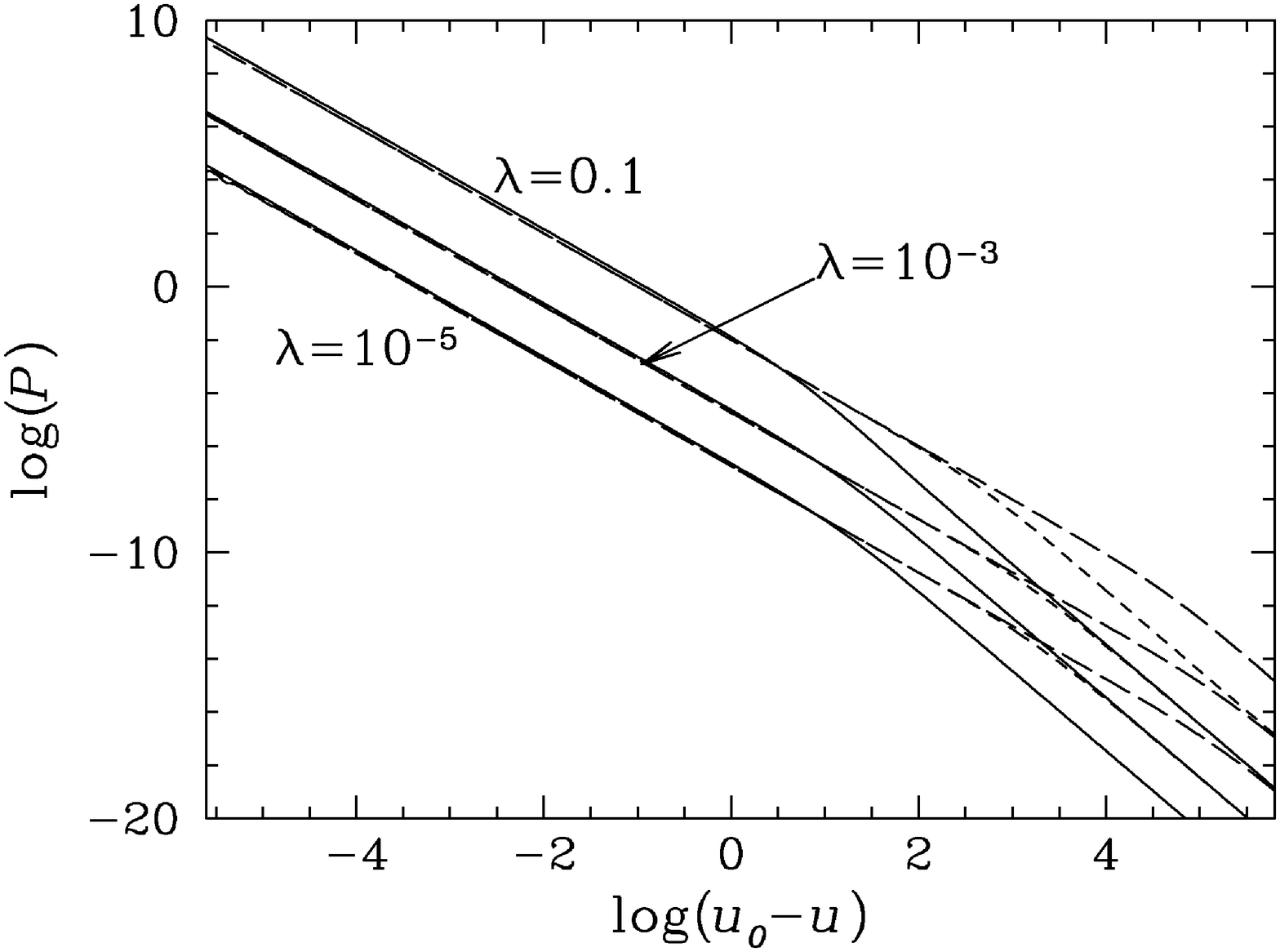}}    
    \subfigure[Energy]{
    \epsfxsize=66mm \epsfysize=66mm\epsfbox{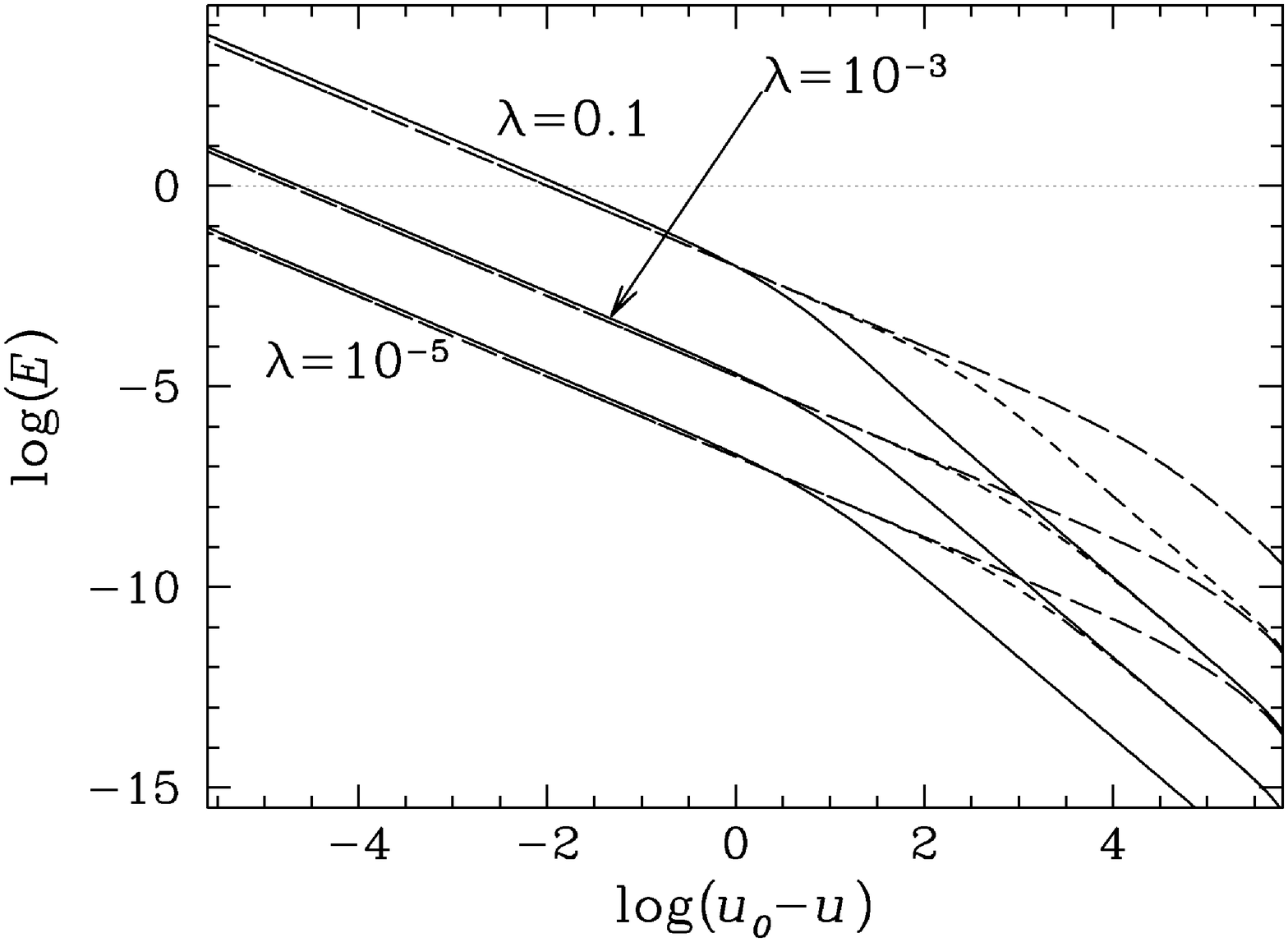}}    
  \end{tabular}
  \caption{
   (a) Radiated power and (b) 
   integrated energy received at infinity in the self-similar 
  LTB spacetime.
  Upper, middle and 
  lower lines correspond to
  $\lambda=0.1$ ,$10^{-3}$ and $10^{-5}$, respectively. 
  Solid, dashed and long-dashed lines
  correspond to $r_b=10$, $10^3$ and $10^5$, respectively.
   }
	 \label{fig:infty_flux}
  \end{center}
 \end{figure}

Next we consider the totally radiated  energy received at infinity.
The radiated power of quantum flux can be expressed as
\begin{equation}
 \label{P}
 P = \langle T_{\mu\nu} \rangle u^\mu n^\nu = 
    \alpha'^2 F_U(\beta') + F_u(\alpha')
\end{equation}
where $u^\mu$ is considered as the velocity of a static observer.
We plot this in Fig.~\ref{fig:infty_flux}(a). We reproduce the results
of the geometric optics estimates for the four-dimensional model, i.e., 
the inverse square
dependence for the retarded time in an approach to the Cauchy horizon.

The totally radiated energy is obtained from the integration of $P$ with respect to 
the retarded time 
at the future null infinity. The results are shown in 
Fig. \ref{fig:infty_flux}(b). We see that the totally emitted energy is
much smaller than the mass of the original star in the range that the
semi-classical approximation can be trusted. 
We can see that the positive divergence of the radiated 
power comes from the term $\alpha'^2 F_U(\beta')$ which has
strong dependence on the spacetime geometry in the interior of the star. 

Let us turn to the inspection for the black hole
formation. For larger $u$ there
appears difference between the naked singularity case and the covered
one. This difference mainly depend on the behavior of $\alpha'$. In an
approach to the event horizon, $w$ goes to unity and then $\alpha'$
goes to zero. It is easy to see that only the second term of 
the radiated power is left in this limit and it becomes
\begin{equation}
F_u(\alpha')=\frac{1}{192\pi\left(2M\right)^2}.
\end{equation}
This means that $F_u(\alpha')$ is interpreted as the Hawking radiation 
contribution. From Eq.~(\ref{F_u=F_v}) we obtain
\begin{equation}
 \langle T_{vv}\rangle = -\frac{1}{192\pi\left(2M\right)^2}
\end{equation}
at the event horizon. This ingoing negative flux crossing into the
black hole is balanced with the energy loss by the positive Hawking flux.

There is similar balance between ingoing and outgoing parts at the Cauchy
horizon for the naked singularity explosion. In the asymptotic region 
positive diverging flux is observed in an approach to the Cauchy
horizon. It has been shown above that there is negative ingoing flux
crossing the Cauchy horizon. This negative flux diverge only at the
central naked singularity. This diverging negative ingoing flux is 
balanced with the diverging positive outgoing flux which propagates 
along the Cauchy horizon.

\subsection{Naked singularities and quantum gravity}
The above discussions are based on quantum field theory
in curved spacetime.
It may lose its validity if we consider a time interval which 
is shorter than the Planck time $t_{Planck}$.
Therefore, the expressions for the energy radiated 
from a forming naked singularity
can be conceivable at least at $u\lesssim u_{0}-t_{Planck}$.
For self-similar dust collapse, it is easily found that 
the emitted energy does not exceed the Planck energy 
$\sim 10^{16}\mbox{erg}$ until this moment.
For the analytic dust collapse, the discussion is not so straightforward,
but the result is the same.~\cite{hinstv2001} 

It is noted that $\omega_{s}$ may exceed the Planck frequency $\omega_{Planck}$ because $\omega_{s}$ is determined
by some combination of two macroscopic scales, the free-fall time and the scale of inhomogeneity.
Nevertheless, we can find that the totally radiated energy for this late-time radiation cannot exceed the
Planck energy
before the Planck time. 
The proof is the following. If $\omega_{s}>\omega_{Planck}$, then it is found that the late-time
radiation is only for $u_{0}-u<\omega_{s}^{-1}<t_{Planck}$. 
Therefore, we only have to consider $\omega_{s}<
\omega_{Planck}$. For this case, it is found
\begin{eqnarray}
E_{u_{0}-u\sim t_{Planck}}&\sim& (\omega_{s}\omega_{Planck})^{1/2}< \omega_{Planck}, \\
\hat{E}_{u_{0}-u\sim t_{Planck}}&\sim &\omega_{Planck}
\left(\frac{\omega_{s}}{\omega_{Planck}}\ln \frac{\omega_{Planck}}{\omega_{s}}\right) < \omega_{Planck},
\end{eqnarray}
where the second inequality comes from the fact that the function $-x \ln x$ has a maximum value $e^{-1}$
in the domain $0<x<1$.
Therefore, we can conclude that the totally radiated energy 
of the late-time emission does not exceed the Planck energy.   

It follows that the further evolution of the star
will depend on quantum gravitational effects, 
and without invoking quantum gravity
it is not possible to say whether the star radiates away
on a short time or settles down into a black hole state.

\subsection{Discussion}

Quantum particle creation during the formation process
of naked singularity in dust collapse 
results in diverging flux emission
within a semi-classical theory of quantum fields 
in curved space.
The discussions so far have been based on some specific models of 
naked singularities. However, 
calculations are
dependent not on the matter fields but on
the spacetime geometry around the singularity.
Therefore, we can conjecture that
the discussions and results will apply 
to a rather wider class of naked singularities which develop
from regular initial data.
Although it is not certain whether this phenomenon 
may be relevant to any observations,
several authors have speculated naked singularities
as a possible central engine of gamma-ray bursts.~\cite{hin2000b,jdm2000}
Any way, to answer the question may be out of scope of 
a semi-classical theory of quantum fields in curved space.
Apart from the observational aspects, 
the contribution of emitted quantum particles
as a source of gravitational field 
may become important in more realistic situations.
The argument that it may prohibit the formation of 
naked singularities leads to the quantum version of 
cosmic censorship conjecture.
In particular, the balance relation between positively diverging
outgoing flux to infinity and
negatively diverging ingoing flux to naked singularity 
may suggest the quantum version of 
``naked singularity evaporation conjecture''
proposed by Nakamura, Shibata and Nakao.~\cite{Nakamura:1993xr}

\section{Summary}
We have several examples in which naked singularity forms
as a result of time development of regular initial 
Cauchy data,
as we have seen in \S \ref{sec:intro}.
However, we should not conclude that they alone disprove
the cosmic censorship conjecture in the exact meaning.
In all of these examples, some kind of high symmetries,
such as spherical symmetry,
axisymmetry, cylindrical symmetry and self-similarity,
are assumed.
Our knowledge on 
the effects of deviations from these symmetries 
on the naked singularities
has been quite restricted so far.
In critical collapse, it is known that 
the occurrence of a zero-mass black hole,
which is regarded as naked singularity,
is realized as 
a result of the exact fine-tuning. 
This means that it is a zero-measure event even within 
the assumed symmetry.
In other cases,
it is quite difficult in general 
to answer the question 
whether the appearance of naked singularity
is generic or not.
It should be also noted that 
there is no generally accepted 
standard measure of the space of initial data sets.
Apart from genericity, 
in some of the examples of naked singularities, 
matter fields cannot be considered
as realistic, such as a dust fluid and null dust.
It is also difficult to answer the question 
what matter model is physically reasonable.
Some people consider that elementary fields such as
a scalar field will be physically reasonable matter models.
However, generally speaking, it is only at the stage
in which quantum effects become so
important that such field descriptions of matter fields 
will be useful.
Others consider that a system of infinitely many self-gravitating 
collisionless particles, i.e., the 
Einstein-Vlasov
system will be suitable for the proof or disproof
of the cosmic censorship conjecture.
This proposition seems to be reasonable from the 
standing point
that singularities that form also in Newtonian gravity
should not be taken seriously,
because
of the global existence 
theorem of regular solutions with
nonsingular initial data sets 
in the corresponding Newtonian system.
However, unfortunately, we do not know any 
realistic matter
field which is expected to behave like
the Einstein-Vlasov system
in an 
extremely high-density regime. 
    
However,
these examples of naked singularities 
strongly motivate us to imagine, as a 
possible stage of gravitational 
collapse, extremely high-density
and/or high-curvature regions uncovered by horizons.
In these strongly curved 
regions, general relativistic effects
will become crucial.
In fact, it is considered that classical 
general relativity cannot apply
beyond some energy scale, which will be characterized by
the Planck scale.
Therefore, the existence of classical solutions with naked singularity
formed from regular initial data will imply not only
the limitation of future predictability of general relativity
but also the possibility that 
new frontier beyond general relativity 
may be exposed to us.
Of course, if a naked singularity 
is a generic outcome of gravitational collapse, 
the importance to examine its consequences
is much more increased. 

If naked singularity is a possible outcome of
gravitational collapse, it may provide an exotic 
source of gravitational wave burst.
Gravitational wave generation reflects the dynamical motion of
the spacetime in strongly curved regions and they will 
directly reach us without 
obstructions unlike electromagnetic waves. 
As we have seen in \S \ref{sec:perturb},
for linearized nonspherical perturbations 
of spherical dust collapse,
not the energy flux of gravitational waves but
the perturbation of the spacetime curvature
blows up unboundedly
as one approaches the Cauchy horizon 
of the background spacetime.
This conclusion is also derived 
semi-analytically using the 
quadrupole formula for the gravitational wave emission
from the dust collapse in Newtonian gravity.
From this fact, we may imagine that some 
sort of perturbations concerning spacetime curvature
will be considerably amplified near the 
Cauchy horizon of a forming 
naked singularity,
although it is not clear if this amplification 
can be related to any observable of a distant observer.
This result also implies weak instability
of the Cauchy horizon. 
Since it means the breakdown of the linearized
perturbation scheme, 
some nonlinear analysis is necessary to make
the conclusion clear.
As we are having high sensitivity to gravitational waves
by laser interferometric gravitational wave detectors,
gravitational wave observations will provide
interesting information of strongly curved regions.

In addition to classical gravitational radiation, 
we have seen quantum particle creation from 
a forming naked singularity 
in the spherical dust collapse in \S \ref{sec:quantum}.
The calculation of this emission is based on
quantum field theory in curved spacetime.
The calculation goes parallel to that of 
the Hawking radiation
of black holes.
The only difference is the form of the mapping 
function between
the advanced and retarded time coordinates, which comes
from the behavior of ingoing and outgoing 
radial null geodesics in the
curved spacetime.
The result is that a forming naked singularity 
emits explosive radiation at the final epoch. 
This conclusion comes from the fact that 
the mapping function has only lower differentiability
at the Cauchy horizon.
It is expected that this property of the mapping function
will be common to a wide class of naked singularities.
In the case of the two dimensional self-similar dust collapse, 
the back reaction effects of quantum radiation 
on the spacetime geometry will be negligible
at least until the singularity formation. 
The radiated positive energy originates 
from the vacuum 
energy in the region surrounding the center.
A large amount of positive energy 
is gradually concentrated and accumulated 
around the center in the course of the  
singularity formation.
After that, the positive energy turns into 
positive outgoing flux just before the
singularity formation.
In addition to the back reaction effects, there appears 
a rather subtle problem between quantum field theory and 
quantum gravity.
If we accept the Planck length a priori 
as the shortest length scale 
of quantum field theory, the total amount of radiated energy
from a forming naked singularity cannot exceed the Planck 
energy as long as quantum field theory is justified.
Actually, the total amount of radiated energy is 
greatly dependent on 
the shortest length scale of quantum field theory. 

In conclusion, the cosmic censorship conjecture has provided 
a strong motivation for researchers 
in this field.
This conjecture was intended to conserve 
the completeness of 
future predictability
of classical theory.
However, we are now convinced of the limitation of 
classical theory as theory of everything.
In this point of view, 
known examples of naked singularities imply 
that the future predictability
of classical theory is violated
in a finite-measure set of initial data sets.
Then, we may pay our attention to more general features
of gravitational collapse.
In fact, classical gravitational wave 
radiation and quantum particle 
emission from a forming naked singularity,
which have been discussed in this paper,
are both irrespective of the final outcome of the forming 
naked singularity,
because these calculations are basically independent
of the spacetime geometry in the 
causal future of the singularity.
In this respect, the forming naked singularity 
can be regarded as an example of inhomogeneously 
runaway collapsing systems.
It could be said that 
our understanding of gravitational collapse 
might have been too much based on the spherically symmetric 
homogeneous collapse, such as the Oppenheimer-Snyder model.
Gravitational collapse in more general 
situations is now and will be an active and attractive field
in gravitational physics.

\acknowledgements

This work was partly supported by Grants-in-Aid (Nos. 05540 and 11217)
from the Japanese Ministry of
Education, Culture, Sports, Science and Technology.

\appendix
\section{Gauge-Invariant Perturbations} 
\label{sec:gauge}
In this appendix we give a brief introduction to the formalism of Gerlach 
and Sengupta \cite{Gerlach:1979rw,Gerlach:1980tx} 
for perturbations around the most general spherically
symmetric spacetime.

We consider a general spherically symmetric spacetime with metric
\begin{equation}
  g_{\mu \nu}dx^{\mu}dx^{\nu} \equiv g_{ab}(x^d)dx^a dx^b
  +R^2(x^d)\gamma_{AB} (x^D)dx^A dx^B
\end{equation}
and stress-energy tensor
\begin{equation}
  t_{\mu \nu}dx^{\mu}dx^{\nu} \equiv t_{ab}(x^d)dx^a dx^b
  +\frac{1}{2}t_A^{~A}R^2(x^d)\gamma_{AB} (x^D)dx^A dx^B,
\end{equation}
where $\gamma_{AB} dx^A dx^B = d\theta ^2 + \sin ^2 \theta d \phi ^2$. 
Lowercase Latin indices refer to radial and time coordinates, while
uppercase Latin indices refer to $\theta$ and $\phi$.

Now we introduce arbitrary perturbations of this spacetime. The
angular dependence of perturbations is decomposed into series of
tensorial spherical harmonics. The scalar spherical harmonics are
$Y_l^m(x^A)$. A basis of vector harmonics is formed by 
$Y_{l~:A}^m$ and  $S_{l~A}^m\equiv \epsilon_A^{~B}Y_{l~:B}^m$. A basis of
symmetric rank-two tensor harmonics is formed by
$Y_l^m\gamma_{AB}$, $Z_{l~AB}^m\equiv Y_{l~:AB}^m + (l(l+1)/2)
Y_l^m \gamma_{AB}$ and $S_{l~A:B}^m+S_{l~B:A}^m$. For $l=0$ and $1$, 
the last
two tensors vanish identically. Linear perturbations with different $l$
and $m$ decouple. In the following,
we suppress these indices. We also suppress the explicit summation
over them. Perturbations with different values of $m$ for the same $l$
have the same dynamics on a spherically symmetric background, 
and therefore
$m$ does not appear in the field equations. Spherical harmonics are
called even parity if they have parity $(-1)^l$ under spatial inversion 
and odd parity if
they have parity $(-1)^{l+1}$. Even and odd perturbations decouple.

The odd-parity perturbations are expressed as
\begin{equation}
  \label{GS-metodd}
   h_{\mu\nu}dx^{\mu}dx^{\nu} = h_a(x^c)S_B(dx^adx^B+dx^Bdx^a)
                                  +h(x^c)S_{(A:B)}dx^Adx^B
\end{equation}
for the metric and
\begin{equation}
  \label{GS-matodd}
   \Delta t_{\mu\nu}dx^{\mu}dx^{\nu} = \Delta t_a(x^c)S_A(dx^adx^A+dx^Adx^a)
                                  +\Delta t(x^c)S_{(A:B)}dx^Adx^B 
\end{equation}
for matter.
The even-parity perturbations are 
\begin{eqnarray}
  \label{GS-meteven}
  h_{\mu\nu}  &=& h_{ab}(x^d)Ydx^adx^b+h_a (x^d)
               Y_{:B}(dx^adx^B+dx^Bdx^a)\nonumber \\
               &&+[K(x^d) R^2 \gamma_{AB}Y+ G(x^d) R^2 Z_{AB}]dx^Adx^B 
\end{eqnarray}
for the metric and
\begin{eqnarray}
  \label{GS-mateven}
  \Delta t_{\mu\nu}  &=& \Delta t_{ab}(x^d)Ydx^adx^b+\Delta t_a (x^d)
               Y_{:B}(dx^adx^B+dx^Bdx^a)\nonumber \\
               &&+[\Delta t^3(x^d) R^2 \gamma_{AB}Y+ \Delta t^2(x^d) R^2
               Z_{AB}]dx^Adx^B 
\end{eqnarray}
for matter.
Here covariant derivatives are distinguished as follows:
\begin{equation}
  \gamma_{AB:C} \equiv 0, ~~~~~ g_{ab|c}\equiv 0.
\end{equation}
For convenience, we introduce 
\begin{equation}
  v_a \equiv \frac{R_{,a}}{R}, 
\end{equation}
and
\begin{equation}
  p_a \equiv h_a - \frac{1}{2}R^2 G_{,a}.
\end{equation}

We then introduce gauge-invariant 
variables to eliminate gauge ambiguities in perturbations. The gauge
transformation is induced by the infinitesimal vector fields
\begin{equation}
  \xi_\mu dx^\mu = M(x^c)S_Adx^A ,
\end{equation}
for odd parity and
\begin{equation}
  \xi_\mu dx^\mu  = \xi_a(x^c)Ydx^a+\xi(x^c)Y_{:A}dx^A 
\end{equation}
for even parity.
The odd-parity gauge-invariant metric variables are given by 
\begin{equation} 
  \label{ka} 
  k_a \equiv h_a-\frac{1}{2}R^2\partial_a\left(\frac{h}{R^2}\right).
\end{equation} 
The odd-parity gauge-invariant matter variables are given by the combinations 
\begin{eqnarray}
  \label{la}
  L_a &\equiv& t_a-\frac{1}{2}\bar{T}_B^{~B}h_a ,\\
  \label{l}
  L &\equiv& t-\frac{1}{2}\bar{T}_B^{~B}h .
\end{eqnarray}
A set of even-parity gauge-invariant metric perturbations is defined as
\begin{eqnarray}
  \label{gi-kab}
  k_{ab} &\equiv& h_{ab}-(p_{a|b}+p_{b|a}) ,\\
  \label{gi-k}
  k &\equiv& K + \frac{l(l+1)}{2}G - 2v^a p_a.
\end{eqnarray}
A set of even-parity gauge-invariant matter perturbations is defined as
\begin{eqnarray}
  T_{ab} &\equiv& \Delta t_{ab}- t_{ab|c}p^c -t_a^{~c}p_{c|b} 
  -t_b^{~c}p_{c|a} ,\\
  T_a &\equiv& \Delta t_a -t_a^{~c}p_c -R^2(t_A^{~A}/4)G_{,a} ,\\
  T^3 &\equiv& \Delta t^3 -(p^c/R^2)(R^2 t_A^{~A}/2)_{,c} 
  +l(l+1)(t_A^{~A}/4)G ,\\
  T^2 &\equiv&  \Delta t^2 -(R^2 t_A^{~A}/2)G.
\end{eqnarray}

The linearized Einstein equations, expressed only in 
terms of gauge-invariant
perturbations,  are
\begin{eqnarray}
  \label{GS-9a}
  k^a_{~|a} &=& 16 \pi L,~~~~~~(l\geq 2) 
\end{eqnarray}
\begin{eqnarray}
  \label{GS-9b}
  -\left[R^4\left(\frac{k_a}{R^2}\right)_{|c}-R^4\left(\frac{k_c}{R^2}\right)_{|a}\right]^{|c} +(l-1)(l+2)k_a &=& 16 \pi R^2L_a,~~(l\geq 1)
\end{eqnarray}
for odd parity and
\begin{eqnarray}
\label{le1}
2v^c\left(k_{ab|c}-k_{ca|b}-k_{cb|a}\right)
-\left[\frac{l(l+1)}{R^2}+G_c^{~c}+G_A^{~A}+2\cal{R}\right]k_{ab} &&\nonumber\\
-2g_{ab}v^c\left(k_{ed|c}-k_{ce|d}-k_{cd|e}\right)g^{ed}
+g_{ab}\left(2v^{c|d}+4v^cv^d-G^{cd}\right)k_{cd} && \nonumber\\
+g_{ab}\left[\frac{l(l+1)}{R^2}+\frac{1}{2}\left(G_c^{~c}+G_A^{~A}\right)
+\cal{R}\right]k_d^{~d}+2\left(v_ak_{,b}+v_bk_{,a}+k_{,a|b}\right)&&\nonumber\\
-g_{ab}\left[2{k_{,c}}^{|c}+6c^ck_{,c}-\frac{(l-1)(l+2)}{R^2}k\right]=
-16\pi T_{ab},&& 
\end{eqnarray}
\begin{eqnarray}
\label{le2}
    k_{,a}-{k_{ac}}^{|c}+{{k_c}^c}_{|a}-v_ak_c^{~c} &=& -16\pi T_a, 
\end{eqnarray}
\begin{eqnarray}
\label{le3}
  -\left({k_{,c}}^{|c}+2v^ck_{,c}+G_A^{~A}k\right)
+\left[{k_{cd}}^{|c|d}+2v^c{k_{cd}}^{|d}+2(v^{c|d}+v^cv^d)k_{cd}\right]
  && \nonumber\\
-g_{ab}\left[{{{k_c}^c}_{|d}}^{|d}+v^c{{k_d}^d}_{|c}
+{\cal{R}}k_c^{~c}-\frac{l(l+1)}{R^2}k\right] =-16\pi T^3, &&
\end{eqnarray}
\begin{eqnarray}
\label{le4}
  k_c^{~c} &=&-16\pi T^2
\end{eqnarray}
for even parity,
where $\cal{R}$ is the Gaussian curvature of the 2-dimensional
submanifold $M^2$ spanned by $x^a$, and
\begin{eqnarray}
  G_{ab} &\equiv& -2\left(v_{a|b}+v_av_b\right)+g_{ab}\left(2v_a^{~|a}
                +3v_av^a-\frac{1}{R^2}\right), \\
  G_A^{~A}  &\equiv& 2\left(v_a^{~|a}+v_av^a -{\cal{R}}\right).
\end{eqnarray}

The linearized conservation equation 
[$\delta(T^{\mu\nu}_{~~;\nu})=0$]
reduces to 
\begin{equation}
  \label{GS-14}
  \left(R^2L^a\right)_{|a} =
  \left(l-1\right)\left(l+2\right)L,~~~~~~~~~(l\geq 1)
\end{equation} 
for odd parity and
\begin{eqnarray}
  \frac{\left(R^2T^a\right)_{|a}}{R^2}+T^3-\frac{\left(l-1\right)
  \left(l+2\right)}{2R^2}T^2&=&\frac{1}{2}t_A^{~A}\left(k-
  \frac{1}{2}k_c^{~c}\right)+\frac{1}{2}t^{ab}k_{ab},\\
  \frac{\left(R^2T_{ab}\right)^{|b}}{R^2}-\frac{T_al\left(l+1\right)}{R^2}
    -2v_aT^3&=&\frac{1}{2}k_{bc|a}t^{bc}+k_{cb}^{|b}t^c_{~a}
     -\frac{1}{2}k_{c~|b}^{~c}t^b_{~a}\nonumber \\
   &&-k_{,c}t^c_{~a}+\frac{1}{2}\left(k_{,a}-kv_a\right)t_A^{~A}\nonumber \\
   &&  +2v^bk_{bc}t^c_{~a}+k^b_{~c}t^c_{~a|b}
\end{eqnarray}
for even parity.

\section{Power of Gravitational Radiation}
\label{chap:power}

In this appendix we examine the asymptotic behavior of the
gauge-invariant variables in asymptotically flat spacetime with
an outgoing wave condition. Then
we can calculate the radiated power of the gravitational waves and thereby
grasp the physical meaning of the gauge-invariant
quantities.

Note that in vacuum at
large distance, the spherically symmetric background metric is identical 
to the Schwarzschild solution, which is the following 
(adopting the Schwarzschild coordinates):  
\begin{equation}
  \label{Schmetric}
  ds^2 = -\left( 1-\frac{2M}{R}\right)d\tau ^2 +\left
  ( 1-\frac{2M}{R}\right)^{-1} dR^2 + R^2 \left( 
  d\theta^{2}+\sin^{2}\theta d\phi^{2}\right) . 
\end{equation}
To relate the perturbation of the metric to the radiated
gravitational power, it is useful to specialize 
the gauge to the radiation gauge,
in which the tetrad components $h_{(\theta)(\theta)}-h_{(\phi)(\phi)}$
and $h_{(\theta)(\phi)}$ fall off as $O(1/R)$, and all other tetrad
components fall off as $O(1/R^2)$ or faster with respect to the following
background tetrad basis:
\begin{eqnarray}
  e^a_{(\tau)}&=&\left(1-\frac{2M}{R}\right)^{1/2}(d\tau)^a,\\
  e^a_{(R)}&=& \left(1-\frac{2M}{R}\right)^{-1/2}(dR)^a,\\
  e^a_{(\theta)}&=& R(d\theta)^a,\\
  e^a_{(\phi)}&=& R\sin\theta (d\phi)^a.
\end{eqnarray}

In this radiation gauge, the metric perturbations in
Eqs. (\ref{GS-metodd}) and (\ref{GS-meteven}) behave as 
\begin{eqnarray}
  h_0, h_1 &=& O\left(\frac{1}{R}\right), \\
  h_2 &=& w(\tau - R_*) + O(1)
\end{eqnarray}
for odd parity and as 
\begin{eqnarray}
  h_{ab} &=& O\left(\frac{1}{R^2}\right), \\
  h_a &=& O\left(\frac{1}{R}\right), \\
  K &=& O\left(\frac{1}{R^2}\right), \\
  G &=& \frac{g(\tau-R_*)}{R} + O\left(\frac{1}{R^2}\right) 
\end{eqnarray}
for even parity, where 
\begin{equation}
  R_* = R + 2M \ln \left( \frac{R}{2M} - 1\right) + \mbox{const},
\end{equation}
and the outgoing wave condition is satisfied.
Then, the gauge-invariant metric perturbations (\ref{gi-kab}) and
(\ref{gi-k}) are calculated as 
\begin{eqnarray}
  k_0 &=& -\frac{1}{2}w^{(1)}R + O(1), \\
  k_1 &=& \frac{1}{2}w^{(1)}R + O(1)
\end{eqnarray}
for odd parity and 
\begin{eqnarray}
  k_{\tau \tau} &=& g^{(2)}R+O(1), \\
  k_{\tau R} &=& -g^{(2)}R+O(1), \\
  k_{RR} &=& g^{(2)}R+O(1), \\
  k &=& -g^{(1)}+O\left(\frac{1}{R}\right)
\end{eqnarray}
for even parity, where $w^{(n)}$ and  $g^{(n)}$ denotes the $n$-th
derivatives of $w$ and $g$, respectively.

In this radiation gauge, the radiated power $P$ per solid angle is
given by the formula derived by Landau and Lifshitz
\cite{Landau:1975} from their stress-energy pseudo-tensor:
\begin{equation}
  \frac{dP}{d\Omega}=\frac{R^2}{16\pi}\left[\left(\frac{\partial
  h_{(\theta)(\phi)}}{\partial \tau}\right)^2
  +\frac{1}{4}\left(\frac{\partial h_{(\theta)(\theta)}}{\partial
  \tau}-\frac{\partial h_{(\phi)(\phi)}}{\partial \tau}\right)^2\right].
\end{equation}
For the axisymmetric mode, (i.e. $m=0$), the above formula is reduced to 
\begin{equation}
  \frac{dP}{d\Omega}=\frac{1}{64\pi}(w^{(1)})^2A_l (\theta)  
\end{equation}
for odd parity and 
\begin{equation}
  \frac{dP}{d\Omega}=\frac{1}{64\pi}(g^{(1)})^2A_l (\theta)  
\end{equation}
for even parity, where
\begin{equation}
  A_l (\theta) \equiv \frac{2l+1}{4\pi}\sin^4\theta\left(\frac{d^2P_l(\cos\theta)}{(d\cos\theta)^2}\right)^2.
\end{equation}
It is found that for the monopole and dipole modes, the radiated power
exactly vanishes.
By using gauge-invariant quantities and integrating over the 
$4\pi$ solid angle, the formula for the power of the gravitational radiation
for $l\ge 2$ is obtained as
\begin{eqnarray}
  \frac{dP}{d\Omega}&=&\frac{1}{16\pi}\frac{k_0^2}{R^2}A_l
  (\theta)=\frac{1}{16\pi}\frac{k_1^2}{R^2}A_l (\theta),\\
  \label{oddpower}
  P &=& \frac{1}{16\pi} B_l \frac{k_0^2}{R^2} = \frac{1}{16\pi} B_l
  \frac{k_1^2}{R^2}
\end{eqnarray}
for odd parity and
\begin{eqnarray}
  \frac{dP}{d\Omega}&=&\frac{1}{64\pi}k^2 A_l (\theta),\\
  P &=& \frac{1}{64\pi} B_l k^2   
\end{eqnarray}
for even parity, where
\begin{equation}
  B_l \equiv \frac{(l+2)!}{(l-2)!}.
\end{equation}

\section{Equations for Even-Parity Perturbations}
\label{chap:eqn}

For marginally bound LTB spacetime linearized quadrupole Einstein
equations  are 
\begin{eqnarray}
\label{00}
& &    \frac{4}{R^2}q +\frac{1}{RR'}q' +\frac{\dot{R}}{R}\dot{q}   \nonumber\\
& &  \quad        -\frac{6}{R^2}{k} +\left(\frac{2}{RR'}-\frac{R''}{R'^3}\right)k' 
          -\left(2\frac{\dot{R}}{R}+\frac{\dot{R'}}{R'}\right)\dot{k}    
        +\frac{1}{R'^2}k''   \nonumber\\ 
& & \quad +2\left(\frac{\dot{R}}{R^2R'}
          -\frac{\dot{R}R''}{RR'^3} 
          +\frac{\dot{R'}}{RR'^2}\right){k_{01}}
         +2\frac{\dot{R}}{RR'^2}{k_{01}}' = -8 \pi \delta \rho, 
\end{eqnarray}
\begin{eqnarray}
  \label{01}
 -\frac{\dot{R}}{R}q' +\frac{R'}{R}\dot{q} 
   +\left(2\frac{\dot{R}}{R} -\frac{\dot{R'}}{R'}\right)k' +\dot{k'}  
   -\frac{3}{R^2}{k_{01}} &=& -8 \pi \bar{\rho} V_1,  
\end{eqnarray}
\begin{eqnarray}
  \label{11}
&&   2\frac{R'^2}{R^2}q  -\frac{R'}{R}q' 
     -\frac{R'^2\dot{R}}{R}\dot{q} +4\frac{R'^2\dot{R}}{R}\dot{k}  
                  +R'^2\ddot{k}        \nonumber\\
&&\quad    -2\frac{R'\dot{R}}{R^2}{k_{01}} -2\frac{R'}{R}\dot{k_{01}} 
     = 0,  
\end{eqnarray}
\begin{eqnarray}
  \label{02}
  -2\frac{\dot{R'}}{R'}q -\dot{q} +2\frac{\dot{R'}}{R'}{k} +2\dot{k} 
  +\frac{R''}{{R'}^3}{k_{01}}   -\frac{1}{R'^2}{k_{01}}'      
  &=& -16 \pi \bar{\rho} V_2,  
\end{eqnarray}
\begin{eqnarray}
  \label{12}
  q' +\dot{k_{01}} +\frac{\dot{R'}}{R'}{k_{01}} &=& 0,  
\end{eqnarray}
\begin{eqnarray}
  \label{22}
&&   \left(\frac{R^2R''}{R'^3} -2\frac{R}{R'}\right)q' 
                  -\left(2R\dot{R}+3\frac{R^2\dot{R'}}{R'}\right)\dot{q} 
                  -\frac{R^2}{R'^2}q'' -R^2\ddot{q}   \nonumber\\
&& \quad                   +4\left(R\dot{R}+\frac{R^2\dot{R'}}{R'}\right)\dot{k} 
                  +2R^2\ddot{k} 
  +2\left(\frac{RR''\dot{R}}{R'^3}-\frac{R\dot{R'}}{R'^2}\right)k_{01}  \nonumber\\
&& \quad                 -2\frac{R\dot{R}}{R'^2}{k_{01}}'
                  +2\left(\frac{R^2R''}{R'^3}-\frac{R}{R'}\right)\dot{k_{01}}
                  -2\frac{R^2}{R'^2}\dot{k_{01}'} = 0,    
\end{eqnarray}
\begin{eqnarray}
  \label{z22}
  -k_{00} + \frac{1}{R'^2}k_{11} &=& 0. 
\end{eqnarray}
Here we can use Eq.\ (\ref{z22}) to eliminate $k_{11}$ in 
Eqs.~(\ref{00})--(\ref{22}).

\section{Characteristic Frequency of a Naked Singularity}
\label{sec:conserved_quantity}
In \S \ref{sec:LTB}, we set the initial data at $t=0$
and specified the solution.
At an arbitrary time $t<t_{0}$, 
we can expand the mass function $F(r)$ 
in terms of the circumferential radius $R$ 
for marginally bound collapse with an 
analytic initial density profile as
\begin{equation}
F(r)=F_{3}\left(\frac{t_{0}-t}{t_{0}}\right)^{-2}R^3
+F_{5}\left(\frac{t_{0}-t}{t_{0}}\right)^{-13/3}R^5+\cdots.
\end{equation} 
Although the quantities $F(r)$, $(t_{0}-t)$ and $R$ have 
physical meanings, $t_{0}$ itself has no physical meaning, 
because it explicitly depends on the choice of the origin of the 
time coordinate $t$.
This implies that neither $F_{3}$ nor $F_{5}$,
but, rather, the combination $F_{3}^{13}/F_{5}^{6}$ 
characterizes the time evolution of the density profile 
in the neighborhood of the center,
because the dependence on $t_{0}$ vanishes only through
this combination of $F_{3}$ and $F_{5}$. 

Observing that the free-fall time and 
the scale of the inhomogeneity at the time $t=0$ 
are
\begin{eqnarray}
t_{0}&\equiv& \frac{2}{3}F_{3}^{-1/2}, \\
l_{0}&\equiv& \left(\frac{-F_{5}}
{F_{3}}\right)^{-1/2},
\end{eqnarray}
we can easily find that
the following constant, which has the dimension of frequency,
characterizes naked singularity formation
in the LTB spacetime:
\begin{equation}
\omega_{s}\equiv\frac{l_{0}^{6}}{t_{0}^7}.
\end{equation}

\end{document}